\pdfoutput=1
\RequirePackage{ifpdf}
\ifpdf 
\documentclass[pdftex]{sigma}
\else
\documentclass{sigma}
\fi

\usepackage{tikz}
\usetikzlibrary{calc,trees,positioning,arrows,chains,shapes.geometric,%
 decorations.pathreplacing,decorations.pathmorphing,shapes,%
 matrix,shapes.symbols,decorations.markings}

\usepackage{ctable}
\newcommand{\bem}{\begin{pmatrix}}
\newcommand{\eem}{\end{pmatrix}}

\usepackage{multirow}

\usepackage{longtable}
\def\sdprod{{\times\!\vrule height5pt depth0pt width0.4pt\,}}

\def\j{\psi}

\def\l{\lambda}
\def\m{\mu}

\def\t{\tau}

\newcommand{\RR}{{\mathbb R}}
\newcommand{\CC}{{\mathbb C}}
\newcommand{\ZZ}{{\mathbb Z}}
\newcommand{\N}{{\mathbb N}}
\newcommand{\HH}{{\mathbb H}}

\newcommand{\ex}{\operatorname{e}} 

\newcommand{\PSL}{\operatorname{{\rm PSL}}} 

\newcommand{\G}{\Gamma}	

\newcommand{\sm}[4]{\bigl( \begin{smallmatrix} #1&#2\\ #3&#4 \end{smallmatrix} \bigr)}

\def\cA {{\mathcal{A}}}

\def\cD {{\mathcal{D}}}

\def\cF {{\mathcal{F}}}

\def\cH {{\mathcal{H}}}

\def\cM {{\mathcal{M}}}
\def\cN {{\mathcal{N}}}
\def\cO {{\mathcal{O}}} \def\ccO{{\mathcal{O}}}
\def\cP {{\mathcal{P}}}
\def\cR {{\mathcal{R}}}
\def\cS {{\mathcal{S}}}
\def\cT {{\mathcal{T}}}
\def\cU {{\mathcal{U}}}
\def\cV {{\mathcal{V}}}

\def\cY {{\mathcal{Y}}}
\def\cZ {{\mathcal{Z}}}

\def\bbC {\mathbb{C}}

\def\bbH {\mathbb{H}}

\def\bbP {\mathbb{P}}

\def\bbR {\mathbb{R}}
\def\bbT {\mathbb{T}}

\def\var{\varphi}

\numberwithin{equation}{section}

\newtheorem{Theorem}{Theorem}[section]
\newtheorem*{Theorem*}{Theorem}
\newtheorem{Corollary}[Theorem]{Corollary}

 { \theoremstyle{definition}

 }

\begin{document}

\newcommand{\arXivNumber}{2002.11125}

\renewcommand{\PaperNumber}{013}

\FirstPageHeading

\ShortArticleName{Modular Exercises for Four-Point Blocks -- I}

\ArticleName{Modular Exercises for Four-Point Blocks~--~I}

\Author{Miranda C.N.~CHENG~$^{\rm a}$, Terry GANNON~$^{\rm b}$ and Guglielmo LOCKHART~$^{\rm cd}$}

\AuthorNameForHeading{M.C.N.~Cheng, T.~Gannon and G.~Lockhart}

\Address{$^{\rm a)}$~Korteweg--de Vries Institute for Mathematics and Institute of Physics,\\
\hphantom{$^{\rm a)}$}~University of Amsterdam, Amsterdam, The Netherlands}
\EmailD{\href{mailto:mcheng@uva.nl}{mcheng@uva.nl}}
\Address{$^{\rm b)}$~Department of Mathematics, University of Alberta, Canada}
\EmailD{\href{mailto:tjgannon@ualberta.ca}{tjgannon@ualberta.ca}}
\Address{$^{\rm c)}$~Institute of Physics, University of Amsterdam, The Netherlands}
\EmailD{\href{mailto:glockhar@uni-bonn.de}{glockhar@uni-bonn.de}}
\Address{$^{\rm d)}$~CERN, Theory Department, Geneva, Switzerland}

\ArticleDates{Received May 08, 2024, in final form February 12, 2025; Published online February 28, 2025}

\Abstract{The well-known modular property of the torus characters and torus partition functions of (rational) vertex operator algebras (VOAs) and 2d conformal field theories (CFTs) has been an invaluable tool for studying this class of theories. In this work we prove that sphere four-point chiral blocks of rational VOAs are vector-valued modular forms for the groups $\Gamma(2)$, $\Gamma_0(2)$, or $\mathrm{SL}_2(\mathbb{Z})$. Moreover, we prove that the four-point correlators, combining the holomorphic and anti-holomorphic chiral blocks, are modular invariant. In particular, in this language the crossing symmetries are simply modular symmetries. This gives the possibility of exploiting the available techniques and knowledge about modular forms to determine or constrain the physically interesting quantities such as chiral blocks and fusion coefficients, which we illustrate with a few examples. We also highlight the existence of a~sphere-torus correspondence equating the sphere quantities of certain theories~${\mathcal T}_s$ with the torus quantities of another family of theories ${\mathcal T}_t$. A companion paper will delve into more examples and explore more systematically this sphere-torus duality.}


\Classification{81T40; 17B69; 11F03}

\section{Introduction}
Modular forms have been useful since they were known to humankind. Specifically, in the study of two-dimensional conformal field theories (CFTs) and closely related vertex operator algebras (VOAs, often referred to as chiral algebras), modular forms have played an important role. On the VOA side, the torus characters of $\mathcal{V}$-modules are well known to transform as components of a vector-valued modular form (vvmf), when the VOA is rational and satisfies the so-called $C_2$ co-finite condition.
 Moreover, the $S$-matrix of the corresponding modular group representation determines certain crucial data of the VOA -- the fusion coefficients -- via the celebrated Verlinde formula~\eqref{Verl}.
 When combining the chiral and anti-chiral theories into a~physical CFT, the torus partition function is famously modular invariant.
 The role of modular forms in this case can be understood through the role of
the modular group ${\rm SL}_2(\ZZ)$
as the mapping class group of the torus with one puncture. The modularity poses stringent constraints on the spectrum of the theories and has been very valuable in exploring the existence of potential theories~\cite{Afkhami-Jeddi:2019zci,Bae:2017kcl, Cappelli:1986hf,Cappelli:1987xt,Collier:2016cls,Friedan:2013cba,Gannon:1992ty,Hellerman:2009bu}.

On the other hand, four-point conformal blocks (or {\em chiral blocks} more generally) are key quantities of a VOA, as the braiding and {fusing} matrices associated to them essentially determine the representation theory (i.e., the modular tensor category) of the VOA. For physical CFTs, the crossing symmetry of the four-point correlation functions together with the modular invariance on the torus guarantees the consistency of the theory on arbitrary Riemann surfaces. Again, crossing symmetries pose valuable constraints on the theory and form the cornerstone of the conformal bootstrap method.

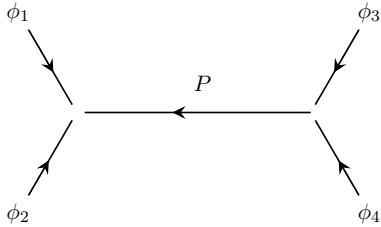
\begin{figure}[t]\centering
\resizebox{.3\textwidth}{!}{
\begin{tikzpicture}
\tikzset{myarr/.style={decoration={markings,mark=at position 1 with {\arrow[scale=2,>=stealth]{>}}},postaction={decorate}}}
\node (M1) {$\phi_1$};
\node (p1) at ($ (M1) + (300:2) $) {};
\node (p11) at ($ (M1) + (300:1.2) $) {};
\node (M2) at ($ (p1) + (240:2) $) {$\phi_2$};
\node (p12) at ($ (p1) + (240:0.8) $) {};
\node (p2) at ($ (p1) + (0:4) $) {};
\node (M3) at ($ (p2) + (60:2) $) {$\phi_3$};
\node (M4) at ($ (p2) + (300:2) $) {$\phi_4$};
\node (p21) at ($ (p2) + (60:0.8) $) {};
\node (p22) at ($ (p2) + (300:.8) $) {};
\node (p3) at ($ (p1) + (0:1.5) $) {};
\node (p4) at ($ (p1) + (0:1.0) $) {};
\node (P) at ($ (p4) + (1.15,.5) $) {$P$};
\draw[myarr] (M1) -- ($(p11)+(0,0)$);
\draw[myarr] (M2) -- (p12);
\draw[-,thick] (p1) -- (M1);
\draw[-,thick] (p1) -- (M2);
\draw[-,thick] (p2) -- (M3);
\draw[-,thick] (p2) -- (M4);
\draw[-,thick] ($(p2)+(0,0)$) -- (p1);
\draw[myarr] (M3) -- ($(p21)+(0,0)$);
\draw[myarr] (M4) -- (p22);
\draw[myarr] (p2) -- (p3);
\end{tikzpicture}}
\caption{The four-point chiral block.}\label{fig:4pt}
\end{figure}

In this work, we first focus on the chiral blocks and make explicit their relation to modular forms. Namely, we prove that the corresponding {\em core blocks}, which we define as the part of the chiral blocks that is not
 dictated by conformal symmetry and that is invariant under the Dehn twists about punctures, form vector-valued modular forms.
They are modular forms for the modular group ${\rm SL}_2(\ZZ)$ when at least three of the external insertions are identical; when two of the external insertions are identical, they are modular forms for its subgroup $\Gamma_0(2)$, and when none are, they are modular forms for its subgroup $\Gamma(2)$. To be more precise, for a~given set of operators $\phi_1\in M_1$, $\phi_2\in M_2$, $\phi_3\in M_3$, $\phi_4\in M_4$, inserted at the four punctures, we obtain a vector-valued modular form, with each component labelled by a module $P$ (the \emph{``internal channel''}) -- cf.\ Figure \ref{fig:4pt}.

 Note that we consider not just the cases of Virasoro blocks but also general chiral blocks when the chiral algebra is an extension of the Virasoro algebra, as is almost always the case. In the case of Virasoro blocks, the relation between four-point conformal blocks and modular forms is already implicit in the work by Zamolodchikov~\cite{Zamolodchikov}. For generic CFTs, each chiral block will involve the contributions from infinitely many Virasoro families (Virasoro primaries and their Virasoro descendents) which are related through the action of the chiral algebra. Here, the appearance of the modular group can again be understood via its relation with the relevant mapping class group (cf.\ Section~\ref{sec:groups}), or alternatively via the geometric picture of a torus as a~double cover of a sphere with four punctures (cf.\ Section~\ref{sec:forms}).
Subsequently, we consider the four-point correlation function by assembling the chiral and anti-chiral sides and appropriately summing over internal channels. Using the language of modular forms, what we do is to combine holomorphic and anti-holomorphic vector-valued modular forms into a modular invariant, non-holomorphic function. In other words, in the language of modular forms the crossing symmetry is nothing but the modular symmetry of the correlation function (with a universal factor removed). This is of course extremely reminiscent of the familiar relation between 2d CFT and modular forms when torus blocks and torus partition functions are considered.

In fact, in some instances the analogy to torus modularity runs deeper. Namely, we find that for certain VOAs $\mathcal{V}_s$ the four-point core blocks associated to a specific choice of external operators $M_*$ coincide with the characters of a different VOA $\mathcal{V}_t$. Moreover, one can combine a~chiral- and anti-chiral copy of the VOAs $\cV_s$ and $\cV_t$ into full CFTs $\cT_s$ and $\cT_t$, and arrive at a CFT version of the correspondence that is also outlined in Table \ref{tab:corr}, where we write $\ast:=\bigl(M_\ast,\widetilde{M_\ast}\bigr)$. We stress that for the same $\mathcal{V}_s$ the core blocks for different choices of ``twist field'' $\mathcal{M}_*$ may be in correspondence with the characters of different VOAs $\mathcal{V}_t$. This is the case for the Ising model, for which the four-point core block associated to the $h=\tfrac{1}{2}$ operator $\epsilon$ coincides with the characters of the $E_{8,1}$ Kac--Moody algebra, while the four-point blocks associated to the $h=\tfrac{1}{16}$ operator $\sigma$ coincide with the characters of the $A_{1,1}$ Kac--Moody algebra.

Clearly, the versatile and powerful arsenal of modular forms should find applications in the computation of sphere four-point chiral blocks and the four-point correlators, just like it has been applied to the torus characters/partition functions to draw conclusions on a wide range of properties of 2d VOAs and CFTs. Indeed, while different techniques such as Zamolodchikov's recursion relation and the BPZ/KZ equations can be employed to determine the conformal blocks for different classes of theories, one of the major advantages of our approach is that it applies uniformly to Virasoro minimal models as well as to any rational conformal field theory with extended chiral algebras, and even to non-rational examples such as Liouville theory with one degenerate external operator, as we will see in Section \ref{sec:examples}.

\newpage

While we focus on laying out and proving the formalism in the present paper, in the companion paper~\cite{followup} we will give more examples to illustrate the consequences of the modularity property.
We will also examine the aforementioned sphere-torus connection between $\cT_t$ and $\cT_s$ more systematically.

\begin{table}[ht!]\centering
\caption{\label{tab:corr}A summary of the sphere-torus correspondence.}
\begin{tabular}{ccc}\toprule
torus && four-punctured sphere \\ \midrule
torus chiral algebra ${\mathcal V}_t$ &$\leftrightarrow$& sphere chiral algebra ${\mathcal V}_s$ \\ \midrule
irreducible module $N$ &$\leftrightarrow$& irreducible module $M(N)$ \\ \midrule
characters $\chi_{N}(\tau)$ &=& core blocks $\var_{M_{\ast},M_{\ast},M_{\ast},M_{\ast}}^{M(N)}(\tau)$\\\midrule
torus S and T matrices ${\bf T}$, ${\bf S}$&=&
sphere S and T matrices ${\mathcal T}$, ${\mathcal S}$ (Theorem \ref{thm:sl2z})\\\midrule
 CFT ${\mathcal T}_t$ &$\leftrightarrow$& sphere CFT ${\mathcal T}_s$\\\midrule
partition function $Z_{T^2}(\tau,\bar\tau)$ &=& correlator ${F}_{\ast,\ast,\ast,\ast}(\tau,\bar\tau)$ \eqref{corr_def2}\\ \midrule
multiplicities ${\mathcal Z}_{N,\tilde N}$ &=& OPE coefficient $({\mathcal A}_{\ast,\ast,\ast,\ast})_{M{(N)}, \tilde M(\tilde N)} $ \eqref{def:cA}\\\bottomrule
\end{tabular}
\end{table}

The plan of the paper is as follows.
In Section~\ref{sec:not}, we list the various notations and symbols used in the paper.
In Section~\ref{sec:groups}, we introduce the various discrete groups related to the mapping class group of spheres with four punctures and clarify the relation among them.
In Section~\ref{sec:bkg}, we review the basic properties of chiral blocks.
In Section~\ref{sec:reps}, we define the notion of {\em core blocks} and investigate their properties.
In particular, in Theorem \ref{thm:corerepmapping} we demonstrate that, unlike the chiral blocks, the core blocks furnish representations of the mapping class group.
In Section~\ref{sec:forms}, we show that the core blocks with given chiral data form vector-valued modular forms, with $S$ and $T$ matrices determined by the braiding and fusing matrices (Theorem \ref{mod_form}).
In Section~\ref{sec:corr}, we turn our attention to the physical correlators and define the core correlators, capturing non-trivial information of the correlators that is not just dictated by conformal symmetry. We then show that the core correlators are non-holomorphic modular forms (see Corollary \ref{cor_corecor}).
In Section~\ref{sec:examples}, we provide some examples illustrating the relevance of modular forms in four-point chiral blocks and correlators.
In Section~\ref{sec:spheretorus}, we elaborate on the correspondence relating sphere and torus blocks.\looseness=-1

 We also include three appendices.
In Appendix \ref{app:BPZ}, the relation between modular differential equations and BPZ equations is exemplified by a detailed treatment in the case of the second-order BPZ equation.
In Appendix \ref{sec:brafu}, we collect several facts about the braiding and fusing operators that we make use of in the main text. In Appendix \ref{app:generalSL2Z}, we collect details about the vector-valued modular forms for general chiral blocks.

While we were preparing this manuscript we learned that~\cite{Maloney:2016kee} has also developed an approach that exploits modularity to constrain the correlators of Virasoro minimal models (the same approach has recently also been extended to WZW models~\cite{Mahanta:2019xnx}). We thank the authors of~\cite{Maloney:2016kee} for considering to coordinate the arXiv submission. While the basic idea of exploiting the relation of the mapping class group to modular groups is the same as for us, the approach and scope of these papers differs significantly. On the one hand, in this paper we develop an approach to determine (holomorphic) conformal blocks of arbitrary rational vertex operator algebras by exploiting their transformations as vector-valued modular forms; physical CFT correlators are then constructed from bilinear combinations of conformal blocks by imposing modular invariance on them. On the other hand, \cite{Mahanta:2019xnx,Maloney:2016kee} work directly at the level of the full CFT, and determine candidate, crossing symmetry invariant, correlation functions by taking a specific choice of conformal blocks as input data and averaging over its images under ${\rm PSL}(2,\mathbb{Z})$.
The focus there is on determining structure constants consistent with crossing symmetry. We comment further on the overlap between these approaches in Section \ref{sec:examples} when we consider specific examples.

\section{Summary of notation}
\label{sec:not}

For convenience to the reader, we summarize our notation in the following table:

\medskip

\centerline{
\begin{tabular}{l l@{\,}}
\toprule
$\mathcal{V}$& The chiral algebra.
\\$\Upsilon=\Upsilon(\mathcal{V})$& The set of irreducible $\mathcal{V} $-modules, also referred to\\
& as the conformal families.
\\$M(n)$& The homogeneous component of $\cV$-module $M$ of $L_0$-grading $n$.
\\$\Sigma_{g,n}$& The Riemann surface of genus $g$ and with $n$ punctures.
\\$\mathcal{D}=(g,n;M_1,\dots,M_n)$& The chiral datum for a surface $\Sigma_{g,n}$, with insertions\\& $M_1,\dots, M_n\in \Upsilon$.
\\$\mathbb{T}_{g,n}$& The Teichm\"uller space of $\Sigma_{g,n}$.
\\$\Gamma^{g,n}$ & The pure mapping class group of $\Sigma_{g,n}$.
\\${\rm F}\Gamma^{g,n}$& The full mapping class group of $\Sigma_{g,n}$.
\\${\rm PBr}_n$& The pure braid group on $n$ strands.
\\${\rm Br}_n$& The full braid group on $n$ strands.
\\$\mathcal{M}_{g,n}$& The Deligne--Mumford compactification of the moduli space\\& of $\Sigma_{g,n}$.
\\$\widehat{\mathcal{M}}_{g,n}$& The extended moduli space.
\\$\widehat \Gamma^{g,n}$& The mapping class group extended by the action of Dehn twists.
\\$\widehat{\mathbb{T}}_{g,n}$& The extended Teichm\"uller space of $\Sigma_{g,n}$.
\\$\widehat{E}_{\mathcal{D}}$ & The vector bundle on $\widehat{\mathcal{M}}_{g,n}$ associated to the chiral datum $\cD$.
\\$E_{\mathcal{D}}$ & The vector bundle on $\mathcal{M}_{g,n}$ associated to the chiral datum $\cD$.
\\$\widehat{R}_{\mathcal{D}}$ & The representation of $\widehat{\Gamma}^{g,n}$ corresponding to the bundle $\widehat{E}_{\mathcal{D}}$.
\\$R_{\mathcal{D}}$ & The representation of ${\Gamma}^{g,n}$ corresponding to the bundle ${E}_{\mathcal{D}}$.
\\$\widehat{\Psi}_{\mathcal{D}}$&
The space of meromorphic sections of $\widehat{E}_{\mathcal{D}}$.
\\${\Psi}_{\mathcal{D}}$&
The space of meromorphic sections of ${E}_{\mathcal{D}}$.
\\$\widehat{\mathfrak{F}}_{\mathcal{D}}(\phi_1,\dots,\phi_n)$&
The space of chiral blocks associated to $\phi_1\in M_1,\dots,\phi_n\in M_n$.
\\$\cN_{\mathcal{D}}$& The rank of $\widehat{E}_\mathcal{D}$.
\\$T,S$& The generators of ${\rm SL}_2(\mathbb{Z})$.
\\$\mathbf{T},\mathbf{S}$& The $T$ and $S$ matrices {acting on the torus blocks}.
\\$\mathbf{C}$& The charge conjugation matrix {acting on the torus blocks}.
\\$(0,n;M_1,\dots,M_n)$& The chiral datum for an $n$-punctured sphere. Also noted\\& as $(M_1,\dots,M_n)$.
 \\
$h_{M}$& The conformal weight (lowest $L_0$ eigenvalue) of $M$.
\\$\Phi^{M_1}_{M_2,M_3}(z)$ & An intertwiner from $M_2\otimes M_3$ into $M_1$.
\\
$\mathcal{Y}^{M_1}_{M_2,M_3}$& The space of intertwiners from $M_2\otimes M_3$ into $M_1$.
\\
\bottomrule
\end{tabular}}

\centerline{\begin{tabular}{l l}
\toprule
$\cN_{(0,3;M_1^*,M_2,M_3)}$& The fusion coefficient.
\\$M^*$ & The $\mathcal{V}$-module dual to $M$.
\\$Y(\phi,z)$& The vertex operator in $\mathcal{Y}^{\cV}_{\cV,\cV}$.
\\$Y^M(\phi,z)$ & The vertex operator in $\mathcal{Y}^M_{\cV,M}$.
\\$\chi_M(\phi,\tau)$ & The torus one-point block (graded character).
\\$\underline{\chi}(\phi,\tau)$ & The vector of graded characters.
\\$\overline{h}_i$& The conformal weight of $\phi_i\in M_i$. \\
$c^\Phi_{\phi_1\phi_2\phi_3}$ & The structure constant. See \eqref{def:3point_intertwiners}.
\\$ w = \displaystyle{z_{12}z_{34}\over z_{13}z_{24}}$ & The cross-ratio.
\\$ \tilde w$ & The coordinate on the universal cover of $\mathbb{P}^1\backslash\{0,1,\infty\}$.
\\$h=\sum_{i=1}^4 \overline{h}_i$& The sum of the conformal weights of external operators.
\\$\mu_{ij}=\frac{h}{3}-\overline{h}_i-\overline{h}_j$& The specific linear combinations of conformal weights of external\\& operators.
\\$\cF^{P}_{\phi_1,\phi_2,\phi_3,\phi_4}$& The four-point chiral block.
\\$\widehat{\Psi}^P_{(M_1,M_2,M_3,M_4)}$& The subspace of $\widehat{\Psi}_{(M_1,M_2,M_3,M_4)}$ isomorphic to $\cY_{M_2,P}^{M_1^*}\otimes \cY_{M_3,M_4}^P$.
\\$\underline{\varphi}$& The vector of core blocks.
\\$\overline \sigma_i$& The generator of ${\rm F}\Gamma^{0,4}$ moving the $i$-th strand over the\\&$(i+1)$-th strand.
\\$\overline \sigma_T$, $\overline \sigma_S$& The generators of ${\rm PSL}_2(\mathbb{Z})\in {\rm F}\Gamma^{0,4}$. See \eqref{PSL2}.
\\$\tilde \sigma_i$& The generators of ${\rm F}\Gamma^{0,4}$, related to $\overline \sigma_i$ by conjugation \eqref{def:changeofbasis}.
\\$B_{PQ}\bigl[\begin{smallmatrix}N_2&N_3\\N_1&N_4\end{smallmatrix}\bigr]\!(\epsilon)$ &The braiding matrices.
\\$F_{PQ}\bigl[\begin{smallmatrix}N_2&N_3\\N_1&N_4\end{smallmatrix}\bigr]$& The fusing matrices.
\\$\xi^{N_1}_{N_2,N_3}$& The skew symmetry \eqref{def:xi} of the intertwiner spaces.
\\$\zeta^{N_1}_{N_2,N_3}$& The isomorphism \eqref{def:zeta} of the intertwiner spaces.
\\$R_i$& The representation matrices \eqref{eq:corerep} corresponding to the action\\& of $\tilde{\sigma}_i\in {\rm F}\Gamma^{0,4}$ on the core blocks.
\\$R_T$, $R_S$& The $T$ and $S$ matrices acting on the core blocks. See\\& \eqref{eq:tpq} and \eqref{eq:spq}.
\\$\cS,\cT$& The $S$ and $T$-matrices of ${\rm PSL}_2(\mathbb{Z})$ acting on the core blocks.
\\${\rm P}\Gamma_0(2)$ & The group $\Gamma_0(2)/\{\pm\}$. See \eqref{eq:projective} and Section \ref{sec:gamma02}.
\\${\rm P}\Gamma(2)$ & The group $\Gamma(2)/\{\pm\}$. See \eqref{eq:projective} and Section \ref{sec:gamma2}.
\\$\cR$&The representation matrix corresponding to the action\\& of $\bigl(\begin{smallmatrix}1 &-1\\2 & -1\end{smallmatrix}\bigr)\in {\rm P}\Gamma_0(2)$ on the core blocks.
\\$\cU$&The representation matrix corresponding to the action\\& of $\bigl(\begin{smallmatrix}1 &0\\-2 & 1\end{smallmatrix}\bigr)\in {\rm P}\Gamma(2)$ on the core blocks.
\\$j(\tau)$& The modular $j$-function.
\\$E_n(\tau)$& The weight-$n$ Eisenstein series.
\\$\Delta(\tau)$& The modular discriminant function.
\\$\eta(\tau)$& The Dedekind eta function.
\\$D_{(k)}: = \frac{1}{2\pi {\rm i}}\frac{{\rm d}}{{\rm d}\tau}-\frac{k}{12}E_2(\tau) $& The modular derivative acting on weight-$k$ modular forms.
\\$D_{(0)}^\ell$& The modular derivative taking weight-0 vvmf's to weight-$2\ell$\\& vvmf's. See \eqref{def:differential operator}.
\\
$\cM^!_{0}(\rho)$& The space of weakly-holomorphic weight-zero vvmf's with\\& multiplier $\rho$.
\\$\cM^!_{0}(\rho;\lambda)$& The subspace of $\cM^!_{0}(\rho)$ whose growth as $\tau\to {\rm i} \infty$ is bounded\\& by $q^\lambda$.
\\$\lambda$& The modular lambda function. See Section \ref{sec:gamma2}.
\\$\theta_i$& The Jacobi theta functions.\\
\bottomrule
\end{tabular}}

\newpage

\begin{tabular}{l l}
\toprule
$\kappa$& The specific Hauptmodul for $\Gamma_0(2)$. See Section \ref{sec:gamma02}.
\\
$\mathcal{H}$& The Hilbert space of a (non-chiral) CFT. \\
$\tilde{\cV}$& The chiral algebra of right-movers of a (non-chiral) CFT.
\\ $\mathcal{Z}_{M,\tilde{M}}$& The multiplicity of the module $M\otimes \tilde{M}$ in $\mathcal{H}$.
\\$\Omega$& The isomorphism between the maximally-extended chiral algebras $\cV$ and $\tilde \cV$.
\\ $\mathbf{\Phi}^i_{j,k}$& A physical vertex operator of the RCFT.
\\$d^i_{jk}$& The OPE coefficients of the RCFT.
\\$\cA_{1234}$& The crossing matrix with entries $d_{12p}d^{p}_{34}$. See \eqref{def:cA}.
\\$F_{\phi^1,\phi^2,\phi^3,\phi^4}$& The four-point core correlator. See \eqref{corr_def2}.
\\$Z_{T^2}$& The torus partition function of the (non-chiral) CFT. \\
\bottomrule
\end{tabular}

\section{Groups}
\label{sec:groups}
The moduli space of genus $g$ surfaces with $n$ punctures is naturally the global orbifold $\bbT_{g,n}/\Gamma^{g,n}$ where $\Gamma^{g,n}$ is the (pure) \textit{mapping class group}, and the universal cover $\bbT_{g,n}$ is the Teichm\"uller space. The group $\Gamma^{g,n}$ fixes the ordering of the $n$ points; if we allow the points to be permuted, then the appropriate group is the \textit{full mapping class group} ${\rm F}\Gamma^{g,n}$. Then $\Gamma^{g,n}$ is to ${\rm F}\Gamma^{g,n}$ as the pure braid group ${\rm PBr}_n$ is to the braid group ${\rm Br}_n$: ${\rm F}\Gamma^{g,n}/\Gamma^{g,n}\cong {\rm Br}_n/{\rm PBr}_n\cong {\rm Sym}(n)$. In~this introductory section we spell out the relation between the mapping class groups $\G^{0,4}$ {and~${\rm F}\Gamma^{0,4}$} and various groups acting naturally on the upper-half plane $\mathbb H$. From this relation we will determine in Section \ref{sec:reps} the modular properties of core blocks.

\begin{figure}[h!]\centering
\includegraphics[width=0.2\textwidth]{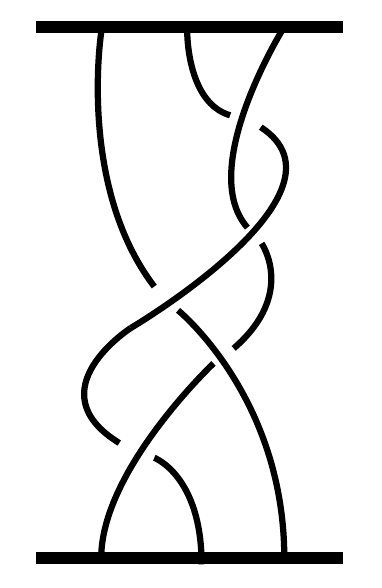}
\caption{A braid $\beta\in {\rm Br}_3$ giving rise to the permutation $(13)\in {\rm Sym}(3)$.}
\label{fig:braid}
\end{figure}

Recall the definitions of the full braid group ${\rm Br}_n$ and the pure braid group ${\rm PBr}_n$ on $n$ strands. The \textit{full} braid group allows any braids. Each braid gives rise to a permutation. For example, in the braid $\beta\in {\rm Br}_3$ shown in Figure \ref{fig:braid} the strand in position 1 at the bottom of the diagram ends up in position 3 at the top of the diagram and vice versa, while the one in position 2 ends up in the same position. In other words, $\beta$ gives rise to the permutation $(13)\in {{\rm Sym}}(3)$. The \textit{pure} braid group ${\rm PBr}_n$ by definition consists of all braids in ${\rm Br}_n$ giving rise to the trivial permutation. It is a normal subgroup of ${\rm Br}_n$, with quotient ${\rm Br}_n/{\rm PBr}_n\cong {{\rm Sym}}(n)$.

We are mainly interested in the case $n=4$:
\begin{equation}
\label{def:br4}
{{\rm Br}}_4
= \langle \overline{\sigma}_1,\overline{\sigma}_2,\overline{\sigma}_3 | \overline{\sigma}_i\overline{\sigma}_{i+1}\overline{\sigma}_i=
\overline{\sigma}_{i+1}\overline{\sigma}_i\overline{\sigma}_{i+1},\,\overline{\sigma}_1\overline{\sigma}_3 = \overline{\sigma}_3\overline{\sigma}_1\rangle,
\end{equation}
where $i=1, 2$ in the above definition,
as it is related to
the full (unextended) mapping class group of a sphere with four punctures in the following way.
Consider the normal subgroup $N$ of ${{\rm Br}}_4$, which is the normal closure of the elements $(\overline{\sigma}_1\overline{\sigma}_2\overline{\sigma}_3)^4$ and
$\overline{\sigma}_1\overline{\sigma}_2\overline{\sigma}_3^2\overline{\sigma}_2\overline{\sigma}_1$. Then~\cite{birman1976braids}
\begin{gather}
\begin{split}
{\rm F}\Gamma^{0,4}&={{\rm Br}}_4/N.
\label{GammaPres}
\end{split}
\end{gather}

This is the \textit{spherical} braid group on four strands, quotiented by its centre (which is $\mathbb{Z}_2$). The generator $\overline{\sigma}_i$ moves the $i$-th strand over the $(i+1)$-th strand,
 and the relations $\overline{\sigma}_1\overline{\sigma}_2\overline{\sigma}_3^2\overline{\sigma}_2\overline{\sigma}_1={(\overline\sigma_1\overline\sigma_2\overline\sigma_3)^4}=1$ reflect the non-trivial topology of the sphere.
On the Riemann sphere with four punctures,
 $\overline{\sigma}_i$ acts by interchanging $z_{i+2}\leftrightarrow
z_{i+3}$ in the orientation
\begin{equation}\label{def:perm_sigma}
z_{i+2,i+3} \mapsto \ex\biggl({1\over2}\biggr)z_{i+2,i+3},\end{equation}
while leaving other two points $z_j$ invariant, where we let the label of the four points $z_i$ be valued in $i \in {\ZZ_4}$. For instance, $\overline{\sigma}_2\colon z_{41} \mapsto \ex\bigl({1\over2}\bigr)z_{41}$.

The above presentation facilitates translating between $\beta\in {\rm F}\Gamma^{0,4}$ and the braid group action.
Moreover,
the mapping class group ${\rm F}\Gamma^{0,4}$ is isomorphic to $(\mathbb{Z}_2\times\mathbb{Z}_2)\sdprod \PSL_2(\mathbb{Z})$, where
the group multiplication is given by $(u,A)(v,B)=(u+Av,AB)$. This fact will be crucial for the rest of the paper. In what follows we write $u,v\in\mathbb{Z}_2\times\mathbb{Z}_2$ as column vectors and we recall that $\PSL_2(\mathbb{Z})={\rm SL}_2(\mathbb{Z})/\{\pm I\}$.
The isomorphism~\cite{farb2011primer}
\begin{gather*}
{\rm F}\Gamma^{0,4} \xrightarrow[]{\cong} (\mathbb{Z}_2\times\mathbb{Z}_2)\sdprod \PSL_2(\mathbb{Z})
\end{gather*}
is realised by the map
\begin{equation}
\label{Matrixreal}\overline{\sigma}_1\mapsto \biggl(\begin{pmatrix} 0 \\ 0 \end{pmatrix},\begin{pmatrix} 1 & 1 \\ 0 & 1 \end{pmatrix}\biggr),\qquad
\overline{\sigma}_2\mapsto\biggl(\begin{pmatrix} 0 \\ 1 \end{pmatrix},\begin{pmatrix} \hphantom{-}1 & 0 \\ -1 & 1 \end{pmatrix}\biggr),\qquad
\overline{\sigma}_3\mapsto\biggl(\begin{pmatrix} 1 \\ 0 \end{pmatrix},\begin{pmatrix} 1 & 1 \\ 0 & 1 \end{pmatrix}\biggr).
\end{equation}
The above presentation of the full mapping class group
${\rm F}\Gamma^{0,4}$ is in some respects more convenient than the one in terms of the braid group. For instance, it is immediate from~\eqref{Matrixreal} that $\overline\sigma_1^2=\overline\sigma_3^2$, a fact which is not manifest in the presentation~\eqref{GammaPres}.

The subgroup $(0,\PSL_2(\mathbb{Z}))$ is generated by the $T$ and $S$ elements, given by
\begin{equation}
\label{PSL2}
\overline{\sigma}_T:=\overline{\sigma}_1\mapsto \biggl(0,\begin{pmatrix} 1 & 1 \\ 0 & 1 \end{pmatrix}\biggr),\qquad \overline{\sigma}_S:= \overline{\sigma}_3\overline{\sigma}_2\overline{\sigma}_3\mapsto \biggl(0,\begin{pmatrix} 0 & -1 \\ 1 & \hphantom{-}0 \end{pmatrix}\biggr),
\end{equation}
where we denote simply by $0$ the identity element of $\ZZ_2\times \ZZ_2$.
This subgroup corresponds
to the elements of $\Gamma^{0,4}$ which fix $z_1$ while permuting the three points $z_2$, $z_3$, $z_4$ arbitrarily: $\overline{\sigma}_1$~permutes~${z_3\leftrightarrow z_4}$ while $\overline{\sigma}_3\overline{\sigma}_2\overline{\sigma}_3$ permutes $z_2\leftrightarrow z_4$.
In terms of the cross ratio, the above generators act as
\begin{equation} \label{STw}
{\overline{\sigma}_T \colon \ w\mapsto { w\over w-1},\qquad
\overline{\sigma}_S \colon \ w\mapsto 1- w}.
\end{equation}
Of course, more important is the fact that $\overline{\sigma}_i$s also act on the universal cover of $\bbP^1\setminus\{0,1,\infty\}$. From this point of view,~\eqref{STw} says that, e.g., $\overline{\sigma}_T$ maps the fibre above $w$ to that above $\frac{w}{w-1}$. We~describe this explicitly in Section~\ref{sec:forms} below.

Finally, the pure mapping class group $\Gamma^{0,4}$, which is the subgroup of ${\rm F}\Gamma^{0,4}$ that does not permute the four points $z_i$, is isomorphic to $(0,{\rm P}\Gamma(2))$, where ${\rm P}\Gamma(2)=\Gamma(2)/\{\pm I\}$.
Recall that~${\rm P}\Gamma(2)$ is the free group
generated by $T^2$ and $S^{-1}T^2S$, represented respectively by
\begin{equation}
\overline{\sigma}_1^2=\overline{\sigma}^2_3\mapsto \biggl(0,\begin{pmatrix} 1 & 2 \\ 0 & 1 \end{pmatrix}\biggr),\qquad
\overline{\sigma}_2^2\mapsto \biggl(0,\begin{pmatrix} \hphantom{-} 1 & 0 \\ -2 & 1 \end{pmatrix}\biggr)\label{Gamma2}.
\end{equation}

Note that ${\rm Sym}(4)\cong {\rm F}\Gamma^{0,4}/\Gamma^{0,4}$ is an extension of
\[{\rm Sym}(\{z_2,z_3,z_4\})\cong {\rm Sym}(3)\cong \PSL_2(\mathbb{Z})/{\rm P}\Gamma(2)\] by
$\mathbb{Z}_2\times\mathbb{Z}_2=\langle (12)(34),(13)(24)\rangle$, which is generated by
\begin{equation}
\overline{\sigma}_1^{-1}\overline{\sigma}_3\mapsto \biggl(\begin{pmatrix} 1 \\ 0 \end{pmatrix},I\biggr)\qquad \text{and} \qquad
\overline{\sigma}_L:= \overline{\sigma}_2\overline{\sigma}_1^{-1}\overline{\sigma}_3\overline{\sigma}_2^{-1}\mapsto \biggl(\begin{pmatrix} 1 \\ 1\end{pmatrix},I\biggr).\label{Z2Z2}
\end{equation}

Here we have made the slightly unconventional notational choice~\eqref{def:perm_sigma}. This choice was made
 so that our choice of $\PSL_2(\ZZ)\subset {\rm F}\Gamma^{0,4}$, described in~\eqref{PSL2}, leaves $z_1$ fixed. This is related to the more common convention in which $\tilde{\sigma}_i$ interchanges $z_{i}\leftrightarrow
z_{i+1}$ via a simple conjugation using the above $\overline\sigma_L$. In other words, we have
\begin{equation}
\label{def:changeofbasis}
\tilde \sigma_{i} := \overline\sigma_L^{-1} \overline\sigma_{i} \overline\sigma_L.
\end{equation}
In particular,
\begin{gather*}
\tilde \sigma_{1}= \overline\sigma_{3}, \qquad
\tilde \sigma_{3}= \overline\sigma_{1}.
\end{gather*}
Finally, we note that $\overline \sigma_S$ is invariant under this conjugation, and we have
\begin{gather*}
\overline{\sigma}_S:= \overline{\sigma}_3\overline{\sigma}_2\overline{\sigma}_3 = \tilde{\sigma}_3\tilde{\sigma}_2\tilde{\sigma}_3.
\end{gather*}

The subgroup $\PSL_2(\mathbb{Z})$ in ${\rm F}\Gamma^{0,4}$ is related to the ${\rm SL}_2(\mathbb{Z})$ in $\Gamma^{1,0}=\Gamma^{1,1}$ through a construction which will play a large role in Section \ref{sec:spheretorus} below. Consider the torus $\bbC/(\mathbb{Z}\tau+\mathbb{Z})$, with fundamental domain the parallelogram with vertices at $0$, $1$, $\tau+1$, $\tau$. Folding the torus by the map $z\mapsto-z$, we get a fundamental domain with vertices $0$, $1$, $\frac{\tau}{2}+1$, $\frac{\tau}{2}$, where the edge $0\rightarrow\frac{1}{2}$ is identified with the edge
$\frac{1}{2}\rightarrow 1$, $ 0 \rightarrow\frac{\tau}{2}$ is identified with $1\rightarrow\frac{\tau}{2}+1$, etc. Performing these identifications, we obtain a sphere. There are 4 fixed-points of this $\mathbb{Z}_2$ action: the classes [0], $\bigl[\frac{1}{2}\bigr]$, $\bigl[\frac{\tau}{2}\bigr]$ and~$\bigl[\frac{\tau}{2}+1\bigr]$. Put another way, the torus is a double cover of the 4-punctured sphere.

An element ${\gamma}\in{\rm SL}_2(\mathbb{Z})={\rm F}\Gamma^{1,0}$ commutes with that $\mathbb{Z}_2$-action, and so projects to an element $\bar{\gamma}\in\PSL_2(\mathbb{Z})\subset {\rm F}\Gamma^{0,4}$. The subgroup $\mathbb{Z}_2\times\mathbb{Z}_2\subset {\rm F}\Gamma^{0,4}$ is generated by the maps \smash{$z\mapsto z+\frac{1}{2}$} and~\smash{$z\mapsto z+\frac{\tau}{2}$}; these conformal maps permute the fixed points so certainly cannot be deformed to the identity. They lift to the (conformal) maps on the torus given by the same formulas. However, on the torus these translations can be continuously deformed to the trivial one, and so $\mathbb{Z}_2\times\mathbb{Z}_2$ lift to the trivial element of the mapping class group of the torus.

\section{The theory of chiral blocks}
\label{sec:bkg}

The \emph{core blocks} of 2d conformal field theories which are the main object of this paper are closely related to the more standard notion of chiral blocks,\footnote{The chiral blocks are also often called conformal blocks, but we use the term \emph{chiral blocks} to stress that they are usually not Virasoro conformal blocks and can in general correspond to {fields} with respect to an extended chiral algebra {(VOA)}.} and roughly speaking are obtained from them by isolating the part that is not dependent on Dehn twists around punctures. In this section, we therefore provide a pedagogical review of the theory of chiral blocks, setting up along the way our conventions and notations. As usual, we will focus on chiral blocks and correlation functions of primary operators of the chiral algebra.


For now, we focus on the chiral half of the theory and let $\cV$ be a \textit{rational vertex operator algebra}, or RCFT. Let $\Upsilon=\Upsilon(\cV)$ denote the
(finite) set of irreducible $\cV$-modules.
For example, we have $\cV\in\Upsilon$, corresponding to the vacuum sector consisting of the identity operator and its descendants. Any $\cV$-module $M$ decomposes into a direct sum of $L_0$-eigenstates. In other words, we write
\begin{equation}\label{eqn:decomposition}
M=\bigoplus_{n=0}^\infty M(n),\qquad L_0|_{M(n)}=(n+h_M){\rm id},
\end{equation}
where $M(0)\ne 0$ and $h_M$ is called the \textit{conformal weight} of $M$. Here, $L_0$ is an operator coming from a copy of the Virasoro algebra contained in $\cV$.

 By a \textit{chiral datum} we mean a choice
$\cD=(g,n;M_1,\dots,M_n)$, where $g\ge 0$ refers to the~{genus} of a surface $\Sigma_{g,n}$, $n\ge 0$ refers to the
number of (distinct) punctures on $\Sigma_{g,n}$, which we name~1 to~$n$, and where the ``insertion'' $M_i\in\Upsilon$ is assigned to the $i$-th point. The moduli space of genus~$g$ surfaces with~$n$ punctures is compactified by including surfaces with nodal (``pinched'') singularities. This is called the Deligne--Mumford compactification, which we will denote by~$\cM_{g,n}$. These compactification points play the same role here as cusps do in the theory of modular forms, but with the additional feature of the factorization formulas which describe what happens to the chiral blocks as we tend to these ``cusps''; we will return to this point later in this section.

In CFT, a fundamental operation is sewing together surfaces at the punctures. This requires us to specify additional data. This can be done in many ways, but most convenient for the VOA language is Vafa's picture~\cite{Vafa:1987} (described in fuller detail in~\cite{HuangBook}), which includes a local coordinate $f_i$ at each puncture. In this paper we focus on the surface $\Sigma$ being the Riemann sphere ${\bbP}^1=\bbC\cup\{\infty\}$, in which case the $f_i$ can be expressed as functions of the local coordinate~$z$ of the sphere:
\begin{gather*}
f_i(z)=\sum_{n=1}^\infty a^{(i)}_n (z-z_i)^n \qquad \text{when $z_i\in\bbC$ and} \\
f_i(z)=\sum_{n=1}^\infty a^{(i)}_n z^{-n} \qquad \text{when $z_i=\infty$},
\end{gather*}
where all \smash{$a_n^{(i)}\in\bbC$} and \smash{$a^{(i)}_1\ne 0$}.

Conformal equivalences between these extended surfaces (i.e., surfaces with local coordinates at punctures) can be defined in the obvious way, giving an extended moduli space \smash{$\widehat{\cM}_{g,n}$}. The relevant mapping class group
 then becomes the \textit{extended mapping class group} $\widehat{\Gamma}^{g,n}$, which includes the Dehn twists corresponding to rotating about the punctures. Thus $\widehat{\Gamma}^{g,n}$ is an extension of~${\Gamma}^{g,n}$ by $n$ copies of $\mathbb{Z}$. The extended Teichm\"uller space $\widehat{\mathbb{T}}_{g,n}$ is now infinite-dimensional, namely~$\mathbb{T}_{g,n}$ with a copy of the set $\{\sum_{n=1}^\infty a_n z^n \mid a_n\in\bbC,\,a_1\ne 0\}$ attached to each puncture, and \smash{$\widehat{\cM}_{g,n}$} is~given~by~\smash{$\widehat{\mathbb{T}}_{g,n}/\widehat{\Gamma}^{g,n}$}.

To any chiral datum $\cD$, the CFT associates a finite rank holomorphic flat vector bundle \smash{$\widehat{E}_\cD\rightarrow\widehat{\cM}_{g,n}$}, whose
 rank $\cN_\cD$ is given by \textit{Verlinde's formula}~\cite{Verlinde:1988sn} which for arbitrary $\cD$ reads~\cite{Moore:1989vd}:%
\begin{equation}
\label{Verl}
\cN_{(g,n;M_1,M_2,\dots,M_n)}:=\sum_{N\in\Upsilon}(\mathbf{S}_{\cV N})^{2(1-g)}\frac{\mathbf{S}_{M_1N}}{\mathbf{S}_{\cV N}}\cdots\frac{ \mathbf{S}_{M_nN}}{ \mathbf{S}_{\cV N}}
,
\end{equation}
where $\mathbf{S}$ is the $S$-matrix associated with the torus chiral blocks
which will be defined shortly. The bundle \smash{$\widehat{E}_{\cD}$} is trivial over
\smash{$\widehat{\mathbb{T}}_{g,n}$}, and corresponds to a representation \smash{$\widehat{R}_\cD$} of the {extended} mapping class group \smash{$\widehat{\Gamma}^{g,n}$}. We denote by \smash{$\widehat{\Psi}_\cD$} the space of meromorphic sections of \smash{$\widehat{E}_\cD$}. These sections can be lifted to \smash{$\widehat{\mathbb{T}}_{g,n}$}, where they become a vector-valued meromorphic function on {$\widehat{\mathbb{T}}_{g,n}$}, covariant (automorphic) with respect to \smash{$\widehat{\Gamma}^{g,n}$}. In fact, in this paper we will mostly focus on the closely-related, but much simpler holomorphic flat vector bundle $E_\cD\rightarrow \cM_{g,n}$ of the same rank, corresponding to a representation $R_\cD$ of $\Gamma^{g,n}$ whose meromorphic sections we denote by $\Psi_\cD$.

A chiral block is, among other things, a multilinear map $M_1\otimes_\bbC\cdots\otimes_\bbC M_n\rightarrow {\widehat{\Psi}_\cD}$.
For a fixed choice of states $\phi_i\in M_i$, denote the space of chiral blocks at $(\phi_1,\dots,\phi_n)$ by $\widehat{{\mathfrak F}}_\cD(\phi_1,\dots,\phi_n)$.
Hence the dimension of {$\widehat{\mathfrak F}_\cD(\phi_1,\dots,\phi_n)$}
 is bounded above (there may be accidental linear dependencies depending on specific choices of $\phi_i$) by the rank of the vector bundle
\begin{gather*}
{\dim \bigl(\widehat{\mathfrak F}_\cD(\phi_1,\dots,\phi_n)\bigr)\leq {\rm rank}\bigl(\widehat{E}_\cD\bigr)= \cN_\cD}
\end{gather*}
 computed as in~\eqref{Verl}. Chiral blocks obey certain factorization formulas that dictate the behavior as Riemann surface approaches a nodal surface.
A standard way to determine the space \smash{$\widehat{\mathfrak F}_\cD$} of chiral blocks in a CFT is as solutions to differential equations such as the BPZ \cite{Belavin:1984vu} or KZ~\cite{Knizhnik:1984nr} equations, which result from the presence of null states in $M_i$ and have the interpretation of modular differential equations in our contexts as discussed in Section \ref{sec:forms} and Appendix \ref{app:BPZ}. From this point of view, the corresponding representation {$\widehat{R}_\cD$} for the mapping class group is given by the monodromy of the differential equations.

Let $M$ be an irreducible $\cV$-module; as such, $M$ restricts to a representation of the Virasoro algebra contained in $\cV$. By a Virasoro-primary we mean
a lowest-weight state $\phi\in M$ of some Virasoro-subrepresentation of $M$, i.e., $L_k\phi=0$ for all $k>0$, and $\phi$ is an eigenvalue for $L_0$. Given a Virasoro-subrepresentation of $M$, the primary is uniquely determined up to a~nonzero scalar factor. {By analogy,} we say that $\phi\in M$ is a $\cV$-primary if it is a lowest {conformal} weight state for the action by $\cV$, i.e., $\phi$ is a nonzero vector in $M(0)$. Regardless of the chiral algebra~$\cV$, chiral blocks with arbitrary insertions $\phi_i\in M_i$ can be determined as linear combinations of chiral blocks with Virasoro-primary insertions and their derivatives. For this reason it suffices to restrict the insertions to be Virasoro-primaries.
 Note however that each $M$ will have infinitely many inequivalent Virasoro primaries, {with ever-increasing conformal weights,} unless $\cV$ is a~Virasoro minimal model. {On the other hand, there are only finitely many linearly independent $\cV$-primaries.
Unfortunately,} knowing the chiral blocks only for $\cV$-primaries is not sufficient {in general} to know the chiral blocks for all insertions -- we give an example of this in a few paragraphs.

As mentioned earlier, the full mapping class group permutes the punctures. So if for example, the modules $M_i$ are pairwise distinct, it may be more convenient to restrict to the (pure) mapping class group.
On the other hand, in the special cases where all $M_i$ are equal, the relevant mapping class group will be the full mapping class group ${\rm F}\Gamma^{g,n}$, since in these cases the punctures can clearly be taken as indistinguishable. More generally, when some but not all $M_i$ will be equal, the most natural group will be between the full and the pure.

The key ingredient of chiral blocks are {\em intertwiners}~\cite{FHL:1993}. Choose any irreducible ${\mathcal V}$-modules
 $M_i\in\Upsilon$, with conformal weights $h_{M_i}$ respectively, for $i\in\{1,2,3\}$. Schematically, we can represent the intertwiner \smash{$\Phi_{M_2,M_3}^{M_1}(z)$} by the diagram of Figure \ref{fig:3pt}. For each $\phi_2\in M_2$, it has decomposition
\begin{equation}
\Phi_{M_2,M_3}^{M_1}(\phi_2,z) = z^{h_{M_1}-h_{M_3}-h_{M_2}}\sum_{n\in\mathbb{Z}}\bigl(\Phi_{M_2,M_3}^{M_1}(\phi_2)\bigr)_n z^{-n-1},
\label{eq:inter}
\end{equation}
 where the mode \smash{$\bigl(\Phi_{M_2,M_3}^{M_1}(\phi_2)\bigr)_n$} is a linear map $M_3(k)\rightarrow M_1({j+k-n-1})$ when $\phi_2\in M_2(j)$, and where we have used the grading by the $L_0$-eigenvalues~\eqref{eqn:decomposition}.
 We denote by \smash{$\mathcal{Y}_{M_2,M_3}^{M_1}$} the space of all such intertwiners.\footnote{To make the index conventions consistent throughout the paper, our definition of the intertwiners differs slightly from the one given, e.g., in~\cite{Moore:1989vd}.} Its dimension is given by the rank
\smash{$\cN_{M_2,M_3}^{M_{1}}:=\cN_{(0,3;M_1^\ast,M_2,M_3)}$}, which bears the name \textit{fusion coefficient}. The fusion coefficients define the
associative, commutative fusion ring on the $\mathbb{Z}$-span of $\Upsilon$, by
\begin{gather*}
M\otimes M'=\sum_{M''\in\Upsilon}\cN_{M,M'}^{M''}M'',
\end{gather*}
and can be computed by the Verlinde formula~\eqref{Verl}:
\begin{equation}
\label{Verfor:3point}
\cN^{M''}_{M,M'}=\sum_{N\in\Upsilon}\frac{\mathbf{S}_{MN}\mathbf{S}_{M'N}{\bigl(\mathbf{S}^{-1}\bigr)_{M''N}}}{\mathbf{S}_{\cV N}}.
\end{equation}

Examples of intertwiners are the vertex operators \smash{$Y(\phi,z)\in \mathcal{Y}_{\cV,\cV}^\cV$} and \smash{$Y^M(\phi,z)\in\mathcal{Y}_{\cV,M}^M$} corresponding to the VOA itself and to its modules $M$. Indeed, \smash{$\cN^\cV_{\cV,\cV}=\cN^M_{\cV,M}=1$}.

\begin{figure}[t]\centering
\begin{tikzpicture}
\tikzset{myarr/.style={decoration={markings,mark=at position 1 with %
 {\arrow[scale=2,>=stealth]{>}}},postaction={decorate}}}
\node (M2) {$M_2$};
\node[below left=2cm of M2] (M1) {$M_1$};
\node[below right =2cm of M2] (M3) {$M_3$};
\draw[-,thick] (M3) -- (M1);
\draw[-,thick] (M2) -- ($(M1)+(M3)+(0,2)$);
\draw[myarr] (M2) -- ($(M1)+(M3)+(0,2.8)$);
\draw[myarr] (M3) -- ($(M1)+(M3)+(1,2)$);
\draw[myarr] (M3) -- ($(M1)+(M3)+(-1,2)$);
\end{tikzpicture}
\caption{Diagram corresponding to the intertwiner $\Phi_{M_2,M_3}^{M_1}$.}
\label{fig:3pt}
\end{figure}

We will shortly see several examples of how chiral blocks are built from the intertwiners. Here we recall some of their basic properties.
Being (components of) quantum fields, they are operator-valued distributions on the sphere, and as such cannot be evaluated at a point.
{For instance, strictly speaking $\Phi(\phi,0)$ or $\Phi(\phi,\infty)$ does not exist. An important role here is played by the local coordinates $f_i$: an intertwiner $\Phi\in\mathcal{Y}_{M_2,M_3}^{M_1}$ will arise in our expressions as the formal expression $\Phi(\phi_2,f_2(z))$ where $f_2(z)$ is the chosen local coordinate about the puncture $z_2$.}
Creating the state $\phi_2$ at time $t=-\infty$, which then propagates to the scattering virtual event, corresponds to attaching to the world-sheet a semi-infinite tube; Vafa's picture conformally maps a world-sheet with $n$ semi-infinite tubes, to a surface with $n$ punctures. The end of the~$i$-th tube corresponds to the puncture $z_i$, and $f_i$ is the local coordinate on the tube.
Now, we can relate $\Phi(\phi_2,f_2(z))$ to $\Phi(\phi_2,z-z_2)$, using the transformation laws derived in, e.g.,~\cite{Gaberdiel:1994}: \smash{$\Phi(\phi_2,f_2(z))(f_2'(z))^k=T_{f_2}\circ\Phi(\phi_2,z-z_2)\circ T_{f_2}^{-1}$}.
This transformation $T_{f_2}$ induces a change of VOA structure, including Virasoro representation, to a different but equivalent one --
we will recall the famous example with $f(z)=\mathrm{e}(z)-1$ in the next subsection.
 This ability to change local coordinates allows us to restrict attention from now on to the standard choices of local coordinates, namely $f_i(z)=z-z_i$ when $z_i\in\bbC$ and $f_i(z)=1/z$ when $z_i=\infty$. This should be very natural and familiar: we study a function locally at a point $z_0$ say through expansions in powers of $z-z_0$.
Eventually, we will restrict attention to $f_i\to 0$.

In the remainder of this section, we review concrete examples of chiral blocks on the torus and on the sphere.

\textbf{Torus one-point functions.} The most familiar example of chiral blocks are the 1-point functions on the torus, i.e., the case $\cD=(1,1;\cV)$. The surface $\Sigma$ here is the torus $\bbC/(\mathbb{Z}+\tau\mathbb{Z})$. We can take the puncture $z_1$ to be at $[0]\in\Sigma$, and take the local coordinate to be the standard $f_1(z)= [z]\in\Sigma$ mapping from a small disc about $0\in\bbC$ to a small patch about $[0]\in\Sigma$. Since $h_\cV=0$, the action of the Dehn twist about the single puncture is trivial, and as a result the chiral
 blocks transform under a representation of the mapping class group $\Gamma^{1,1}\cong {\rm SL}_2(\mathbb{Z})$.
In~this~case, the dimension of the space of the chiral blocks is given by~\eqref{Verl} to be $\sum_{M\in\Upsilon} 1=\|\Upsilon\|$.
This suggests that a basis of the space \smash{$\widehat{\mathfrak F}_{1,1;\cV}(\phi)$}, for each state $\phi\in\cV$, is given as follows:
for each~${M\in\Upsilon}$, we take the corresponding basis element to be the 1-point function
\begin{equation}
\label{1ptfunc}\chi_{M}(\phi,\tau)=
\ex(kz)\operatorname{tr}_M Y^M(\phi,\ex(z)-1)q^{L_0-c/24}=q^{h_M-c/24}\sum_{n=0}^\infty \operatorname{tr}_{M(n)}o(\phi)q^n.
\end{equation}

Here and elsewhere, we write $\ex(x):={\rm e}^{2\pi {\rm i} x}$ for $x\in \mathbb C$. Moreover, $q=\ex(\tau)$ where $\bbC/(\mathbb{Z}+\mathbb{Z}\tau)$ is our torus, $c$ is the central charge of ${\mathcal V}$, and $h_{M}$ is the conformal weight as before. $Y^M$ is the vertex operator.
Implicit in~\eqref{1ptfunc} is the change of variables $z\mapsto \ex(z)-1$, which is the source of the pervasive $-c/24$. Let $L[n]$ denote the new Virasoro operators, whose corresponding conformal vector is proportional to $\omega-c/24$. Equation~\eqref{1ptfunc} requires $\phi$ to be an eigenvector of the energy operator $L[0]$, i.e., $L[0]\phi=k\phi$. The zero-mode $o(\phi)=\phi_{k-1}$ of the state $\phi$ acts linearly on each space $M(n)$, and ``tr'' denotes its trace.
 The appearance of $z$ in the middle of~\eqref{1ptfunc} is just the ever-present dependence of the chiral blocks on the local coordinates, although here the dependence turns out to be trivial.

These 1-point functions $\chi_M(\phi,\tau)$, also sometimes referred to as graded characters, are described in more detail in~\cite[Section 5.3]{GannonBook}: like finite group characters, they are built from traces of matrices, and uniquely identify the modules $M$ up to equivalence.
They are the components of a
weakly holomorphic vector-valued modular form for ${\rm SL}_2(\mathbb{Z})$ of weight $k$. This means each~$\chi_M(\phi,\tau)$
is holomorphic for $\tau\in\bbH$,
grows at worst exponentially in $\operatorname{Im}(\tau)$ as $\tau\to {\rm i} \infty$
 (i.e., when $q\to 0$), and transforms into a linear combination of $\chi_{M'}(\phi,\tau)$, $M'\in \Upsilon$, under ${\rm SL}_2(\mathbb{Z})$.
Organising $\chi_M$, $M\in \Upsilon$ into a vector {$\underline{\chi}$},
the resulting $S$-matrix $\mathbf{S} = R_{(1,1;\cV)}\bigl(\bigl(\begin{smallmatrix} 0 & -1 \\ 1 & \hphantom{-} 0 \end{smallmatrix}\bigr)\bigr)$, satisfying
\begin{gather*}\label{chiS}
{\underline{\chi}(\phi,\t)=\tau^{-k}\mathbf{S}^{-1} \underline{\chi}\bigl(\phi,-\tfrac{1}{\t}\bigr),}
\end{gather*}
 is precisely that appearing in Verlinde's formula~\eqref{Verl}, and the $T$-matrix $\mathbf{T}=R_{(1,1;\cV)}\bigl(\bigl(\begin{smallmatrix}1 & 1 \\ 0 & 1 \end{smallmatrix}\bigr)\bigr)$ is diagonal with {$M$-th entry equal to $\ex\bigl(h_{M}-\tfrac{c}{24}\bigr)$}.

The familiar special case $\phi={\mathbf{1}} \in {\mathcal V}$, the vacuum, recovers the so-called
characters ({although ``graded dimensions'' is arguably a more accurate name for them})
\begin{equation}
\label{char}
\chi_M(\tau)=\chi_M({\mathbf{1}},\tau)=q^{h_M-c/24}\sum_{n=0}^\infty \dim ({M}(n)) q^n
\end{equation}
of the irreducible $\cV$-modules $M$. However, it is important to allow for other choices for $\phi$. For one thing, there is in general no way to recover the other $\chi_M(\phi,\tau)$ from the $\chi_M(\tau)$, except in the very special case of the Virasoro minimal models. Moreover, just as the dimension of a Lie algebra or finite group representation rarely uniquely identifies it whereas its character always does, it is the 1-point functions $\chi_M(\phi,\tau)$ and not the graded dimensions $\chi_M(\tau)$ which uniquely determine $M$ as a ${\mathcal V}$-module. For example, the representation $R_{(1,1;\cV)}$ decomposes into a direct sum $R_+\oplus R_-$, where $R_\pm(-I)=\pm I$, and $\underline{\chi}(\phi,\tau)$ only sees $R_{(-1)^k}$, so the graded dimensions~$\underline{\chi}(\tau)$, like that of $\underline{\chi}(\phi,\tau)$ for any fixed $\phi\in\cV$, can only see roughly half of $R_{(1,1;\cV)}$.
More precisely, all $\underline{\chi}(\phi,\tau)$ transform with respect to the same representation $R_{(1,1;\cV)}$, but there exists a basis which diagonalizes the charge-conjugation matrix $\mathbf{C}=\mathbf{S}^2$ (with $\dim R_+$ +1's and $\dim R_-$ $-1$'s as eigenvalues). When expressed in this basis, the $\dim R_-$ (resp.\ $\dim R_+$) components of $\underline{\chi}(\phi,\tau)$ will vanish when the insertion field $\phi$ has even (resp.\ odd) conformal weight.

The origin of the aforementioned modular property can be understood in the following way.
The extended mapping class group $\widehat{\Gamma}^{1,1}$ is the braid group ${\rm Br}_3=\langle\sigma_1,\sigma_2| \sigma_1\sigma_2\sigma_1=\sigma_2\sigma_1\sigma_2\rangle$.
It relates to the modular group $({\rm P}){\rm SL}_2(\mathbb{Z})$ by
${\rm Br}_3/\langle Z\rangle\cong\PSL_2(\mathbb{Z})$ and ${\rm Br}_3/\bigl\langle Z^2\bigr\rangle\cong {\rm SL}_2(\mathbb{Z})$, where $Z=(\sigma_1\sigma_2)^3$ generates the center of ${\rm Br}_3$. The Dehn twist about the puncture equals $Z^{-2}$. Because the Dehn twists act trivially on the chiral blocks, the action of~$\widehat{\Gamma}^{1,1}$ collapses to that of~$\Gamma^{1,1}$. See Figure \ref{fig:braid} and Section~\ref{sec:groups} for definitions and more detailed discussions about the braid groups.

Let us consider the 2d Ising model as a concrete example. It has central charge $c=1/2$ and
 three irreducible modules, which we label $\Upsilon=\{1,
\sigma,\epsilon\}$. Note that what we denote by 1 is really the chiral algebra $\cV$ itself.
The nontrivial fusion coefficients of the Ising model are given by
\begin{gather*}
\epsilon\otimes \epsilon=1,\qquad \epsilon\otimes \sigma=\sigma,\qquad \sigma\otimes\sigma=1+\epsilon.
\end{gather*}
 For the Ising model, as for any Virasoro minimal model, it suffices to consider their graded dimensions (as opposed to more general torus 1-point block), which are
 \begin{gather*}\underline{\chi}(\tau)=\begin{pmatrix}\chi_1(\tau)\\ \chi_\sigma(\tau)\\ \chi_\epsilon(\tau)\end{pmatrix}=\begin{pmatrix}q^{-1/48}\bigl(1+q^2+q^3+2q^4+2q^5+3q^6+\cdots\bigr)\vspace{1mm}\\
q^{1/24}\bigl(1+q+q^2+2q^3+2q^4+3q^5+4q^6+\cdots\bigr)\vspace{1mm}\\
q^{23/48}\bigl(1+q+q^2+q^3+2q^4+2q^5+3q^6+\cdots\bigr)\end{pmatrix}
\end{gather*}
From this, we can read off the conformal weights $h_1=0$, $h_\sigma=1/16$,
$h_\epsilon=1/2$ {(cf.\ \eqref{char})}, and the $T$-matrix is given by
\begin{gather*}
\mathbf{T}=\begin{pmatrix}\ex\bigl(-\tfrac{1}{48}\bigr)&0&0\\[0.5mm] 0&\ex\bigl(\tfrac{1}{24}\bigr) &0\\[0.5mm] 0&0&\ex\bigl(\tfrac{23}{48}\bigr)\end{pmatrix},
\end{gather*}
while the $S$-matrix is
\begin{gather*}
\mathbf{S}={1\over 2}\begin{pmatrix}1&\hphantom{-}\sqrt{2}&\hphantom{-}1\\ \sqrt{2}&\hphantom{-} 0&-\sqrt{2}\\ 1&-\sqrt{2}
&\hphantom{-}1\end{pmatrix}.
\end{gather*}
By comparison, $\underline{\chi}(\phi,\tau)=0$ for any $\phi$ of odd conformal weight, and
$\underline{\chi}(\omega,\tau)=\bigl(q\frac{{\rm d}}{{\rm d}q}+\frac{c}{24}\bigr)\underline{\chi}(\tau)$ where $\omega\in\cV(2)$ is the conformal vector. Note that $\underline{\chi}(\omega,\tau)$ is not a vector-valued modular form, since $\omega$ is not an eigenstate of $L[0]$.

Next we turn to the punctured spheres. To avoid clutter we write $(M_1,\dots,M_n)$ instead of~$(0,n;M_1,\dots,M_n)$ throughout the paper. As mentioned already, we restrict to the standard local coordinates $z-z_i$ and $1/z$, where $z$ is the global coordinate on $\bbC$. As a vector space, the homogeneous space $M^\ast(k)$ is naturally identified with the dual vector space $M(k)^*$. For~${x\in M^\ast(k)}$ and~$y\in M(k)$, it is common to write $\langle y|x\rangle$ for the evaluation map $x(y)\in\bbC$. The permutation $M\mapsto M^\ast$ coincides with the involution $\mathbf{C}=\mathbf{S}^2\colon \Upsilon \to \Upsilon.$ Now define ${\mathbf{1}}^*\in\cV(0)^*$ so that~${{\mathbf{1}^*(\mathbf{1})}=1}$ and ${\mathbf{1}}^*(\cV(k))=0$ for $k>0$. Then for any composition $F$ of intertwiners sending~$\cV$ to~$\cV$, we will write $\langle \mathbf{1},F\mathbf{1}\rangle=\mathbf{1}^*(F(\mathbf{1}))$. Likewise, for any composition $F$ of intertwiners mapping $M'$ to $M^\ast$, we also write $\langle y|F|x\rangle=(F(x))(y)$ for any $x\in M'$ and $y\in M$.

Fix a choice of $N_i \in \Upsilon$ with $N_1=M_1^\ast$, $N_{n-1}=M_n$, and
 \smash{$\cN_{M_1,N_1}^\cV,\cN_{M_2,N_2}^{N_1},\dots,\allowbreak \cN_{M_n,\cV}^{N_{n-1}}$} all nonzero (or else the chiral blocks are identically zero), and let
\[
\Phi_1\in \cY^\cV_{M_1,N_1} , \qquad \Phi_i\in \cY^{N_{i-1}}_{M_i,N_i} \quad \text{for~${1<i<n}$}\qquad \text{and}\qquad
\Phi_{n}\in \cY^{N_{n-1}}_{M_n,\cV}.
\]
 With this choice, the chiral blocks for $\cD=(M_1,\dots,M_n)$, with insertions $\phi_i\in M_i(k_i)$ at $z_i\in \bbC$, can be written
\begin{equation}\bigl\langle \mathbf{1},\Phi_1\bigl(\phi_1,z'_1\bigr)\circ\Phi_2\bigl(\phi_2,z'_2\bigr)\circ\cdots\circ \Phi_n\bigl(\phi_n,z'_n\bigr)\mathbf{1}\bigr\rangle\,{{\rm d}z_1^{\prime\, \bar{h}_1}\cdots {\rm d}z_n^{\prime\, \bar{h}_n}},\label{nptsphere}
\end{equation}
where $z_i'=z-z_i$ {and $\bar{h}_i=h_{M_i}+k_i$.} As always, if one of the punctures has $z_i=\infty$, then we~replace $z'_i$ with $1/z$.
Note that $\Phi_1$ can be fixed by requiring that $\langle\mathbf{1},\Phi_1(\phi_1,z)x\rangle=\smash{z^{-2\bar{h}_1}x(\phi_1)}$ for all~${x\in M_1^\ast(k_1)}$; similarly, $\Phi_n$ is fixed by requiring that $\langle y,\Phi_n(\phi_n,z)\mathbf{1}\rangle=\langle y,\phi_n\rangle$ for all~${y\in M_n(k_n)}$.
Note that $\cY_{M_1,M_1^\ast}^\cV$ and $\cY_{M_n,\cV}^{M_n}$ are both one-dimensional, with a canonical~basis~-- namely \smash{$\langle\mathbf{1},\Phi_1(\phi_1,z)x\rangle=z^{-2\bar{h}_1}x(\phi_1)$} for all ${x\in M_1^\ast(k_1)}$. Similarly, $\Phi_n$ chosen by skew symmetry applied to the module vertex operator (see \cite[equation (5.4.33)]{FHL:1993}).

With this canonical choice, \eqref{nptsphere} simplifies to
\begin{equation}
z_1^{\prime -2\bar{h}_1}\bigl\langle \phi_1\big|\Phi_2\bigl(\phi_2,z'_2\bigr)\circ\cdots\circ \Phi_{n-1}\bigl(\phi_{n-1},z'_{n-1}\bigr)\phi_n\bigr\rangle\, {\rm d}z_1^{\prime\, \bar{h}_1}\cdots {\rm d}z_n^{\prime\, \bar{h}_n}. \label{nptsphere2}
\end{equation}
In VOA language, $\Phi(\phi,z)$ is merely a formal expansion, though the combination~\eqref{nptsphere2} is a~meromorphic multi-valued function of the $z'_i$, with branch-points and poles only when some $z_i'=z_j'$ {(equivalently, $z_i=z_j$). These singularities arise because the intertwiners, being distributions, cannot be multiplied at the same points.}
The global conformal group $\PSL_2(\bbC)$ of the sphere acts on~\eqref{nptsphere2} as follows: for a Virasoro-primary field $\phi\in M(k)$, each $\gamma=\bigl(\begin{smallmatrix} a & b \\ c & d \end{smallmatrix}\bigr)$ gives
\begin{gather*}
\Phi\bigl(\phi,z'\bigr)=\Phi\bigl(\phi,\gamma(z')\bigr)\bigl(cz'+d\bigr)^{-2\bar{h}}
\end{gather*}
so~\eqref{nptsphere2} in the limit as the global parameter $z\to0$ is M\"obius-invariant.
This limit exists, and indeed this is what is commonly meant by chiral blocks. This specialization is enough to recover everything (including the full VOA structure and its module structures~\cite{GaGo}, hence the blocks for any local coordinate $f_i(z)$). For this reason, we will eventually restrict to $z=0$, i.e., identify the $z_i'$ with the punctures $z_i$.\footnote{The observant reader will have noticed that $z=0$ gives $z'_i=-z_i$. But that is conformally equivalent to~${z_i'=z_i}$. In our definition of core blocks we choose the specialization $z_i'=z_i$.}

Factorization of chiral blocks on the sphere is called cluster decomposition:
\begin{gather}
\lim\limits_{\lambda\to\infty} \lambda^{\sum_{i=1}^kh_{\phi_i}} \label{cluster}\\
\quad{}\times \langle\phi_1| \Phi_2(\phi_2,\lambda z_2)\circ\cdots\circ\Phi_k(\phi_k,\lambda z_k)\circ\Phi_{k+1}(\phi_{k+1},z_{k+1})\circ\cdots\circ \Phi_{n-1}(\phi_{n-1},z_{n-1})\phi_n \rangle \nonumber\\
 = \sum_\phi \langle \phi_1|\Phi_2(\phi_2,z_2)\circ\cdots\circ\Phi_k(\phi_k,z_k)\phi \rangle \langle\phi^*\Phi_{k+1}(\phi_{k+1},z_{k+1})\circ\cdots\circ \Phi_{n-1}(\phi_{n-1},z_{n-1})\phi_n \rangle,\nonumber
\end{gather}
where the sum is over a basis $\phi$ of $\cV$-primaries (lowest conformal weight states) in the internal module $N_k$, and $\phi^*$ is the dual basis.
 We will use this in the following to relate four-point functions at the cusp to structure constants. Let us now turn to concrete examples.

 \textbf{Sphere $\boldsymbol{n}$-point functions with $\boldsymbol{n < 4}$.} The moduli spaces of spheres with $n<4$ punctures are trivial, i.e., single points,
and their
full mapping class group is ${\rm Sym}(n)$. This is because the~M\"obius symmetry on $\bbP^1$
is triply transitive.
Consider first $n=1$, the 1-point function on the sphere, with insertion $\phi\in M\in\Upsilon$. Without loss of generality we can take $z_1=0$. Now, $\cN_{\cV,M}^\cV=0$ unless $M=\cV$. When $\phi\in\cV(k)$,
\[ \langle\mathbf{1},Y(\phi,z)\mathbf{1}\rangle= \biggl\langle\mathbf{1},\sum_n\phi_nz^{-n-1}\mathbf{1}\biggr\rangle =z^{-k}\mathbf{1}^*(\phi_{k-1}(\mathbf{1}))=0\]
unless $k=0$. Thus $\widehat{\mathfrak{F}}_{(0,1;M)}(\phi)$ equals 0 unless $M=\cV$ and $k=0$, when it has the basis 1 and the chiral block is a $\CC$ number.

Slightly less trivial are the 2-point blocks. We keep $z_1$ and $z_2$ arbitrary, though one may fix them to, e.g., $z_1=\infty$ and $z_2=0$. Here and everywhere else we use the notation ${z_{ij}:= z_i - z_j}$. The extended mapping class group \smash{$\widehat \Gamma^{0,2}$}
is $\mathbb{Z}^2$, generated by Dehn twists around the two punctures.
The Dehn twist about $z_1'$ sends ${\rm d}z_1^{\prime\,x}$ to
$\mathrm{e}(x){\rm d}z_{1}^{\prime\,x}$ but fixes both ${\rm d}z_2^{\prime\,x}$ and $(z_{12})^{y}$. Note that $\cN_{M_1,M}^\cV \cN^M_{M_2,\cV}=0$ unless $M=M_2=M_1^\ast$, in which case it equals 1. Choose $\phi\in M(k)$ and $\psi^*\in M^\ast(\ell)$; then we find $\mathfrak{F}_{(M,M^\ast)}(\phi,\psi^*)$ is 0, unless $k=\ell$ in which case the chiral block becomes
\begin{equation}
\label{eq:2pt}
z_{21}^{-2h_M-2k}\psi^*(\phi)\,{\rm d}z_1^{\prime\,\bar{h}}{\rm d}z_2^{\prime\,\bar{h}}.
\end{equation}

Consider next the case $(g,n)=(0,3)$. Using conformal symmetry, it is possible to fix the punctures to be $z_1=\infty$, $z_2=1$, $z_3=0$, so $z'_1=1/z$, $z'_2=z-1$, $z'_3=z$, but this will not be necessary for our analysis.
The chiral block with insertions $M_1$, $M_2$, $M_3$ will be 0 unless~\smash{$\cN_{M_2,M_3}^{M_1^*}>0$}.
As mentioned above, the moduli space $\cM_{0,3}$ is a point and conformal symmetries dictate the 3-point chiral blocks corresponding to $\phi_i\in M_i(k_i)$ to have the following~form:
\begin{gather}
\label{def:structure_constant}
 {\mathcal F}_{\phi_1,\phi_2,\phi_3}\bigl(z'_1,z'_2,z'_3\bigr)\prod_{i=1}^3 {\rm d}z_i^{\prime\,{\bar{h}_i}}
 = c_{\phi_1,\phi_2,\phi_3} \prod_{1\leq i< j\leq 3}z_{ij}^{\bar h-\bar h_i-\bar h_j} \prod_{i=1}^3 {\rm d}z_i^{\prime\,\bar{h}_i},
\end{gather}
where $\bar h = \bar h_1+\bar h_2 + \bar h_3$. As before, note that {the above chiral block} carries a nontrivial action of the extended mapping class group $\widehat{\Gamma}^{0,3}\cong\mathbb{Z}^3$ through the factors involving ${\rm d}z_i'$.
Using~\eqref{nptsphere2}, we can compute the \emph{structure constants} $c_{\phi_1,\phi_2,\phi_3}$ as follows. For any \smash{$\Phi\in \cY_{M_2,M_3}^{M_1^*}$}, we get a~core~block%
\begin{equation}
\label{def:3point_intertwiners}
c_{\phi_1,\phi_2,\phi_3}^{\Phi}={z_2^{\prime \bar{h}_2+\bar{h}_3-\bar{h}_1}}\bigl(\Phi\bigl(\phi_2,z_2'\bigr)(\phi_3)\bigr)(\phi_1)={z_2^{\prime \bar{h}_2+\bar{h}_3-\bar{h}_1}}\bigl\langle\phi_1\big|\Phi\bigl(\phi_2,z_2'\bigr)\phi_3\bigr\rangle.
\end{equation}
Choosing a basis of $\cY_{M_2,M_3}^{M_1^*}$, we can collect these structure constants into a vector-valued one
\[
\underline{c}_{\phi_1,\phi_2,\phi_3}\in\bbC^{\cN_{M_2,M_3}^{M^*_1}},
\] though there is a priori no canonical basis. Note that there is no well-defined structure constant for a given choice of states $\phi_i\in M_i$, $i=1,2,3$. Rather, there are precisely \smash{$\cN_{M_2,M_3}^{M_1^*}$} independent structure constants, one for each intertwiner \smash{$\Phi\in\cY_{M_2,M_3}^{M_1^*}$}.

 \textbf{Sphere $\boldsymbol{4}$-point functions.} Our main interest lies in the case $(g,n)=(0,4)$.
The moduli space ${\mathcal M}_{0,4}$ is one-complex dimensional, and is parametrized by the \emph{cross ratio}
\begin{gather*}
w:={z_{12} z_{34} \over z_{13} z_{24}}.
\end{gather*}
More precisely, there is a unique M\"obius transformation which moves say $z_1\mapsto\infty$, $z_2\mapsto1$, and~$z_4\mapsto 0$, in which case $z_3$ is sent to the cross ratio $ w$, which can take any value on $\bbP^1$ except for $0$, $1$, $\infty$.
 The four-point blocks are multi-valued with respect to $w$, {with} monodromies around the points $0$, $1$ and $\infty$. To make them single-valued, we have to replace $w\in\bbP^1\setminus\{0,1,\infty\}$ with a parameter {$\widetilde{w}$} running over its universal cover.
 In Section \ref{sec:forms} (cf.\ \eqref{lambda_function}), we will see that this universal cover is most naturally identified with the upper-half plane $\HH$.

Fix any Virasoro-primaries $\phi_i\in M_i(k_i)$. Conformal symmetry dictates the following form for a four-point chiral block:
\begin{gather}\label{def:coreblock4point}
{\mathcal F}_{\phi_1,\phi_2,\phi_3,\phi_4}\bigl(z'_i\bigr) \prod_{i=1}^4{\rm d}z_i^{\prime\,\bar{h}_i} =\var_{\phi_1,\phi_2,\phi_3,\phi_4}(\widetilde{w})\prod_{1\le j< k\le 4}z_{jk}^{\mu_{jk}} \prod_{i=1}^4{\rm d}z_i^{\prime\,\bar{h}_i},
\end{gather}
where
\begin{equation}
\label{eq:muconstr}
\sum_{j<i}\mu_{ji}+\sum_{i<j}\mu_{ij} = -2\bar{h}_i.
\end{equation}
Choosing different solutions to the constraints~\eqref{eq:muconstr} is equivalent to rescaling $\var_{\phi_1,\phi_2,\phi_3,\phi_4}(\widetilde{w})$ by factors of the form
\begin{equation}
\label{eq:corepref}
\tilde w^a(1-\tilde w)^b.
\end{equation}
We make the following natural choice:
\begin{equation}
{\mu_{ij} = \frac{h}{3}-\bar{h}_i-\bar{h}_j, \qquad h=\sum_{i=1}^4 \bar{h}_i},\label{eq:muij}
\end{equation}
which is fixed by imposing that $\mu_{ij}$ coincides with $\mu_{i'j'}$ upon permuting $\bar{h}_i\leftrightarrow\bar{h}_{i'}$, $\bar{h}_j\leftrightarrow\bar{h}_{j'}$.

Using again the fact that $\mathbf{S}$ is a symmetric matrix, \eqref{Verl} reduces to
\begin{equation} \label{dim_4point}
\dim \bigl({\widehat{E}}_\cD\bigr)=\sum_{P\in \Upsilon}
\cN_{M_2,P}^{M_1^\ast}\cN_{M_3,M_4}^{P},
\end{equation}
and analogous formulas hold for the sphere with $n$ punctures.
This dimension formula~\eqref{dim_4point} suggests a natural decomposition of
 the space $\widehat{\Psi}_{(M_1,M_2,M_3,M_4)}$ of sections of the bundle $\widehat{E}_\cD$ into a~direct sum of subspaces \smash{${\widehat{\Psi}}_{(M_1,M_2,M_3,M_4)}^P$}, for $P\in\Upsilon$, each of which is naturally isomorphic to the tensor product $\cY_{M_2,P}^{M_1^*}\otimes \cY_{M_3,M_4}^P$ of intertwiner spaces.
 For each $P\in\Upsilon$ with ${\cN_{M_2,P}^{M_1^\ast}\cN_{M_3,M_4}^{P}>0}$, and each choice \[
 \Phi\in\cY_{M_2,P}^{M_1^*} \qquad \text{and}\qquad
\Phi'\in\cY_{M_3,M_4}^P,
\]
 we obtain a chiral block \smash{${\mathcal F}_{\phi_1,\phi_2,\phi_3,\phi_4}^{P;\Phi,\Phi'}(z'_i)$} as in~\eqref{nptsphere2}. We compute from~\eqref{eq:inter}:
\begin{align}
\nonumber
{\mathcal F}^{P;\Phi,\Phi'}_{\phi_1,\phi_2,\phi_3,\phi_4}\bigl(z'_i\bigr)&{}=z_1^{\prime -2\bar{h}_1}z_2^{\prime\,\bar{h}_1-\bar{h}_2-h_P}z_3^{\prime\,h_P-\bar{h}_3-\bar{h}_4}\\
&\quad{}\times \sum_{m=0}^\infty \biggl(\frac{{z_3'}}{{z_2'}}\biggr)^m \bigl\langle \phi_1 | \Phi(\phi_2)_{m+k_2-k_1-1}\circ\Phi'(\phi_3)_{k_3+k_4-m-1}\phi_4\bigr\rangle.
\label{eq:fp2}
\end{align}

A basis for the space \smash{$\widehat{{\mathfrak F}}_{M_1,M_2,M_3,M_4}$} of chiral blocks for the sphere with 4 punctures, labelled {by}
$M_1,M_2,M_3,M_4\in\Upsilon(\cV)$, can be read off from the following description of the space \smash{$\widehat{\Psi}_{M_1,M_2,M_3,M_4}$} of sections:
\begin{equation}
\label{def:spaceof4poitn}
\bigoplus_{P}\cY^\cV_{M_1,M_1^*}\otimes \cY^{M_1^*}_{M_2,P}\otimes \cY^P_{M_3,M_4}\otimes \cY^{M_4}_{M_4,\cV},
\end{equation}
where \smash{$\cY_{M, M'}^{M''}$} is the space of intertwiners introduced earlier. We can express the action of the generators of the braid group ${\rm Br}_4$ on the space of chiral blocks in terms of {\rm braiding operators}.
Here, the most convenient choice of generators are $\tilde \sigma_{i}$, conjugates of $\overline \sigma_{i}$, as given in~\eqref{def:changeofbasis}:
\begin{gather}
\begin{split}
\label{def:brdgroup_action}
\tilde\sigma_1\mapsto &\bigoplus_{Q} B_{M_1^* Q}\begin{bmatrix} M_1&M_2\\ \cV&P\end{bmatrix}\!(+)
\otimes I\otimes I, \\
\tilde\sigma_2\mapsto &\bigoplus_{Q}I\otimes B_{PQ}\begin{bmatrix} M_2&M_3\\ M_1^*&M_4\end{bmatrix}\!(+)
\otimes I,\\
\tilde\sigma_3\mapsto &\bigoplus_{Q} I\otimes I\otimes B_{M_4Q}\begin{bmatrix} M_3&M_4\\ P&\cV\end{bmatrix}\!(+).
\end{split}
\end{gather}
In the first line, the first factor acts on the first two factors in each summand in~\eqref{def:spaceof4poitn} and the identity operators act on the third and fourth factors, and similarly for the rest.
The~braiding operator is a standard quantity associated to any vertex
operator algebra (see, e.g.,~\cite{Huang:2008} and references therein, which follows from, but gives a more careful treatment than, \cite{Moore:1989vd}). See~Appendix~\ref{sec:brafu} for some of its key properties. It gives the isomorphism\footnote{Note that, conforming with the vast amount of literature starting from \cite{Moore:1989vd}, our $B\bigl[\begin{smallmatrix}N_2&N_3\\N_1&N_4\end{smallmatrix}\bigr](+)$ treats $N_1$ differently from $N_{2,3,4}$ as it has a reversed arrow, in accordance with the notation of Figure \ref{fig:3pt}.}
\begin{gather*}
B\begin{bmatrix}N_2&N_3\\N_1&N_4\end{bmatrix}\!(+)\colon \ \bigoplus_P \cY^{N_1}_{N_2,P}\otimes \cY^{P}_{N_3,N_4}\xrightarrow[]{\cong} \bigoplus_Q \cY^{N_1}_{N_3,Q}\otimes \cY^{Q}_{N_2,N_4}.
\end{gather*}
Note that the summands on the left- and right-hand side above are the spaces relevant for the core blocks
\[
\var^P_{\phi_1,\phi_2,\phi_3, \phi_4}(\widetilde{w}) \qquad \text{and}\qquad \var^Q_{\phi_1,\phi_3,\phi_2, \phi_4}(\widetilde{w}),
\]
 respectively.
Given a basis $\bigl\{\Phi^{N_1,a}_{N_2,P}\bigr\}_a$ for $\cY^{N_1}_{N_2,P}$ and similarly for \smash{$\cY^{P}_{N_3,N_4}$}, \smash{$\cY^{N_1}_{N_3,Q}$}, \smash{$\cY^{Q}_{N_2,N_4}$}, one can express the above isomorphism in terms of the tensors $B_{P Q}\bigl[\begin{smallmatrix}N_2&N_3\\N_1&N_4\end{smallmatrix}\bigr](+)$:
\begin{equation} \label{braiding_tensor}
\Phi^{N_1,a}_{N_2,P}{(z) \Phi^{P,b}_{N_3,N_4}(z')\mapsto \sum_Q {\sum_{c,d}} B_{P Q}\begin{bmatrix}N_2&N_3\\N_1&N_4\end{bmatrix}^{a,b}_{c,d}\!\!\!\!\!(+)\,\,\Phi^{N_1,c}_{N_3,Q}(z')\Phi^{Q,d}_{N_2,N_4}(z)},
\end{equation}
which holds when {${\rm Im}(z-z')>0$; for ${\rm Im}(z-z')<0$}, one must use a different braiding operator
\begin{gather*}
B{\begin{bmatrix}N_2&N_3\\N_1&N_4\end{bmatrix}}\!(-),
\end{gather*}
which obeys
\begin{gather*}
B{\begin{bmatrix}N_3&N_2\\N_1&N_4\end{bmatrix}\!(+)B\begin{bmatrix}N_2&N_3\\N_1&N_4\end{bmatrix}}\!(-) = I.
\end{gather*}

\section{The core blocks}
\label{sec:reps}
In Section \ref{sec:core}, we introduce the notion of core blocks and explain their relation to the chiral blocks, focussing primarily on the case of spheres with four punctures, given the fundamental role four-point blocks and correlators play in conformal field theory~\cite{Moore:1989vd}. In Section~\ref{sec:mcgrep}, we then discuss the action of the mapping class groups $\Gamma^{0,4}$ and ${\rm F}\Gamma^{0,4}$ on the core (resp.\ chiral) blocks and prove that these give rise to ordinary (resp.\ projective) representations of the mapping class groups. In Section \ref{subsec:mod}, we then exploit the relation between mapping class groups and modular groups of Section~\ref{sec:groups} to determine the modular transformation properties of core blocks.

\subsection{Definition and basic properties}
\label{sec:core}
Given a chiral block, we take its corresponding {\it core block} to be the part capturing the non-trivial information that only depends on the conformal equivalence classes of Riemann surfaces, and is independent of the Dehn twists about the punctures. In particular, a core block will be a~multilinear map $\varphi\colon M_1\otimes_\bbC\cdots\otimes_\bbC M_n\rightarrow \Psi_\cD$ to the space of meromorphic sections of $E_\mathcal{D}$. In~analogy to the chiral blocks, for a choice of states $\phi_i\in M_i$, we denote the space of core blocks at $(\phi_1,\dots,\phi_n)$ by ${{\mathfrak F}}_\cD(\phi_1,\dots,\phi_n)$. The core blocks are sensitive only to the most trivial choices of local coordinates $f_i(z)$. In particular, they transform with respect to the mapping class group, and do not see the extra parts of the extended mapping class group corresponding to the Dehn twists.

As we have seen in Section \ref{sec:bkg}, the action of Dehn twists is trivial on the torus one-point functions, and therefore the definition of core and chiral blocks is identical in that case. The~same is true for the sphere one-point functions. We therefore focus on the cases $(g,n)=(0,n)$ with~${n \geq 2}$. The effect on the chiral blocks of the Dehn twist about the $i$-th puncture will be multiplication by $\ex(h_{M_i})$, where $h_{M_i}$ is the conformal weight (the minimal $L_0$-eigenvalue) of $M_i$. These give rise to the factor \smash{$\prod_i {\rm d}z_i^{h_{M_i}}$} in the chiral blocks, which we will drop when considering the core blocks. The $(0,n)$-chiral blocks moreover depend on the local coordinates~$z_i$ at the~punctures through factors of the form $(z_i-z_j)^{d_{ij}}$ for quantities~$d_{ij}$ depending on the~$h_{M_i}$; dropping those factors as well gives the core blocks. We will describe this more explicitly and illustrate it with a few examples.

First of all, in the case $n=2$, we have seen in Section \ref{sec:bkg} that the space of chiral blocks $\mathfrak{F}_{(M_1,M_2)}(\phi,\psi^*)$ has dimension 1 only if $M_1^*=M_2 = M$, $\phi\in M(k)$ and $\psi^*\in M^*(k)$, in which case the chiral block is given by equation~\eqref{eq:2pt}, i.e.,
\begin{gather*}
z_{21}^{-2h_M-2k}\psi^*(\phi)\,{\rm d}z_1^{\prime\,\bar{h}}{\rm d}z_2^{\prime\,\bar{h}}.
\end{gather*}
Otherwise, the space of chiral blocks has dimension zero. The Dehn twist-independent part, which we take as our definition of core block, is then simply
\begin{gather*}
\varphi_{\phi,\psi^\ast} := \psi^*(\phi).
\end{gather*}

Next, consider the case $n=3$. Here, the only terms in the chiral block that do not depend on Dehn twists are the structure constants (equation~\eqref{def:3point_intertwiners}), which we take as our definition of the core blocks
\begin{gather*}
{\var}^\Phi_{\phi_1,\phi_2,\phi_3}:= c_{\phi_1,\phi_2,\phi_3}^{\Phi}={z_2^{\prime \bar{h}_2+\bar{h}_3-\bar{h}_1}}\bigl\langle\phi_1\big|\Phi\bigl(\phi_2,z_2'\bigr)\phi_3\bigr\rangle.
\end{gather*}
We can combine these into a vector{\samepage
\begin{gather*}
\underline{\var}_{\phi_1,\phi_2,\phi_3}:= \underline{c}_{\phi_1,\phi_2,\phi_3} \in \mathbb{C}^{\cN_{M_2,M_3}^{M^*_1}}
\end{gather*}
of three-point core blocks.}

Finally, let us turn to the case of greatest interest, namely the four-point core blocks. Fixing Virasoro-primaries $\phi_i\in M_i(k_i)$, the four-point chiral block is given by equation~\eqref{eq:fp2} as
{{\begin{gather*}\begin{split}
&{\mathcal F}^{P;\Phi,\Phi'}_{\phi_1,\phi_2,\phi_3,\phi_4}\bigl(z'_i\bigr) \prod_{i=1}^4{\rm d}z_i^{\prime\,\bar{h}_i}
\end{split}\end{gather*}}}
with
\begin{align*}
{\mathcal F}^{P;\Phi,\Phi'}_{\phi_1,\phi_2,\phi_3,\phi_4}\bigl(z'_i\bigr)&{}=z_1^{\prime -2\bar{h}_1}z_2^{\prime\,\bar{h}_1-\bar{h}_2-h_P}z_3^{\prime\,h_P-\bar{h}_3-\bar{h}_4}
\\&\quad{}\times\sum_{m=0}^\infty \biggl(\frac{{z_3'}}{{z_2'}}\biggr)^m \langle \phi_1|\Phi(\phi_2)_{m+k_2-k_1-1}\circ\Phi'(\phi_3)_{k_3+k_4-m-1}\phi_4\rangle
\end{align*}
and
\begin{gather*}
{\mu_{ij} = \frac{h}{3}-\bar{h}_i-\bar{h}_j, \qquad h=\sum_{i=1}^4 \bar{h}_i}.
\end{gather*}
Writing the chiral block as in equation~\eqref{def:coreblock4point},
\begin{gather*}
{\mathcal F}^{P;\Phi,\Phi'}_{\phi_1,\phi_2,\phi_3,\phi_4}\bigl(z'_i\bigr) \prod_{i=1}^4{\rm d}z_i^{\prime\,\bar{h}_i} =\var_{\phi_1,\phi_2,\phi_3,\phi_4}(\widetilde{w})\prod_{1\le j< k\le 4}z_{jk}^{\mu_{jk}} \prod_{i=1}^4{\rm d}z_i^{\prime\,\bar{h}_i}
\end{gather*}
we define the corresponding core block to be \smash{$\var^{P;\Phi,\Phi'}_{\phi_1,\phi_2,\phi_3,\phi_4}(\widetilde{w})$}. The choice~\eqref{eq:muij} for the expo\-nents~$\mu_{ij}$ guarantees, among other things, that the core blocks for four identical insertions transform as (vector valued) modular forms for the full modular group $\PSL_2(\ZZ)$; other choices would lead to spurious multiplicative factors of the form~\eqref{eq:corepref} in the definition of the core blocks which would spoil this modular transformation.

We can recover the core block of ${\mathcal F}^{P;\Phi,\Phi'}$ as follows. First,
 one has
\begin{gather*}
{\lim_{u\to \infty} u^{2\bar{h}_{1}} {\mathcal F}^{P;\Phi,\Phi'}(\phi_i;u,z_2,z_3,0) = \langle \phi_1 |\Phi(\phi_2,z_2) \Phi'(\phi_3,z_3) | \phi_4\rangle,}
\end{gather*}
and the (in this notation, multi-valued) core block is obtained by further specializing ${z_2= 1}$, ${z_3= w}$:
\begin{gather*}
{\var^{P;\Phi,\Phi'}_{\phi_1,\phi_2,\phi_3,\phi_4}(w) = (1-w)^{-\mu_{23}} w^{-\mu_{34}}\langle \phi_1 |\Phi(\phi_2,1)\Phi'(\phi_3,w) | \phi_4\rangle.}
\end{gather*}
 Of course it suffices to have $\Phi$ run over a basis of $\cY_{M_2,P}^{M_1^*}$ and $\Phi'$ run over one of $\cY_{M_3,M_4}^P$; for~generic states $\phi_i$ this will yield a basis of ${\mathfrak F}(\phi_1,\phi_2,\phi_3,\phi_4)$, though for some $\phi_i$ the dimension will be smaller as there could be accidental linear dependencies. As before, we can then collect these basis core blocks into a (multi-valued) vector $\underline{\var}(w)$. As we explicitly work out in later sections, to make this single-valued we should lift this to $\widetilde{w}$.

From equation~\eqref{def:3point_intertwiners} it follows that the $w$-expansion of the core blocks has the following leading order behavior as $w\to 0$:
\begin{equation}
\var^P_{\phi_1,\phi_2,\phi_3,\phi_4}(w) =
w^{-h/3+h_P}\bigl(c_0^P(\phi_1,\phi_2,\phi_3,\phi_4)+\ccO(w)\bigr),
\label{eq:coreLO}
\end{equation}
for some function $c_0^P(\phi_1,\dots,\phi_4)$ multilinear in the states $\phi_i$. Using cluster decomposition~\eqref{cluster}, the leading constant $c_0^P$ can be expressed in terms of structure constants, though this expression will be complicated if the space $P(0)$ of $\cV$-primaries in $P$ is large. Likewise, we obtain
\begin{gather*}
 \var^P_{\phi_1,\phi_2,\phi_3,\phi_4}(w) =(1-w)^{-h/3+h_P}
c_1^P(\phi_1,\phi_2,\phi_3,\phi_4) +o\bigl((1-w)^{-h/3+h_P}\bigr),\\
\var^P_{\phi_1,\phi_2,\phi_3,\phi_4}(w) = w^{h/3-h'}
c_\infty^P(\phi_1,\phi_2,\phi_3,\phi_4)+o\bigl(w^{h/3-h'}\bigr)
\end{gather*}
as the leading order terms in the expansions around $w= 1$ and $w = \infty$, respectively.
Here $c_1^P$,~$c_\infty^P$ are likewise multilinear, and $h'$ is the minimum of the $h_Q$ for $Q$ satisfying ${\cN_{M_1,M_4}^Q\cN_{Q,M_2}^{M_3}\ne 0}$. Of course, choosing as we did an $s$-channel basis broke the symmetry between the $w=0$ expansions and those at $w=1$ and $w=\infty$; we could have instead diagonalized the expansions at $w=1$ or $w=\infty$ by instead using the $t$- or $u$-channels.

\subsection{Mapping class group action}
\label{sec:mcgrep}
It has been shown in~\cite{Moore:1989vd} that the action of the braiding matrices on the chiral blocks given in~\eqref{def:brdgroup_action} indeed satisfies the braid group relation~\eqref{def:br4}. In other words, the space of four-point chiral blocks ${\bigoplus_{\pi\in {\rm Sym}(4)}\widehat{\mathfrak{F}}(\phi_{\pi 1},\phi_{\pi 2},\phi_{\pi 3},\phi_{\pi 4})}$ furnishes a representation of ${\rm Br}_4$. In this subsection, we will show that it moreover furnishes a projective representation of the full mapping class group~${\rm F}\Gamma^{0,4}$ (cf.\ \eqref{GammaPres}). Furthermore, the space of core blocks ${\bigoplus_{\pi\in {\rm Sym}(4)}{\mathfrak F}(\phi_{\pi 1},\phi_{\pi 2},\phi_{\pi 3},\phi_{\pi 4})}$ (cf.~\eqref{def:coreblock4point}) furnishes a (true) representation of ${\rm F}\Gamma^{0,4}$.
For notational simplicity, we will sometimes restrict in the following to the {where}
\[
\dim \cY^{M_1^\ast}_{M_2,P} \dim \cY^{P}_{M_3,M_4} \leq 1 \qquad \text{for all}\quad P\in \Upsilon(\cV)
\]
 in which case the braiding tensor {component} $B_{P Q}\bigl[\begin{smallmatrix}M_2&M_3\\M_1&M_4\end{smallmatrix}\bigr](+)$ does not need the additional labels~$a$,~$b$,~$c$,~$d$ and is just a number (cf.\ \eqref{braiding_tensor}). This restriction is not significant: the generalisation to the more general case is straightforward. We will provide an explicit proof of Theorem~\ref{thm:corerepmapping} under this restriction, and explain after that how the proof works in the general~case.

\begin{Theorem}\label{thm:corerepmapping}
The space of core blocks {$\bigoplus_{\pi\in {\rm Sym}(4)}{\mathfrak F}(\phi_{\pi 1},\phi_{\pi2},\phi_{\pi3},\phi_{\pi4})$} furnishes a representation of ${\rm F}\Gamma^{0,4}$.
\end{Theorem}

\begin{proof}
 We represent the action of $\tilde\sigma_i$ on the core blocks in terms of matrices $R_i$ as follows:
\begin{equation}
\label{rep_rho}
\varphi^P_{\phi_1,\phi_2,\phi_3,\phi_4}{(\widetilde{w})} \mapsto \sum_{Q}(R_i)_{PQ}\, \varphi^Q_{\tilde\sigma_i(\phi_1,\phi_2,\phi_3,\phi_4)}({\tilde\sigma_i \widetilde{w}}),
\end{equation}
where the action of the $\tilde\sigma_i$ on $\widetilde{w}$ is determined (in part) by their action on the $z_i$, and the action on the labels $\varphi_1,\dots,\varphi_4$ (hence the sectors $M_1,\dots, M_4$) is by the induced permutation. Explicitly, one has
\begin{gather*}
\tilde\sigma_i(\varphi_1,\varphi_2,\varphi_3,\varphi_4) = \begin{cases} (\varphi_2,\varphi_1,\varphi_3,\varphi_4) &\text{for } i=1, \\(\varphi_1,\varphi_3,\varphi_2,\varphi_4) &\text{for } i=2,\\ ( \varphi_1,\varphi_2,\varphi_4,\varphi_3) &\text{for } i=3 \end{cases}
\end{gather*}
and
\begin{gather}
(R_1)_{PQ} = \delta_{PQ} \ex\biggl(\frac{\mu_{12}}{2}\biggr) B_{M_1^* M_2^*}\begin{bmatrix} M_1&M_2\\ \cV&P\end{bmatrix}\!(+),\nonumber\\
(R_2)_{PQ} = \ex\biggl(\frac{\mu_{23}}{2}\biggr) B_{PQ}\begin{bmatrix} M_2&M_3\\ M_1^*&M_4\end{bmatrix}\!(+),\nonumber\\
(R_3)_{PQ} = \delta_{PQ} \ex\biggl(\frac{\mu_{34}}{2}\biggr) B_{M_4 M_3}\begin{bmatrix} M_3&M_4\\ P&\cV\end{bmatrix}\!(+),\label{eq:corerep}
\end{gather}
where the phases arise from the factors of $z_{ij}$ on the right-hand side of~\eqref{def:coreblock4point}.
Moreover,
using equations~\eqref{eq:B}, one sees that $R_1$ and $R_3$ are given simply by
\begin{gather}
\label{eq:sg}
(R_1)_{PQ} = \delta_{PQ} \zeta^{M_1^*}_{M_2,P}\ex\biggl(\frac{h}{6}-\frac{h_P}{2}\biggr),\qquad
(R_3)_{PQ} = \delta_{PQ} \xi^P_{M_3,M_4}\ex\biggl(\frac{h}{6}-\frac{h_P}{2}\biggr),
\end{gather}
where $h$ is as in~\eqref{eq:muij}, and \smash{$\xi_{M_iM_j}^{M_k}$} and \smash{$\zeta_{M_iM_j}^{M_k}$} are the phases defined in~\eqref{def:xi}--\eqref{def:zeta} and satisfying~\eqref{ord2zetaxi}.

We need to verify that the following five relations, which follow from~\eqref{GammaPres}, are satisfied
\begin{gather*}
(1)\quad R_{1} R_{3} = R_{3} R_{1} ,\\
(2)\quad R_{1} R_{2} R_{1} = R_{2} R_{1} R_{2} ,\\
(3)\quad R_{2} R_{3} R_{2} = R_{3} R_{2} R_{3} ,\\
(4)\quad R_{1} R_{2} R_{3}^2R_{2} R_{1} = I ,\\
(5)\quad (R_{1} R_{2} R_{3})^4 = I .
\end{gather*}
The proof of relations (1)--(3) is straightforward since it has been shown in~\cite{Moore:1989vd}
 that the braiding matrices provide a representation of ${\rm Br}_4$, whose generators satisfy relations analogous to \mbox{(1)--(3)}, as we see in~\eqref{def:br4}. Therefore, all we need to check is that the extra phases, present due to the fact that we are considering the action on the core blocks and not the chiral blocks, are the same on both sides of the equations. Indeed, in relation (1) the right- and left-hand sides have an identical phase factor of $\ex\bigl(\tfrac{\m_{12}+\mu_{34} }{2}\bigr)= \ex(-h/6)$, so (1) holds. Likewise, the left- and right-hand side of $(2)$ have an identical phase factor given by \smash{$\ex\bigl(\tfrac{\m_{12}+\m_{23}+\m_{13}}{2}\bigr)$}, and that of $(3)$ is given~by~\smash{$\ex\bigl(\tfrac{\m_{23}+\m_{34}+\m_{24}}{2}\bigr)$}.

To prove (4), we use~\eqref{eq:corerep} and~\eqref{eq:sg} to write
\begin{gather*}
\bigl(R_{1} R_{2} R_{3}^2R_{2} R_{1}\bigr)_{PQ}=\sum_{R}\zeta_{M_2,P}^{M_1^*}\bigl(\xi_{M_1,M_4}^R\xi_{M_4,M_1}^R\bigr) \zeta_{M_1,Q}^{M_2^*}\ex\biggl(h_2+h_4-\frac{h_P}{2}-h_R-\frac{h_Q}{2}\biggr) \nonumber\\
\hphantom{\bigl(R_{1} R_{2} R_{3}^2R_{2} R_{1}\bigr)_{PQ}=\sum_{R}}{}
\times B_{PR}\begin{bmatrix}M_1&M_3\\M_2^*&M_4\end{bmatrix}\!(+) B_{RQ}\begin{bmatrix}M_3&M_1\\M_2^*&M_4\end{bmatrix}\!(+),
\end{gather*}
which leads to (4) upon plugging in~\eqref{ord2zetaxi} and~\eqref{eq:BB}.
Finally, to show that (5) holds we begin by noting that the following relation holds:
\begin{gather*}
(4')\quad R_3R_2R_1^2R_2R_3=I,
\end{gather*}
the derivation is identical to that of (4). We now make repeated use of relations (1)--(4) and ($4'$) as well as $R_{1}^2=R_3^2$, which follows trivially from equations~\eqref{eq:sg}, to show that
\begin{align*}
(R_1R_2R_3)^4&{}
=(R_1R_2R_1R_3R_2R_3)^2
=(R_2R_1R_2R_3R_2R_3)^2 \\
&{}=R_2R_1R_3R_2R_3R_3R_2R_1R_2R_3R_2R_3
=R_2R_3\bigl(R_1R_2R_3^2R_2R_1\bigr)R_2R_3R_2R_3 \\
&{}=(R_2R_3)^3
=R_3R_2R_3R_3R_2R_3
=R_3R_2R_1^2R_2R_3
=I,
\end{align*}
and therefore relation (5) holds as well.
\end{proof}

Now let us explain why Theorem \ref{thm:corerepmapping} holds in general. As shown in the proof of~\cite[Theorem~5.4.1]{BK}, our assignments give a (true) representation of the extended mapping class group $F{\widehat{\Gamma}}^{0,4}$ on the chiral blocks. To see that the core blocks carry a representation of ${\rm F}\Gamma^{0,4}$, it suffices to verify that the Dehn twists about the punctures act trivially, as the quotient of the extended mapping class group by those Dehn twists is ${\rm F}\Gamma^{0,4}$.

From the definition~\eqref{def:coreblock4point} of core blocks, it follows immediately.
\begin{Corollary}
The space {$\bigoplus_{\pi\in {\rm Sym}(4)}\widehat{\mathfrak{F}}(\phi_{\pi1},\phi_{\pi2},\phi_{\pi3},\phi_{\pi4})$} of chiral blocks furnishes a projective representation of ${\rm F}\Gamma^{0,4}$.
\end{Corollary}

\subsection{Modular group}
\label{subsec:mod}

From the discussion in the previous section, it follows that the core blocks transform under the generators $\bar\sigma_T=\tilde\sigma_3$ and $\bar\sigma_S=\tilde\sigma_3\tilde\sigma_2\tilde\sigma_3$ of $\PSL_2(\mathbb{Z})$ as
\begin{gather}
\var^P_{\phi_1,\phi_2,\phi_3,\phi_4}(\widetilde{w}) = \sum_Q (R_T)_{PQ} \,\var_{\phi_1,\phi_2,\phi_4,\phi_3}^Q(\bar\sigma_T\widetilde{w}),\nonumber\\
\var^P_{\phi_1,\phi_2,\phi_3,\phi_4}(\widetilde{w}) =\sum_Q (R_S)_{PQ}\,\var_{\phi_1,\phi_4,\phi_3,\phi_2}^Q(\bar\sigma_S\widetilde{w}),\label{STsigma}
\end{gather}
where following~\eqref{rep_rho} we write $R_T := R_3$, $R_S := R_3 R_2 R_3 $, and denote their matrix elements by
\begin{equation}
\label{eq:tpq}
 T_{PQ}\begin{bmatrix}M_2&M_3\\M_1&M_4\end{bmatrix} := (R_T )_{PQ} =\delta_{PQ}\xi^P_{M_3,M_4}\ex \biggl(\frac{h}{6}-\frac{h_P}{2}\biggr)
\end{equation}
and
\begin{align}
 S_{PQ}\begin{bmatrix}M_2&M_3\\M_1&M_4\end{bmatrix}:={}&(R_S)_{PQ} \nonumber\\
 ={}& \ex \biggl(\frac{h_{M_1}+h_{M_4}-h_P-h_Q}{2}\biggr) \xi^P_{M_3,M_4}B_{PQ}\begin{bmatrix}M_2&M_4\\M_1^*&M_3\end{bmatrix}(+)\xi^Q_{M_2,M_3}\nonumber\\
={}&F_{PQ}\begin{bmatrix}M_2&M_3\\M_1^*&M_4\end{bmatrix}\xi^{M_1^*}_{Q,M_4}\xi^Q_{M_2,M_3}.\label{eq:spq}
\end{align}
In arriving at the second line of the last equation we have used the identity~\eqref{eq:BF}.

From Theorem \ref{thm:corerepmapping} and since $R_T$ and $R_S$ provide a representation of the $\PSL_2(\mathbb{Z})$ subgroup of {${\rm F}\Gamma^{0,4}$}, it follows that
\begin{equation}
\label{eq:reprel}
R_S^2=(R_TR_S)^3=1.
\end{equation}

Define the following vector of core blocks:
\begin{equation}
\label{eq:phiA}
\underline{\var}{}_{\phi_1,\phi_2,\phi_3,\phi_4}(\widetilde{w}) =
\begin{pmatrix}
\var^{P_1}_{\phi_1,\phi_2,\phi_3,\phi_4}\\
\vdots\\
\var^{P_r}_{\phi_1,\phi_2,\phi_3,\phi_4}
\end{pmatrix}(\widetilde{w}),
\end{equation}
where $P_i \in \Upsilon(\cV)$ runs along the set of internal channels.
By making use of the $\mathbb{Z}_2\times\mathbb{Z}_2$ transformations~\eqref{Z2Z2} we can rearrange $\phi_1,\dots,\phi_4$ so that any given one of them is in the first position. Thus in studying the action of $\PSL_2(\mathbb{Z})$ on conformal blocks we can limit ourselves to the following three cases:
\begin{itemize}\itemsep=0pt
\item[(A)] when at least three of the four states $\phi_i$ are the same. We can arrange this so that $\cD = (M_1,M_2,M_2,M_2) $ and $\phi_1\in M_1$, $\phi_2=\phi_3=\phi_4\in M_2$;

\item[(B)] when two of the four states are the same. We can arrange this so that $\cD \!=\! (M_1,M_2,M_3,M_3) $ and $\phi_1\in M_1$, $\phi_2\in M_2$, $\phi_3=\phi_4\in M_3$;

\item[(C)] when all four states are distinct. \end{itemize}

We want to generate actions of the following groups, with the following generating sets\footnote{See Section \ref{sec:gencom} for general facts about these groups.}
\begin{equation}\renewcommand{\arraystretch}{1.1}
\label{eq:projective}
\begin{array}{c|c}\renewcommand{\arraystretch}{1.2}
\PSL_2(\mathbb{Z})&\pm\sm 1 1 0 1,\pm\sm 0 {-1} 1 {\hphantom{-} 0}\\
{\rm P}\Gamma_0(2)& \pm \sm 1 1 0 1,\pm\sm 1 {-1} 2 {-1}\\
{\rm P}\Gamma(2)& \pm\sm 1 2 0 1,\pm\sm {\hphantom{-} 1} 0 {-2} 1.
\end{array}
\end{equation}
Corresponding to these generators, we define the matrices
\begin{gather*}
\pm\begin{pmatrix} 1& 1\\ 0& 1\end{pmatrix}\mapsto \cT_{}:= T\begin{bmatrix}M_2&M_3\\M_1&M_4\end{bmatrix},\\
\pm\bem 0 &{-1}\\ 1& 0\eem\mapsto \cS_{} := S\begin{bmatrix}M_2&M_3\\M_1&M_4\end{bmatrix},\\
\pm\bem 1& 2\\ 0& 1\eem \mapsto \cT^2_{}:= T\begin{bmatrix}M_2&M_4\\M_1&M_3\end{bmatrix} T\begin{bmatrix}M_2&M_3\\M_1&M_4\end{bmatrix},\\
\pm\bem 1& {-1}\\ 2 &{-1}\eem\mapsto \cR_{}:=T\begin{bmatrix}M_2&M_3\\M_1&M_4\end{bmatrix} S\begin{bmatrix}M_4&M_3\\M_1&M_2\end{bmatrix} T\begin{bmatrix}M_4&M_2\\ M_1&M_3\end{bmatrix}
T\begin{bmatrix}M_4&M_3\\ M_1&M_2\end{bmatrix} S\begin{bmatrix}M_2&M_3\\M_1&M_4\end{bmatrix},\\
\pm\bem 1& 0\\ {-2}& 1\eem\mapsto \cU_{}:= S\begin{bmatrix}M_4&M_3\\M_1&M_2\end{bmatrix} T\begin{bmatrix}M_4&M_2\\M_1&M_3\end{bmatrix} T\begin{bmatrix}M_4&M_3\\M_1&M_2\end{bmatrix}
 S\begin{bmatrix}M_2&M_3\\M_1&M_4\end{bmatrix}.
 \end{gather*}
 In analogy with the definition of $\bar\sigma_T$ and $\bar\sigma_S$ in~\eqref{PSL2}, we define $\bar\sigma_U=\bar\sigma_S\bar\sigma_T^2\bar\sigma_S$ and $\bar\sigma_R=\bar\sigma_1\bar\sigma_U$.

The following theorem then follows directly from the discussion above (see also Appendix~\ref{app:generalSL2Z}):

\begin{Theorem}\label{thm:sl2z}
The group $\PSL_2(\ZZ)$ acts on the vector-valued functions $\underline{\varphi}{}_{A}:=\underline{\varphi}{}_{\phi_1,\phi_2,\phi_2,\phi_2}(\widetilde{w})$, $\underline{\varphi}{}_{B}(\widetilde{w}):=\underline{\varphi}{}_{B;\phi_1,\phi_2,\phi_3,\phi_3}$, and $\underline{\varphi}{}_{C}(\widetilde{w}):=\underline{\varphi}{}_{C;\phi_1,\phi_2,\phi_3,\phi_4}(\widetilde{w})$, defined in equations~\eqref{eq:phiA}, \eqref{eq:phiB}, and~\eqref{eq:phiC}, as follows:
\begin{alignat*}{3}
&\underline{\varphi}{}_{A}(\widetilde{w})= \cT^{-1} \cdot \underline{\varphi}{}_{A}(\bar\sigma_T \widetilde{w}),
&& \qquad \underline{\varphi}{}_{A}(\widetilde{w}) = \cS^{-1} \cdot \underline{\varphi}{}_{A}(\bar\sigma_S \widetilde{w}),& \\
&\underline{\varphi}{}_{B}(\widetilde{w})= \cT^{-1}_B \cdot \underline{\varphi}{}_{B}(\bar\sigma_T \widetilde{w}),
&& \qquad \underline{\varphi}{}_{B}(\widetilde{w}) = \cS^{-1}_B \cdot \underline{\varphi}{}_{B}(\bar\sigma_S \widetilde{w}),& \\
&\underline{\varphi}{}_{C}(\widetilde{w})= \cT^{-1}_C \cdot \underline{\varphi}{}_{C}(\bar\sigma_T \widetilde{w}),
&& \qquad \underline{\varphi}{}_{C}(\widetilde{w}) = \cS^{-1}_C \cdot \underline{\varphi}{}_{C}(\bar\sigma_S \widetilde{w}),&
\end{alignat*}
where the matrices $\cT_{B}$, $\cS_{B}$, $\cT_{C}$, $\cS_{C}$ are defined in Appendix~\ref{app:generalSL2Z}.

In case $(\text{B})$, the group {${\rm P}\Gamma_0(2)$} acts on the vector-valued function $\underline{\varphi}(\widetilde{w})=\underline{\varphi}{}_{\phi_1,\phi_2,\phi_3,\phi_3}(\widetilde{w})$ by $\underline{\varphi}{}_{}(\widetilde{w}) = \cT^{-1} \cdot \underline{\varphi}{}(\bar\sigma_T \widetilde{w})$ and $\underline{\varphi}{}(\widetilde{w}) = \cR^{-1} \cdot \underline{\varphi}{}(\bar\sigma_R \widetilde{w})$. In case $(\text{C})$, the group {${\rm P}\Gamma(2)$} acts on the vector-valued {function} $\underline{\varphi}(\widetilde{w})=\underline{\varphi}{}_{\phi_1,\phi_2,\phi_3,\phi_4}(\widetilde{w})$ by $\underline{\varphi}{}_{}(\widetilde{w}) = \bigl(\cT^2\bigr){}^{-1} \cdot \underline{\varphi}{}(\bar\sigma_T\bar\sigma_T \widetilde{w})$ and $\underline{\varphi}{}(\widetilde{w}) = \cU^{-1} \cdot \underline{\varphi}{}(\bar\sigma_U \widetilde{w})$.
\end{Theorem}

In cases (A) and (B), one verifies straightforwardly that these generators satisfy the $ \PSL_2(\mathbb{Z})$ and ${\rm P}\Gamma_0(2)$ relations
\begin{gather*}
\cS^2= (\cT\cdot\cS)^3=1,\qquad \cR^2=1
\end{gather*}
(the generators $\cT^2$ and $\cU$ for ${\rm P}\Gamma(2)$ satisfy no relations; see Section \ref{sec:gencom}), which follow from equations~\eqref{eq:tpq}--\eqref{eq:reprel}. That $\cS_B$, $\cT_B$ and $\cS_C$, $\cT_C$ satisfy the $\PSL_2(\mathbb{Z})$ relations is automatic from the induction picture presented in Appendix~\ref{app:generalSL2Z}.

\section{Modular forms}
\label{sec:forms}

In this section, we review the basic properties of the relevant modular objects. Subsequently, we discuss the properties of core blocks as modular forms and how such properties can be employed to determine the core blocks.

\subsection{Vector-valued modular forms}
\label{sec:gencom}

From Section \ref{subsec:mod}, it follows that the type of modular object that is most relevant to us is the vector-valued modular forms. In this subsection, we include some general comments on vvmf's that will be useful later.

\subsubsection[Modular forms for SL\_2(Z)]{Modular forms for $\boldsymbol{{\rm SL}_2(\mathbb{Z})}$}

The modular group ${\rm SL}_2(\mathbb{Z})$, or its quotient $\PSL_2(\mathbb{Z})={\rm SL}_2(\mathbb{Z})/\{\pm\}$ which is what we will use, is genus zero. We say that a discrete subgroup $\Gamma$ of ${\rm SL}_2(\RR)$ is {\em{genus zero}} if the quotient $\Gamma\backslash \HH$ is a~Riemann surface of genus zero minus finitely many points. This means that the (meromorphic) modular functions form the field of rational functions $\bbC(j(\tau))$ in the $j$-function, with expansion%
\begin{equation}
j(\tau)=q^{-1}+744+196884q+21493760q^2+\cdots,
\label{eq:jfn}
\end{equation}
in terms of the local parameter $q=\mathrm{e}(\tau)$ near the cusp $\tau= {\rm i} \infty$.
By a \textit{weakly holomorphic} modular form or function,
 we mean one which is holomorphic in $\bbH$ but with poles allowed at the cusps. The weakly holomorphic modular functions form the ring $\bbC[j(\tau)]$ of polynomials in~$j(\tau)$.

 The (holomorphic) modular forms form the polynomial ring $\bbC[E_4(\tau),E_6(\tau)]$,
 where \begin{gather*}
E_n(\tau) = 1-\frac{4n}{B_{2n}}\sum_{k=1}^\infty \frac{k^{2n-1}q^k}{1-q^k}
\label{eq:eisenstein}
\end{gather*}
is the weight-$n$ Eisenstein series satisfying (for $n>2$)
\begin{gather*}
E_n(\tau)=E_n(\tau+1),
\qquad
E_n(\tau)=\tau^{-n}E_n\bigl(-\tfrac{1}{\tau}\bigr).
\end{gather*}
An important modular form is the discriminant form \begin{gather*}\Delta(\tau)=\eta(\tau)^{24}=\frac{E_4(\tau)^3-E_6(\tau)^2}{1728},\end{gather*} where
\begin{gather*}
\eta(\tau) = q^{\frac{1}{24}}\prod_{n=1}^\infty(1-q^n)
\end{gather*}
is the Dedekind eta function, which transforms under $\PSL_2(\mathbb{Z})$ as
\begin{equation}
\eta(\tau) = \mathrm{e}\bigl(-\tfrac{1}{24}\bigr)\eta(\tau+1),
\qquad
\eta(\tau) = (- {\rm i} \tau)^{-{1\over 2}}\eta\bigl(-\tfrac{1}{\tau}\bigr).
\label{eq:ths}
\end{equation}
The importance of $\Delta(\tau)$ lies in the fact that it has no zeroes in $\bbH$.

The modular transformations of $E_2(\tau)$ are a little more complicated -- it is a \textit{quasi-modular} form. It appears in the differential modular (Serre) derivative
\begin{equation}\label{serreder}
D_{(k)}=\frac{1}{2\pi {\rm i}}\frac{{\rm d}}{{\rm d}\tau}-\frac{k}{12}E_2(\tau),
\end{equation}
which takes weight-$k$ (vector valued) modular forms to weight-$(k+2)$ vvmf's. In several equations below we employ the differential operator
\begin{equation} \label{def:differential operator}
D_{(0)}^\ell:=D_{(2\ell-2)}\circ\cdots\circ D_{(2)}\circ D_{(0)},
\end{equation}
 which takes a weight-0 vvmf to a weight-$2\ell$ vvmf.

The Dedekind eta function is a modular form for $\PSL_2(\mathbb{Z})$ \big(of weight $\frac{1}{2}$\big), but with nontrivial multiplier (a projective representation of $\PSL_2(\mathbb{Z})$ with values in 24th roots of 1). More generally, we can speak of a \textit{vector-valued modular form} $\underline{f}(\tau)$ for $\PSL_2(\mathbb{Z})$ of weight $k$ and multiplier $\rho$: here, $\rho$ is a representation of $\PSL_2(\mathbb{Z})$ (merely projective, if $k\not\in2\mathbb{Z}$) by $d\times d$ matrices for some~$d$ called the rank, $\underline{f}(\tau)$ takes values in $\bbC^d$, and they obey{\samepage
\begin{gather*}
\underline{f}(\tau)=(c\tau+d)^{-k}\rho\bigl(\sm a b c d\bigr)^{-1}\underline{f}\biggl(\frac{a\tau+b}{c\tau+d}\biggr).
\end{gather*}
We call $\underline{f}(\tau)$ \textit{weakly holomorphic} if each component $f_i(\tau)$ is weakly holomorphic.}

Corresponding to a representation $\rho$ of $\PSL_2(\mathbb{Z})$, we can define a space of weight-zero weakly holomorphic vvmf's, which we will denote by \smash{$\mathcal{M}^!_0(\rho)$}.
Furthermore, for any $d$-tuple $\lambda\in\bbR^d$, we~denote by $\cM_0(\rho;\lambda)\subset \cM^!_0(\rho)$ the subspace
\begin{equation}\label{def:spaces}
\cM_0(\rho;\lambda) :=	 \bigl\{ \underline{f}\in \cM^!_0(\rho)\mid q^{-\lambda_i} f_i(\tau) = \ccO(1) \text{ for all }i, \text{ as } \tau\to {\rm i} \infty\bigr\},
\end{equation}
where $T=\rho\bigl(\begin{smallmatrix}1 & 1 \\ 0 & 1\end{smallmatrix}\bigr)$ is assumed to be diagonal.
Note that there is a natural filtration among these spaces: $\cM_0(\rho;\lambda') \subset \cM_0(\rho;\lambda)$ whenever $\lambda'\ge \lambda$ component-wise.

The properties of vvmf's can then be employed to help to compute the precise form of the core blocks. In the remainder of this paper we will describe this application in some detail. The scalar case $\dim (\rho)=1$ is particularly simple: any element of $\mathcal{M}^!_0(\rho)$ can be expressed as a~polynomial in the $j$-function~\eqref{eq:jfn} multiplying one of the following functions~\cite{Bantay:2007zz}:
\begin{equation}
1,\qquad \frac{E_4(\tau)}{\eta(\tau)^{8}},\qquad \frac{E_6(\tau)}{\eta(\tau)^{12}},\qquad \frac{E_8(\tau)}{\eta(\tau)^{16}},\qquad \frac{E_{10}(\tau)}{\eta(\tau)^{20}},\qquad \frac{E_{14}(\tau)}{\eta(\tau)^{28}}.
\label{eq:1dimbasis}
\end{equation}

Because $\PSL_2(\mathbb{Z})$ is generated by $\pm \sm 0 {-1} 1 {\hphantom{-} 0}$ and $\pm\bigl(\begin{smallmatrix} 1 & 1 \\ 0 & 1\end{smallmatrix}\bigr)$, the multiplier $\rho$ is uniquely determined by the values $\cS:=\rho\bigl(\sm 0 {-1} 1 {\hphantom{-}0}\bigr)$ and
$\cT:=\rho\bigl(\bigl(\begin{smallmatrix} 1 & 1 \\ 0 & 1\end{smallmatrix}\bigr)\bigr)$. We are primarily interested in the cases where the corresponding $\cT$ matrices are diagonal. Then the condition of weakly holomorphic reduces to requiring that there is a diagonal matrix $\lambda$ such that $\cT=\ex(\lambda)$ and
\begin{gather*}
\underline{f}(\tau)=q^\lambda\sum_{n=0}^\infty \underline{f}[n] q^n.
\end{gather*}
The coefficients $\underline{f}[n]\in\bbC^d$ will in general not be integral -- integrality is expected (see, e.g.,~\cite{AS}) only when the kernel of $\rho$ contains a congruence subgroup
\begin{equation}
\label{pcs}\Gamma(N):=\bigl\{\sm a b c d \in{\rm SL}_2(\mathbb{Z})\mid \sm a b c d = \bigl(\begin{smallmatrix} 1 & 0 \\ 0 & 1\end{smallmatrix}\bigr) \mod N\bigr\}.
\end{equation}

\subsubsection[Modular forms for Gamma(2)]{Modular forms for $\boldsymbol{\Gamma(2)}$}
\label{sec:gamma2}
The congruence group $\Gamma(2)$ (recall~\eqref{pcs}), as well as its quotient ${\rm P}\Gamma(2) = \Gamma(2)/\{\pm 1\}$, are also genus-0. The role of $j(
\tau)$ is played in this case by
\begin{equation}
\lambda(\tau)=\frac{\theta_2(\tau)^4}{\theta_3(\tau)^4}=16q^{1/2}-128q+704q^{3/2}-3072q^2+11488q^{5/2}-38400q^3+\cdots\,\label{eq:lambdainfty},
\end{equation}
where
\begin{gather*}
\theta_2(\tau)=\sum_{n\in\mathbb{Z}}q^{\frac{1}{2}(n+\frac{1}{2})^2},\qquad \theta_3(\tau)=\sum_{n\in\mathbb{Z}}q^{\frac{n^2}{2}},\qquad \theta_4(\tau)=\sum_{n\in\mathbb{Z}}(-1)^n q^{\frac{n^2}{2}},
\end{gather*}
transform under $\PSL_2(\mathbb{Z})$ as a weight $\frac{1}{2}$ vvmf:
\begin{equation}
\begin{pmatrix}
\theta_2\\
\theta_3\\
\theta_4
\end{pmatrix}
(\tau)
=
\begin{pmatrix}
\mathrm{e}\bigl(-\frac{1}{8}\bigr)\theta_2\\
\theta_4\\
\theta_3
\end{pmatrix}
(\tau+1),
\qquad
\begin{pmatrix}
\theta_2\\
\theta_3\\
\theta_4
\end{pmatrix}
(\tau)
= (-{\rm i} \tau)^{-\tfrac{1}{2}}
\begin{pmatrix}
\theta_4\\
\theta_3\\
\theta_2
\end{pmatrix}\bigl(-\tfrac{1}{\tau}\bigr).
\label{eq:tht}
\end{equation}
Recall that $\theta_2$, $\theta_3$, $\theta_4$ do not vanish in $\bbH$, and (holomorphic) modular forms for ${\rm P}\Gamma(2)$ form the polynomial ring $\bbC\bigl[\theta_3^4,\theta_4^4\bigr]$. A useful identity among these functions is $\theta_2^4=\theta_3^4-\theta_4^4$.

The group $\Gamma(2)$ is index 6 in ${\rm SL}_2(\mathbb{Z})$: ${\rm SL}_2(\mathbb{Z})/\Gamma(2)\cong \PSL_2(\mathbb{Z})/({\rm P}\Gamma(2))\cong {\rm Sym}(3)$. The~function $\lambda(\tau)$ is related to the $j$-function
 as follows:
\begin{equation}
j(\tau)=256\frac{\bigl(\lambda(\tau)^2-\lambda(\tau)+1\bigr)^3}{\lambda(\tau)^2(1-\lambda(\tau))^2}.
\label{eq:jlambda}
\end{equation}

The group ${\rm P}\Gamma(2)$ has 3 cusps, namely 0, 1 and ${\rm i} \infty$, and it is necessary to have control on how the modular forms behave at all 3 cusps. One way to do this is to use $\tau\mapsto-1/\tau$ and~${\tau\to-1/(\tau-1)}$ to map the other cusps to ${\rm i} \infty$. For example, applying these to $\lambda$ shows that~${\lambda=1}$ at $\tau=0$ and has a pole at $\tau=1$. For ${\rm P}\Gamma(2)$, weakly holomorphic functions are defined as before: they are holomorphic in $\bbH$ and have at worse poles at the 3 cusps. For~example, the weakly holomorphic modular functions form the ring $\bbC\bigl[\lambda,\lambda^{-1},1/(1-\lambda)\bigr]$. Note that~${\rm P}\Gamma(2)$, like any subgroup of $\PSL_2(\mathbb{Z})$, inherits the same modular derivative $D_{(k)}$.

Next we discuss the multiplier for a ${\rm P}\Gamma(2)$ vector-valued modular form.
The multiplier $\rho$ is a~representation.
Recall that ${\rm P}\Gamma(2)$ is freely generated by $\pm\bigl(\begin{smallmatrix}1 & 2 \\ 0 & 1 \end{smallmatrix}\bigr)$ and $\pm\sm {\hphantom{-} 1} 0 {-2} 1$.
As~a~result,~$\rho$~is uniquely determined by its values $\rho\bigl(\pm\bigl(\begin{smallmatrix}1 & 2 \\ 0 & 1 \end{smallmatrix}\bigr)\bigr)$ and $\rho\bigl(\pm\sm {\hphantom{-}1} 0 {-2} 1\bigr)$. In the one-dimensional case, $\rho=\rho_{\alpha,\beta}$ is specified by a pair $\alpha,\beta\in\bbC^\times$, where
 $\rho\bigl(\pm\bigl(\begin{smallmatrix}1 & 2 \\ 0 & 1 \end{smallmatrix}\bigr)\bigr)=\alpha$ and ${\rho\bigl(\pm\sm {\hphantom{-}1} 0 {-2} 1\bigr)=\beta}$. Writing~${\alpha=\ex(a)}$ and $\beta=\ex(b)$, we see that the space of weakly holomorphic vvmf of weight 0 for~$\rho_{\alpha,\beta}$ takes the form $\lambda^a(1-\lambda)^{b}\bbC\bigl[\lambda,\lambda^{-1},1/(1-\lambda)\bigr]$. To understand this, note that~$\theta_2(\tau)$,~$\theta_3(\tau)$,~$\theta_4(\tau)$ have no zeros or poles in the simply-connected region $\bbH$, and hence they have holomorphic logarithms there. Being the composition of holomorphic functions, functions such as $\lambda(\tau)^a=\exp (4a\log (\theta_2(\tau) )-4a\log (\theta_3(\tau) ) )$ are themselves holomorphic in $\bbH$.
 Finally, in Appendix~\ref{app:generalSL2Z} we explain how a vvmf for ${\rm P}\Gamma_0(2)$ can induce up to one for $\PSL_2(\mathbb{Z})$.

\subsubsection[Modular forms for Gamma\_0(2)]{Modular forms for $\boldsymbol{\Gamma_0(2)}$}
\label{sec:gamma02}
The group $\Gamma_0(2)=\bigl\{\sm a b c d \in{\rm SL}_2(\mathbb{Z})\mid c\in 2\mathbb{Z}\bigr\}$ is index 3 in ${\rm SL}_2(\mathbb{Z})${, and is generated by $\pm\sm 1 1 0 1$ and an order two element $\pm\sm 1 {-1} 2 {-1}$}. As before, we are more interested in ${\rm P}\Gamma_0(2)$. It is genus zero, with the role of $j(\tau)$ being played by \begin{gather*}
\kappa(\tau)=\frac{\theta_3(\tau)^4\theta_4(\tau)^4}{\theta_2(\tau)^8}=\frac{1}{256}q^{-1}-\frac{3}{32}+\frac{69}{64}q-128q^2+\frac{5601}{128}q^3-\cdots.
\end{gather*}
Holomorphic modular forms for ${\rm P}\Gamma_0(2)$ form the polynomial ring $\bbC\bigl[E_4,\theta_3^4+\theta_4^4\bigr]$.

The group ${\rm P}\Gamma_0(2)$ has 2 cusps, namely at 0 and ${\rm i} \infty$. Note that $\kappa=0$ at $\tau=0$, and hence the weakly holomorphic modular functions form the ring $\bbC\bigl[\kappa,\kappa^{-1}\bigr]$. The space of holomorphic vvmf forms a module over that ring.
The group ${\rm P}\Gamma_0(2)$ is generated by $\pm\sm 1 1 0 1$ and $\pm\sm 1 {-1} 2 {-1}$, where the latter has order 2, so the multiplier $\rho$ of any vvmf (of even weight) is uniquely determined by the matrices $\rho\bigl(\pm\sm 1 1 0 1\bigr)$ and $\rho\bigl(\pm\sm 1 {-1} 2 {-1}\bigr)$ (the latter squaring to $I$). In Appendix~\ref{app:generalSL2Z} we explain how a vvmf for ${\rm P}\Gamma_0(2)$ induces up to one for $\PSL_2(\mathbb{Z})$.

The one-dimensional representations are $\rho_{\alpha,\sigma}$ where $\alpha\in\bbC^\times$ and $\sigma=\pm 1$: $\rho_{\alpha,\sigma}\bigl(\pm\sm 1 1 0 1\bigr)=\alpha$ and $\rho\bigl(\pm\sm 1 {-1} 2 {-1}\bigr)=\sigma$. Writing
$\alpha=\ex(a)$ and $\sigma=(-1)^s$ for $s\in\{0,1\}$, we find that the~space~of weakly holomorphic vvmf of weight 0 for $\rho_{\alpha,\sigma}$ are \smash{$\kappa^{-a}\bigl(1+\frac{1}{4\kappa}\bigr)^{s/2}\bbC\bigl[\kappa,\kappa^{-1}\bigr]$}. To see this, note that~$\kappa$ can be raised to arbitrary powers by the same argument as at the end of the last subsection; the minus sign in the factor $\kappa^{-a}$ arises because $\kappa$ has a pole and not a zero at $\tau={\rm i} \infty$. The~elliptic element $\pm\sm 1 {-1} 2 {-1}$ fixes $\tau=\frac{1+{\rm i}}{2}$ and $\kappa\bigl(\frac{1+{\rm i}}{2}\bigr)=-\frac{1}{4}$, so $\kappa+\frac{1}{4}$ must have a double-zero at $\tau=\frac{1+{\rm i}}{2}$. Being a uniformizing function for $\bbH/\Gamma_0(2)$, the only zeros and poles of $\kappa+\frac{1}{4}$ can be at the obvious spots, namely the orbits of $\frac{1+{\rm i}}{2}$ and ${\rm i} \infty$. Thus \smash{$\sqrt{\kappa+\smash{\frac{1}{4}}\vphantom{A^1}}$} is holomorphic in $\bbH$.

\subsection{Core blocks as vector-valued modular forms}\label{section6.2}

As we have seen in the previous section, the core blocks transform in $\PSL_2(\mathbb{Z})$ representations which are completely determined provided that the braiding and {fusing} matrices of the RCFT are given. This ${\rm PSL}_2(\mathbb{Z})$ action is displayed {very} naturally by passing to the double-cover of the four-punctured sphere, which is a torus. By a change of variable, one can express the chiral blocks in terms of the torus complex structure $\tau=\widetilde w$~\cite{Zamolodchikov}, which is related to the cross-ratio as follows:
\begin{equation}
\label{lambda_function}
w = \lambda(\tau).
\end{equation}

The function $w = \lambda(\tau)$ is the Hauptmodul identifying $\bbC\setminus\{0,1\}$ \big($\bbP^1$ with three points removed\big) with $\Gamma(2)\backslash \bbH$, which maps {the three cusps at $\tau = {\rm i}\infty, 0,1$ to the three points $ w=0,1,\infty$, respectively}.

It is easy to see that the identification~\eqref{lambda_function} is consistent with the action of $\Gamma^{0,4}\cong {\rm P}\Gamma(2)$ via
\eqref{Gamma2}, when the action of ${\rm P}\Gamma(2)$ on $\HH$ is taken to be the usual $\tau\mapsto \frac{a\tau+b}{c\tau+d}$. To see this, note that $\bar\sigma_1^2$ sends $z_{12}\mapsto \ex(1) z_{12}$ while leaving the other $z_{ij}$ invariant, and similarly for \smash{$\bar\sigma_2^2$} and $z_{41}$.

As alluded to above, the map~\eqref{lambda_function} has a simple geometric origin in terms of the torus that is the double cover of the four puncture sphere. Namely, the hyper-elliptic equation ${y^2= x(x-1)(x-\lambda)}$ defines an isomorphism class of an elliptic curve together with a basis of the group of its 2-torsion points, and hence {an orbit $[\t] \in \Gamma(2)\backslash\HH$}.
Conversely, given two periods $\omega_1$ and $\omega_2$ with $\tau= \omega_1/\omega_2$, the points $e_1,e_2,e_3\in \CC$ given by
\begin{gather*}
e_1=\wp\biggl(\frac{\omega_2}{2}\biggr),\qquad e_2=\wp\biggl(\frac{\omega_1+\omega_2}{2}\biggr),\qquad e_3=\wp\biggl(\frac{\omega_1}{2}\biggr),
\end{gather*}
together with the point at infinity, have cross-ratio
\begin{gather*}
 \lambda(\tau) = {e_1-e_2 \over e_1-e_3}.
\end{gather*}

A more direct way to motivate the identification~\eqref{lambda_function} of $\widetilde{w}$ with $\tau$, is the realization that the moduli space of 4-punctured spheres is the 3-punctured sphere (namely the location of the 4-th point, {or} the cross-ratio, after the first 3 are moved to $0$, $1$, $\infty$). The 3-punctured sphere has universal cover $\bbH$ and mapping class group $F_2\cong {\rm P}\Gamma(2)$: ${\rm P}\Gamma(2)$ is genus zero with 3 cusps and no elliptic points. Then Theorem \ref{thm:sl2z} becomes:

\begin{Theorem}\label{mod_form}
The core blocks $\underline{\varphi}{}_{\phi_1,\phi_2,\phi_3,\phi_4}$ are single-valued functions on $\mathbb H$ via the identification~\eqref{lambda_function}. Moreover, they are weight zero vector-valued modular forms for multipliers given in Theorem $\ref{thm:sl2z}$.
\end{Theorem}

After establishing the single-valuedness as above, we will often write
\begin{gather*}
\underline{\varphi}{}_{\phi_1,\phi_2,\phi_3,\phi_4}(\widetilde{w})=\underline{\varphi}{}_{\phi_1,\phi_2,\phi_3,\phi_4}(\tau)\qquad \text{and} \qquad \var^{P}_{\phi_1,\phi_2,\phi_3,\phi_4}(\widetilde{w})=\var^{P}_{\phi_1,\phi_2,\phi_3,\phi_4}(\tau)
\end{gather*}
to denote the (vector of) core blocks considered as (vector-valued) functions on $\HH$.

\subsection{Methods for constructing vector-valued modular forms}

For vvmf's of general dimensions, three methods for constructing vector-valued modular forms have appeared in the literature which we review next. In Section \ref{sec:hybrid}, we present a new hybrid method which combines two of the existing techniques and is more effective in more sophisticated examples. For concreteness, we will restrict our discussion to $\PSL_2(\mathbb{Z})$, although all the approaches we discuss should generalize. We also mention that for $d\leq 5$ the (equivalence classes~of) irreducible representations are uniquely determined by their $T$ eigenvalues~\cite{Tuba:2001}, which leads to useful simplifications in explicit calculations.

\subsubsection{Rademacher sum}
The simplest method is the Rademacher sum of~\cite{DMMV,MM} (related are the Poincar\'e series of~\cite{KM}).
Let $\rho$ be a representation for ${\rm SL}_2(\mathbb{Z})$ of dimension $d$, with diagonal $\rho(T)$ with
entries $\rho(T)_{j,j}=\mathrm{e}(\lambda_j)$ for $0\le\lambda_j<1$.
Suppose
\begin{gather*}
X_j(\tau)=q^{\lambda_j}\sum_{n=-\infty}^\infty X_{j}[n] q^n,\qquad 1\leq j \leq d
\end{gather*}
are the components of a weakly holomorphic vector-valued modular form of weight 0 with multiplier $\rho$, with
each $X_{j}[n]\in\bbC$ and $X_{j}[n]\neq 0$ for all but finitely many $n<0$.
For any pair of coprime integers $(c,d)$, there will be infinitely many $\gamma\in{\rm SL}_2(\mathbb{Z})$
with bottom row $(c\;d)$; let~${\gamma_{cd}=\bigl(\begin{smallmatrix}a&b\\ c&d\end{smallmatrix}\bigr)}$ be any one of these. Define the \emph{principal part} of $X_j(\tau)$ to be
\begin{equation}
\cP[X_j(\tau)] = q^{\lambda_j} \sum_{n+\lambda_j<0}X_{j}[n]q^n.\label{eq:RadP}
\end{equation}

Given the principal part of $X_j(\tau)$, one has~\cite{KM, Rad_J,Whalen_vvRad}
\begin{gather}
X_{j}[n]=\sum_{i=1}^d \sum_{\substack{m\in \ZZ, \\ m+\lambda_i<0}}-(m+\lambda_i)\nonumber\\
\hphantom{X_{j}[n]=}{}
\times \sum_{c=1}^\infty \frac{4\pi^2}{c^2} \mathrm{Kl}(n,j,m,i;c) X_{i}[m] {I}\biggl(-\frac{4\pi^2}{c^2}{(m+\lambda_i)
(n+\lambda_j)}\biggr),\label{rade}
\end{gather}
where $I$ is given in terms of the Bessel function as
\begin{gather*}
{I}(x)=\sqrt{x}^{-1}I_1\bigl(2\sqrt{x}\bigr)=\sum_{k=0}^\infty \frac{x^{k}}{k!(k+1)!}=1+\frac{x}{2}+\frac{x^2}{12}+\frac{x^3}{144}+\cdots
\end{gather*}
and where the generalised Kloosterman sum Kl is given by
\begin{gather*}
\mathrm{Kl}(n,j,m,i;c)=\sum_{0<-d<c,\, (d,c)=1}\ex\bigl(\tfrac{1}{c}(d(n+\lambda_j) +a(m+\lambda_i))\bigr)
{\rho(\gamma_{c,d})}^{-1}_{j,i}
\end{gather*}
for $c>1$, while for $c=1$ we define ${\rm Kl}(n,j,m,i;1)={\rho(S)^{-1}_{j,i}}$. It is easy to check that the above quantity is independent of the choice of $\gamma_{cd}$ for a given pair $(c,d)$, and therefore the definition makes sense.

The strength of the Rademacher sum method is that, for a given ${\rm SL}_2(\ZZ)$ representation, the~input is minimal (the principal part) and the formula is universal. The main computational drawback is that it often converges rather slowly. For instance, even if one knows that the~coefficients $X_{j}[n]$ are integers, one must often compute a large number of terms in the $c$-expansion in~\eqref{rade} in order to predict their value.
Also, in more general contexts there are restrictions to the Rademacher method: special care must be taken for certain weights
\cite{Cheng:2011ay,DunFre_RSMG,Kno_ConstMdlrFnsI,Kno_ConstMdlrFnsII, Kno_ConstAutFrmsSuppSeries,Kno_AbIntsMdlrFns,Nie_ConstAutInts}; not all principal parts work, with the obstruction in the weight-zero cases given by weight-two vector-valued cusp forms in the dual representation; and finally the multiplier system $\rho$ is assumed to be unitary. That said, in this paper we are interested in the weight zero unitary representations with polar parts that necessarily correspond to modular forms, and hence these restrictions do not affect us.

\subsubsection{Riemann--Hilbert}
\label{sec:RH}
The second method we discuss~\cite{Bantay:2007zz,Ga} constructs explicit generators for the space $\cM^!_{0}(\rho)$ of weakly holomorphic vector-valued modular forms with multiplier $\rho$ (of dimension $d$).\footnote{Here we focus on weight-zero vvmf's for ${\rm SL}_2(\mathbb{Z})$, though as explained in~\cite{Ga} the results also apply to non-zero weight, and as explained in~\cite{Ga2} the method extends with little change to any genus-0 group.} These generators are determined once certain $d\times d$ matrices $\Lambda$ and $\chi$, satisfying a set of conditions which we will discuss shortly, are found.
These two matrices $\Lambda$ and $\chi$ then determine a basis for $\cM^!_{0}(\rho)$.
In more details, let $\Xi[0]=I$ be the $d\times d$ identity matrix, and $\Xi[1]=\chi$, and for all~${n> 1}$ define $\Xi[n]$ recursively by
\begin{equation}\label{eq:recurs}
[\Lambda,{\Xi}[{n}]]+n\, {\Xi}[{n}]=\sum_{l=0}^{n-1}{\Xi}[{l}]
(f_{n-l}\Lambda+g_{n-l}(\chi+[\Lambda,\chi])),
\end{equation}
where we write
\begin{gather*}
(j(\tau)-984)\frac{\Delta(\tau)}{E_{10}(\tau)}=\sum_{n=0}^\infty f_{n}q^n=1+0q+338328 q^2 +\cdots,\\
\frac{\Delta(\tau)}{E_{10}(\tau)}=\sum_{n=0}^\infty g_{n}q^n=q+240 q^2 + 199044 q^3 + \cdots.
\end{gather*}
Then the columns of the matrix \begin{gather*} \Xi(\tau)=q^\Lambda \sum_{n=0}^\infty \Xi[n]q^n\end{gather*} are the desired generators.
More precisely, any vvmf $X(\tau)\in\cM^!_0(\rho)$ corresponds bijectively to a vector-valued polynomial {$P(j(\tau))\in\bbC^d[j(\tau)]$}, through {$X(\tau)=\Xi(\tau)P(j(\tau))$}. This polynomial can be read off from the principal part of $X(\tau)$, where principal part here means the terms~$q^\Lambda\sum_{n\le 0}
X[n]q^n$. In particular, the number of $X[n]$ in the principal part of $X(\tau)$ coincides with the dimension of $\cM^!_0(\rho)$, when $\dim (\rho)\leq 5$.\footnote{Additional subtleties must be taken into account for $\dim (\rho)>5$; we refer the reader to~\cite{Ga} for a careful analysis of that case.} Note that the definition for the principal part employed in this method does not necessarily coincide with that in the Rademacher sum method (cf.\ equation~\eqref{eq:RadP}); {as opposed to the Rademacher case, here} the principal part isomorphically identifies the vvmf. This method is called the Riemann--Hilbert method because the recursion~\eqref{eq:recurs} comes from a Fuchsian differential equation solving the Riemann--Hilbert problem for $\rho$ on the sphere.

It remains to discuss the conditions on the pair of matrices $\Lambda$ and $\chi$. The method requires $\Lambda$ to be diagonal and to satisfy $\mathrm{e}(\Lambda)=\rho(T)$ as well as
\begin{equation}
\operatorname{Tr}\Lambda=-\frac{7d}{12}+\frac{1}{4}\operatorname{Tr}{\rho(S)}+
\frac{2}{3\sqrt{3}}\mathrm{Re}\bigl(\mathrm{e}(-1/12)
\operatorname{Tr}\rho\bigl(ST^{-1}\bigr)\bigr).
\label{eq:trace}
\end{equation}
Additionally, the matrices $\Lambda$ and $\chi$ must satisfy the equations
\begin{gather*}
a_2\biggl(a_2-\frac{1}{2}I\biggr)=0,\\
a_3\biggl(a_3-\frac{1}{3}I\biggr)\biggl(a_3-\frac{2}{3}I\biggr)=0,
\end{gather*}
where $a_2$ and $a_3$ are $d\times d$ matrices given by
\begin{gather*}
a_2=-\frac{31}{72}\Lambda-\frac{1}{1728}(\chi+[\Lambda,\chi]),\qquad
a_3=-\frac{41}{72}\Lambda+\frac{1}{1728}(\chi+[\Lambda,\chi]).
\end{gather*}
Once a pair of matrices $\Lambda$ and $\chi$ satisfying the above conditions is found, a basis for $\cM_0^!(\rho)$ can be determined recursively.

On the other hand, any $X(\tau) \in \cM_0^!(\rho)$ whose components are linearly independent over $\bbC$ gives us all of $\cM^!_0(\rho)$ (and hence also determines $\Lambda$ and $\chi$) via
\begin{equation}\label{vvmfgen}
\cM^!_0(\rho) = \bbC\biggl[j(\tau), \frac{E_4 E_6}{\Delta} D_{(0)}, \frac{E_4^2}{\Delta}D_{(0)}^2, \frac{E_6}{\Delta}D_{(0)}^3\biggr]X(\tau).
\end{equation}

Recall the space $\cM_0(\rho;\lambda)$ of vvmf defined in~\eqref{def:spaces}. We can assume ${\rm e}^{2\pi {\rm i}\lambda}=T$ without loss of generality. The Riemann--Roch-like expression for the dimension of this space is given in~\mbox{\cite[equation~(51)]{Ga}}.
When $\rho$ is irreducible and of dimension $<6$, this simplifies to (see~\mbox{\cite[Propo\-si\-tion~3.3]{Ga}})
\begin{gather*}
\dim \cM_0(\rho;\lambda)=\max\{0,\operatorname{Tr}(\Lambda-\lambda)+d\},
\end{gather*}
where $\Lambda$ is any diagonal matrix satisfying ${\rm e}^{2\pi {\rm i}\Lambda}=T$ and~\eqref{eq:trace}.

One strength of this method is that, once successfully implemented, it determines $d$ convenient and explicit generators for $\cM^!_0(\rho)$; from them, arbitrarily many components of any $X(\tau)$ can be rapidly computed with complete accuracy from the principal part of $X(\tau)$. In more general contexts, the aforementioned subtlety occurring in the Rademacher method (namely that the Rademacher sum procedure does not always yield a vvmf; it yields a vector-valued mock modular form for some choices of polar parts) is avoided here because the principal part is defined with respect to $\Lambda$. Moreover, this method does not assume $\rho$ is unitary.
The drawbacks are its complexity and the fact that it is not always known how to explicitly determine $\Lambda$ and $\chi$. At~the~present time, $\Lambda,\chi$ are both known whenever $d= 2$ (see~\cite[Section 4.2]{Ga}), or when~$\rho$ contains any congruence subgroup $\Gamma(N)$, where the conductor $N$ is a product of {powers of primes not exceeding 31} \cite{Ga2}. A nontrivial example of this method is included in Section \ref{sec:hybrid}.

\subsubsection{Modular linear differential equation}\label{section6.3.3}
The oldest available method (see, e.g.,~\cite{Anderson:1987ge,Ma,MMS,Mil}) is the method of Wronskians, now more commonly known as the method of modular linear differential equations (MLDE). The idea is that, if one has a $d$-dimensional vector-valued modular form $X(\tau)$ (which for the purposes of our paper we take to be weakly holomorphic and of weight 0), then the vanishing of
\begin{gather*}\renewcommand{\arraystretch}{1.2}
\mathrm{det}\begin{pmatrix}f&f'&f''&\cdots&f^{(d)}\cr X_1&X_1'&X_1''&\cdots&X_1^{(d)}\cr
X_2&X_2'&X_2''&\cdots&X_2^{(d)}\cr \vdots&&&&\vdots\cr X_d&X_d'&X_d''&\cdots&X_d^{(d)}\end{pmatrix}
\end{gather*}
gives an order-$d$ differential equation $a_df^{(d)}+a_{d-1}f^{(d-1)}+\cdots+a_{1}f'+a_0f=0$ satisfied by the $d$ components $X_j$. The coefficients $a_j(\tau)$ are quasi-modular (i.e., they also involves powers of the second Eisenstein series $E_2(\tau)$), but we get something simpler by replacing $\frac{{\rm d}}{{\rm d}\tau}$ and its powers~by the modular derivative~\eqref{serreder}
and its powers. Replacing the \smash{$\frac{{\rm d}^l}{{\rm d}\tau^l}$} in the Wronskian with $D_{(0)}^l$ only changes it by a factor of \smash{$(2\pi {\rm i})^d$}, so the expansion
\begin{gather*}b_dD_{(0)}^df+b_{d-1}D_{(0)}^{d-1}f+\cdots+b_{1}D_{(0)}f+b_0f=0, \end{gather*}
 where we have used the notation~\eqref{def:differential operator},
 is also satisfied by all components, but now the coefficients $b_k(\tau)$ will be weakly holomorphic modular forms of weight $d(d+1)-2k$ with multiplier ${\rm det}(\rho)$, with principal parts controlled by the $X_j$. The spaces of such modular forms are finite-dimensional, so partial information on the $X_j$ (for instance the first few terms of their $q$-expansions) will suffice to determine the coefficients $b_j(\tau)$ exactly. We work out a nontrivial example of this method in the next subsection on our hybrid method.

The strength of this method is its simplicity: any vvmf $X(\tau)$ is associated to a differential equation with scalar modular forms as coefficients, and the solution space is the span over $\bbC$ of the components $X_j(\tau)$. Furthermore, the representation $\rho$ need not be unitary, and this method can be generalized to any Fuchsian group.
A drawback of this method is that, at best, this method determines vector-valued modular forms which transform with some multiplier equivalent to $\rho$; moreover it is unclear in general how one can obtain enough partial information of the $X_j$ to pin down the coefficients $b_j(\tau)$.

We end this section by commenting on the role of modular differential equations in the context of core blocks.
It is very well known that conformal blocks involving primaries which have a~null state among their descendants at level-$n$ is are solutions to an order-$n$ differential equation (the most famous examples being the BPZ equation~\cite{Belavin:1984vu} for Virasoro conformal blocks, and the KZ equation in WZW models~\cite{Knizhnik:1984nr}). On the other hand, as we just reviewed, $\ell$-dimensional vector-valued modular forms are solutions to an MLDE of order $\ell$; one therefore expects that differential equations following from the presence of null states can be rephrased as MLDEs for the core blocks. In Appendix \ref{app:BPZ}, we analyse explicitly this relation in the case of Virasoro blocks, and for a null state at level 2 we rewrite the BPZ equation for the chiral blocks as a~second-order~MLDE.

\subsubsection{A hybrid method}
\label{sec:hybrid}
For simple examples, any of these three methods are effective. For more complicated methods, where $\rho$ does not contain a congruence group in the kernel,
we propose a hybrid of the second and third methods, which we argue is effective. We will describe this method with an example we introduce in Section \ref{sec:mod4dim}.
There, we wish to find a core block with expansion\footnote{The overall factor of $2^{\frac{31}{3}}$ guarantees that the vacuum chiral block is correctly normalized (see~\eqref{eq:cfblockexp}).}
\begin{gather*}\renewcommand{\arraystretch}{1.1}
\underline{\varphi}(\tau)
=
2^{-\frac{31}{3}}
\begin{pmatrix}
q^{-31/24}(1+{\ccO}(q))\\
q^{-7/8}\bigl(n_{(1,3)}+{\ccO}(q)\bigr)\\
q^{11/24}\bigl(n_{(1,5)}+{\ccO}(q)\bigr)\\
q^{65/24}\bigl(n_{(1,7)}+{\ccO}(q)\bigr)
\end{pmatrix}
,
\end{gather*}
whose transformations under $\PSL_2(\mathbb{Z})$ are given by
\begin{gather*}
\cT
=
\begin{pmatrix}
\mathrm{e}\bigl(-\frac{7}{24}\bigr)&0&0&0\\
0&\mathrm{e}\bigl(-\frac{7}{8}\bigr)&0&0\\
0&0&\mathrm{e}\bigl(\frac{11}{24}\bigr)&0\\
0&0&0&\mathrm{e}\bigl(-\frac{7}{24}\bigr)
\end{pmatrix}
\end{gather*}
and
\begin{gather*}
\cS
=
\begin{pmatrix}
\frac{1}{\sqrt{2}}-\frac{1}{\sqrt{6}}& \sqrt{\frac{1}{\sqrt{3}}-\frac{1}{3}} & \hphantom{-}\frac{1}{\sqrt{3}} & \frac{1}{\sqrt{3}}\\
\sqrt{\frac{1}{\sqrt{3}}-\frac{1}{3}} & \frac{\sqrt{6}}{2}-\frac{1}{\sqrt{2}} & 0 & -\sqrt{\frac{2}{3} \bigl(\sqrt{3}-1\bigr)}\\
\frac{1}{\sqrt{3}} & 0 & -\frac{1}{\sqrt{2}} & \frac{1}{\sqrt{6}}\\
\frac{1}{\sqrt{3}} & -\sqrt{\frac{2}{3} \bigl(\sqrt{3}-1\bigr)} & \hphantom{-}\frac{1}{\sqrt{6}} &\frac{1}{\sqrt{2}}-\frac{\sqrt{6}}{3}
\end{pmatrix}.
\end{gather*}

The Rademacher sum method is not very helpful in this case, as it converges very slowly. This is because the desired leading powers are not very equitable: one component starts with~$q^{-31/24}$, whilst another starts with~$q^{65/24}$, which forces some leading coefficients to be very large for some components, adversely affecting convergence.

Our hybrid method begins with the MLDE method. The components $\varphi_i$ will be solutions to the MLDE \smash{$b_0D_{(0)}^4f+b_1D_{(0)}^3f+b_2D_{(0)}^2f+b_3D_{(0)}f+b_4f=0$}. Here, $b_j$ is a (weakly holomorphic) modular form for ${\rm SL}_2(\mathbb{Z})$ of weight $12+2j$ with leading term \smash{$q^{1}$} \big(since the products of~the components of $\underline{\varphi}$ has leading power $q^1$\big) and trivial multiplier \big(since ${\rm det}(\cT)=1$, which also guarantees ${\rm det}(\cS)={\rm det}(\cT)^{-3}=1$\big). Thus $b_0(\tau)=a\Delta(\tau)$, $b_1(\tau)=0$, $b_2(\tau)=bE_4(\tau)\Delta(\tau)$, $b_3(\tau)=c E_6(\tau)\Delta(\tau)$, and $b_4(\tau)=dE_4(\tau)^2\Delta(\tau)$. The constants $a,b,c,d\in\bbC$ satisfy the condition that there are solutions $f_i(\tau)$ to
\begin{gather*}
aD_{(0)}^4f+b E_4(\tau)D_{(0)}^2f+cE_6(\tau)D_{(0)}f+dE_4(\tau)^2f=0
\end{gather*} with expansions $f_1(\tau)=q^{-31/24}(1+{\ccO}(q))$,
$f_2(\tau)=q^{-7/8}(1+{\ccO}(q))$,
$f_3(\tau)=q^{11/24}(1+{\ccO}(q))$,
$f_4(\tau)=q^{65/24}(1+{\ccO}(q))$. Conveniently, there is a unique (up to rescaling by a constant) MLDE with the required properties: take coefficients $a=1$, $b=-\frac{1381}{288}$, $c=-\frac{1435}{864}$, $d=\frac{155155}{110592}$. This uniquely determines the normalized solutions $f_i(\tau)$, except that an arbitrary scalar multiple of~$f_4$ can be added to $f_1$. Choose any such $f_1$.

We are not yet done, as until now we have completely ignored the $S$-transformation: the MLDE method only retain information about {${\rm det}(\rho)$} and the solution $f$ transforms according to~${\tilde\cS=P^{-1}\cS P}$ and $\tilde\cT=P^{-1}\cT P$ under the $S$- and $T$-transformation. In the above we explicitly choose $\underline{f}$ to so that $\tilde \cT = \cT$, but it remains to determine $\cS$ and from there the change of basis $P$. In other words,
we need to determine how the $f_i(\tau)$ transform under $\tau\mapsto-1/\tau$, as this will then tell us how to obtain the desired $\varphi_i$ from these $f_i$. This is where {the second (Riemann--Hilbert) method} appears. First, we need a diagonal matrix $\Lambda$ compatible with $\cT$ and with trace $-2$ (recall~\eqref{eq:trace}): the choice $\Lambda={\rm diag}\bigl(-\frac{31}{24},-\frac{15}{8},-\frac{13}{24},\frac{41}{24}\bigr)$ lets us recycle $\underline{f}$ as the first column of~$\Xi$. The other columns of the corresponding fundamental matrix $\Xi$ can be obtained from $\underline{f}$ by repeatedly applying to it the differential operators \smash{$\frac{E_4E_6}{\Delta}D_{(0)}$},~\smash{$\frac{E_4^2}{\Delta}D^2_{(0)}$},~\smash{$\frac{E_6}{\Delta}D^3_{(0)}$} and taking linear combinations over $j(\tau)$ (recall~\eqref{vvmfgen}). This step is just linear algebra. Each of these columns is~simply required to have a leading $q$-expansion compatible with leading term~$\Xi[0]=I$. In~particular, the remaining three columns of $\Xi$ are linear combinations of~$\underline{f}$,~\smash{$\frac{E_4E_6}{\Delta}D_{(0)}\underline{f}$},~\smash{$\frac{E_4^2}{\Delta}D^2_{(0)}\underline{f}$},~\smash{$\frac{E_6}{\Delta}D^3_{(0)}\underline{f}$}. It suffices to record their first nontrivial coefficients:
\begin{gather*}\renewcommand{\arraystretch}{1.3}
\Xi[1]=\chi=\begin{pmatrix}\frac{41633}{21}&-\frac{9068526720}{2107}&-\frac{93109120}{189}&-\frac{163826816}{1161}\\ 1&-\frac{501885}{301}&\frac{520}{9}&-\frac{608440}{11997}\\ 1&-\frac{577368}{301}&-\frac{5369}{9}&-\frac{121688}{1935}\\ 1&-\frac{1994544}{301}&-\frac{1520}{63}&\frac{16031}{387}\end{pmatrix}.
\end{gather*}
Given $\chi$ and $\Lambda$, the full $q$-expansion of $\Xi$ is easily obtained by the recursion~\eqref{eq:recurs}.
Again, obtaining $\chi$ and hence $\Xi$ is elementary if a little tedious.

The point is that {$\Xi(\tau)=\tilde \cS^{-1}\,\Xi(-1/\tau)$}. Choose any $\tau_0$ close to but different from $\sqrt{-1}$, then~$\Xi(\tau_0)$ will be invertible and \smash{$\tilde\cS=\Xi(-1/\tau_0)\Xi(\tau_0)^{-1}$}. So if we compute the first several terms in the $q$-expansion of $\Xi$, and evaluate it at both $\tau_0$ and $-1/\tau_0$, then we get an approximation of~$\tilde\cS$. Both $|{\rm e}(\tau_0)|$ and $|{\rm e}(-1/\tau_0)|$ can be less than $0.002$, so convergence of the $q$-expansion for~$\Xi$ should be quite fast.

We know that
\begin{align*}
&2^{\frac{31}{3}}\varphi_1(\tau)=f_1(\tau)+\alpha f_4(\tau),\\
&2^{\frac{31}{3}}\varphi_2(\tau)=\beta f_2(\tau),\\
&2^{\frac{31}{3}}\varphi_3(\tau)=\gamma f_3(\tau),\\
&2^{\frac{31}{3}}\varphi_4(\tau)=\delta f_4(\tau),
\end{align*}
for constants $\alpha,\beta,\gamma,\delta\in\bbC$ such that the corresponding $\underline{\varphi}(\tau)$ transforms with respect to $\cS$. Computing $\Xi$ up to $q^9$, and choosing $\tau_0=1.0002{\rm i}$, we obtain $n_{(1,1)}=1$, $n_{(1,3)}\approx 26.7262690$, $n_{(1,5)}\approx 102595.534$, $n_{(1,7)}\approx1.79592621\times10^9$. We will see shortly how accurate this is.

This example illustrates the following point.
The weakness of the second (Riemann--Hilbert) method is in obtaining one vvmf, but that is provided by the third (MLDE) method. Once one vvmf is obtained, the second method quickly determines all (equivalently, the generators $\Xi$). The~weakness of the third method is that it has no way to accurately and effectively determine the $S$-matrix, but this can be obtained to whatever desired accuracy using $\Xi$. The hybrid method successfully blends the strengths of the second and third methods.

Note that the eigenvalue e$(-7/24)$ of $\cT$ has multiplicity 2. If the dimension of an ${\rm SL}_2(\mathbb{Z})$-representation is $\le5$, and some eigenvalue of $\cT$ has multiplicity $>1$, then $\rho$ will be reducible~\cite{Tuba:2001}. Here, $P^{-1}\rho P$ is a direct sum of two 2-dimensional representations, where
\begin{gather*}
P=\begin{pmatrix} 0& \sqrt{3\sqrt{3}-3}& 0& 2\sqrt{3}\\ 3& 0& 0& 0\\ 0& 0& 3& 0\\ 0& -\sqrt{6\sqrt{3}-6}& 0& \sqrt{6}\end{pmatrix}.
\end{gather*}
Because of this, we can solve this example exactly. From~\cite[Section 4.2]{Ga} with the choices $\Lambda_1={\rm diag}\bigl(-\frac{15}{8},\frac{17}{24}\bigr)$ and $\Lambda_2={\rm diag}\bigl(-\frac{13}{24},-\frac{7}{24}\bigr)$, we obtain matrices $\Xi_1$ and $\Xi_2$, and the columns of the matrix product $P\bigl(\begin{smallmatrix}\Xi_1&0\\ 0&\Xi_2\end{smallmatrix}\bigr)$ are free generators (over the polynomial ring $\bbC[j(\tau)]$) of the full space \smash{$\cM^!(\rho)$} of vvmf, which includes the desired $\underline{\varphi}$. In this way we can quickly compute
the exact values
\begin{gather*}n_{(1,3)}={346921\over 32805}2^{\frac16}\sqrt{3+5/\sqrt{3}}\frac{\Gamma(7/12)^3}{\pi\Gamma(3/4)}\approx26.72643460,\\
 n_{(1,5)}=\frac{376960}{9}3^{\frac34} \frac{\Gamma(11/12)\Gamma(7/12)}{\Gamma(3/4)^2}\approx102595.517,\\
 n_{(1,7)}=\frac{491455971328}{387}\sqrt{2}\approx1.79592687\times 10^9.
 \end{gather*}
 We see that the convergence of our hybrid method is quite fast.

\section{Physical correlators}
\label{sec:corr}
We now turn our attention to the correlation functions of the rational CFT, built out of the physical operators which involve both chiral and anti-chiral degrees of freedom. Correlation functions can always be expressed as sesquilinear combinations of chiral and anti-chiral blocks. In a rational CFT, the physical Hilbert space $\cH$ can be decomposed as
\begin{gather*}
\displaystyle{\cH={\bigoplus_{\substack{M\in\Upsilon\\ \tilde{M}\in\tilde\Upsilon}} \cZ_{M,\tilde{M}}M\otimes \tilde{M}}},
\end{gather*}
where the sum is over the finitely many irreducible modules of the left- and right-moving chiral algebras, $\cV$ and $\tilde\cV$, and {$\cZ_{M,\tilde{M}}$} are non-negative integers corresponding to the multiplicities. Here and everywhere else we use the tilde notation to indicate quantities associated to the right chiral algebra.
The multiplicities determine the modular invariant partition function of the RCFT:
\begin{gather*}
Z_{T^2}(\tau,\bar\tau) = {\sum_{M,\tilde{M}} \mathcal{Z}_{M,\tilde{M}} \chi_{M}(\tau)\overline{\chi_{\tilde{M}}(\tau)}},
\end{gather*}
where {$\chi_M(\tau)$ are the graded dimensions defined in~\eqref{char}.}
The modular invariance of $Z_{T^2}$ (under~$\PSL_2(\ZZ)$) requires the matrix $\cZ_{M,\tilde{M}}$ to satisfy
\begin{equation}
\label{torus_mod_inv}
{\cZ_{M,\tilde{M}} = \bigl(\mathbf{S} \mathcal{Z} \mathbf{\widetilde S}^\dag\bigr)_{M,\tilde{M}}=\bigl(\mathbf{T} \mathcal{Z} \mathbf{\widetilde T}^\dag\bigr)_{M,\tilde{M}}},
\end{equation}
where $\mathbf{S},\mathbf{T}$ \smash{\big(resp.\ $\widetilde{\mathbf{S}}$, $\widetilde{\mathbf{T}}$\big)} are the ${\rm SL}_2(\mathbb{Z})$ modular transformation matrices of the chiral (anti-chiral) sectors of the CFT. In particular, for the case $\cV\cong \tilde \cV$, the unitary of the $T$ and $S$ matrices imply that $\cZ$ commutes with $\mathbf{S}$ and $\mathbf{T}$.

It was shown in~\cite{Moore:1988ss} that if $\cV$ and $\tilde \cV$ are the maximally-extended chiral algebras, then there exists an isomorphism $\Omega$ of the fusion algebra of $\tilde\cV$ with that of $\cV$ such that
\begin{equation}\label{OmegaM}
\cZ_{M,\tilde{M}} = \delta_{M,\Omega(\tilde{M})}.
\end{equation}
In fact, $\Omega$ is much more: it is a tensor equivalence between the corresponding modular tensor categories of $\cV$ and $\tilde\cV$, which means that it establishes equivalences between all mapping class group representations, matches up intertwiners of $\cV$ and $\tilde\cV$, etc. Although $\cV$ and $\tilde\cV$ do not~have to be isomorphic as VOAs, $\Omega$ establishes an equivalence of their combinatorial structures.
Without loss of generality, in the following we assume that $\cV$ and $\tilde \cV$ are maximally-extended chiral algebras, which lead
 to simpler expressions.

In this paper, we will mainly focus on $n$-point correlation functions on the sphere. Recall the discussion of chiral blocks on the sphere from Section \ref{sec:bkg}. For each $i=1,\dots,n$, fix any sector $M_i\otimes\Omega(M_i)\in \Upsilon(\cV)\times\Upsilon(\tilde{\cV})$ as in~\eqref{OmegaM}. Choose any quasi-primary states ${\phi^i=\phi_i\otimes\tilde{\phi}_i\in M_i\otimes\Omega(M_i)}$. We say $\phi$ is quasi-primary if it is annihilated by the Virasoro generator $L_1$ and is an eigenvector for $L_0$, the eigenvalue being its conformal weight $h_{\phi_i}$ (similarly for $\tilde{\phi}$). We write
 \begin{gather*}
\bigl\langle \phi^1(z_1,\bar z_1)\cdots
\phi^n(z_n,\bar z_n)\bigr\rangle\prod_{i=1}^n{\rm d}z_i^{h_{\phi_i}}{\rm d}\bar{z}_i^{{h}_{\tilde{\phi}_i}}
\end{gather*}
for the associated correlator. This will be a sesquilinear combination of chiral blocks,
\begin{gather*}
\langle \mathbf{1},\Phi_1(\phi_1,z_1)\circ\cdots\circ \Phi_n(\phi_n,z_n)\mathbf{1}\rangle\overline{\bigl\langle \mathbf{1},\tilde{\Phi}_1\bigl(\tilde{\phi}_1,{z}_1\bigr)\circ\cdots\circ \tilde{\Phi}_n\bigl(\tilde{\phi}_n,{z}_n\bigr)\mathbf{1}\bigr\rangle}\prod_{i=1}^n{\rm d}z_i^{{h}_{\phi_i}}{\rm d}\bar{z}_i^{{h}_{\tilde{\phi}_i}},
\end{gather*}
where the intertwiners $\Phi_i$ and $\tilde{\Phi}_i=\Omega(\Phi_i)$ run through a basis of \smash{$\cY_{N_i,M_i}^{N_{i+1}^*}$} and \smash{$\tilde{\cY}_{\Omega(N_i),\Omega(M_i)}^{\Omega(N_{i+1})^*}$} respectively.

In the rest of this section, we focus on the case where the states $\phi_i$ and $\tilde{\phi}_i$ are $\cV$- and $\tilde{\cV}$-primaries, i.e., states with conformal weight $h_{M_i}$ and \smash{$h_{\Omega({M}_i)}$}, respectively. As a result, we will often label the correlators simply by the choices of the modules $M_i$, $\tilde M_i$ \big(as opposed to $\phi_i \in M_i$, \smash{$\tilde \phi_i \in \tilde M_i$}\big).
In contrast with the chiral blocks, the correlation functions for the full CFT are invariant under the action of the mapping class group. Importantly, this statement of crossing symmetry also applies to non-rational 2d CFTs.

Consider first the two-point correlation functions. These are just given by
\begin{gather*}
\bigl\langle \phi^1(z_1,\bar z_1)\,\phi^2(z_2,\bar z_2)\bigr\rangle \prod_{i=1}^2{\rm d}z_i^{h_{M_i}}{\rm d}\bar{z}_i^{{h}_{{\tilde M}_i}}
=
\phi_{2}(\phi_{1})\,\tilde \phi_{2}\bigl(\tilde \phi_{1}\bigr)
\frac{\delta_{M_1,M_2^*}\delta_{\tilde M_1,\tilde M_2^*}}{z_{12}^{2h_{M_1}}\bar{z}_{12}^{2h_{{\tilde M}_1}}}\prod_{i=1}^2{\rm d}z_i^{h_{M_i}}{\rm d}\bar{z}_i^{{h}_{{\tilde M}_i}},
\end{gather*}
and by a choice of normalization we can set
\begin{gather*}
\phi_{2}(\phi_{1})=\tilde \phi_{2}\bigl(\tilde \phi_{1}\bigr)=1.
\end{gather*}
The two-point correlation functions are invariant under the action of the extended mapping class group generated by the Dehn twist, which is just $\ZZ$, if and only if{\samepage
\begin{gather*}
h_{M_i}= {h_{\Omega({M}_i)}} \mod \mathbb Z.
\end{gather*}
Note that this is also required by the invariance of $Z_{T^2}$ under $\tau\mapsto\tau+1$.}

Next we look at three-point correlation functions. We can introduce \emph{physical vertex operators} which take the form
\begin{equation}\label{eqn:physical vertex op}
{\mathbf\Phi}^i_{j,k}(z,\bar z) = d^{i}_{jk} {\Phi^{M_i}_{M_j,M_k}(\phi_k,z) \tilde\Phi^{\tilde M_i}_{\tilde M_j,\tilde M_k}\bigl(\tilde\phi_k,\bar z\bigr)},
\end{equation}
where
\begin{gather*}
\tilde M_i=\Omega(M_i),\qquad \tilde M_j=\Omega(M_j),\qquad \tilde M_k=\Omega(M_k), \\
 {\Phi_{M_j,M_k}^{M_i}\in\cY_{M_j,M_k}^{M_i}},\qquad {\Phi_{M_j,M_k}^{M_i}=\Omega\bigl(\Phi_{M_j,M_k}^{M_i}\bigr)\in\cY_{\tilde M_j,\tilde M_k}^{\tilde M_i}},
\end{gather*}
 and $\phi_k\in M_k$, \smash{$\tilde\phi_k\in \tilde M_k$}.
Denote the module \smash{$\cV\otimes \tilde \cV$} by the label $0$. The OPE coefficients satisfy, among other things,
\begin{gather*}
d^i_{0i}=d^i_{i0}=1
\qquad \text{and} \qquad
d^0_{i^*i}=\pm 1,
\end{gather*}
where the sign can be shown to be always positive for unitary RCFTs~\cite{Moore:1988ss}. Using
$d^0_{i^*i}$
to raise and lower indices,
we can then write the three-point correlator as follows {(cf.\ \eqref{def:structure_constant})}:
\begin{gather}
\langle {\mathbf\Phi}_1(z_1,\bar z_1){\mathbf\Phi}_2(z_2,\bar z_2){\mathbf\Phi}_3(z_3,\bar z_3)\rangle\nonumber\\
\qquad{}=
\langle 0 \vert {\mathbf\Phi}^0_{1,1^*}(z_1,\bar z_1){\mathbf\Phi}^{1^*}_{2,3}(z_2,\bar z_2) {\mathbf\Phi}^3_{3,0}(z_3,\bar z_3)\vert 0\rangle\nonumber\\
\qquad{}=
d_{123}\cF_{M_1,M_2,M_3}(z_1,z_2,z_3){\tilde \cF}_{{\tilde M}_1,{\tilde M}_2,{\tilde M}_3}(\bar z_1,\bar z_2,\bar z_3)\nonumber\\
\qquad{}=
\biggl(\prod_{{1\leq i< j\leq 3}}z_{ij}^{h-2h_i-2h_j}{\bar z}_{ij}^{\tilde h-2h_i-2h_j}\biggr) d_{123}{\varphi}_{M_1,M_2,M_3}{{\tilde \varphi}}_{{\tilde M}_1,{\tilde M}_2,{\tilde M}_3}.\label{eq:3ptcor}
\end{gather}
In the above, $h= h_{M_1}+h_{M_2}+h_{M_3}$ and similarly $\tilde h = h_{{\tilde M}_1}+ h_{{\tilde M}_2}+ h_{{\tilde M}_3}$. Notice that in principle one has the freedom to rescale the left and right chiral structure constants $\varphi$, $\tilde \varphi$, while simultaneously rescaling the OPE coefficients $d$, such that the three-point correlator~\eqref{eq:3ptcor} remains the same.
A~natural choice for the normalisation of the intertwiners is to choose \smash{$\Phi_{M_2,M_3}^{M_1^\ast} \in \cY_{M_2,M_3}^{M_1^\ast}$} such that (cf.\ \eqref{def:structure_constant} and~\eqref{def:3point_intertwiners})
\begin{gather*}
\lim_{u\to \infty} u^{2h_{M_1}} {\mathcal F}_{M_1,M_2,M_3}(u,z,0) = \langle M_1 | \Phi(z) | M_3\rangle,
\end{gather*}
in which case one has \smash{$\varphi^\Phi_{M_1,M_2,M_3}=c^\Phi_{M_1,M_2,M_3}$} \eqref{def:3point_intertwiners} and similarly for \smash{${\tilde \varphi}_{{\tilde M}_1,{\tilde M}_2,{\tilde M}_3}^{\tilde \Phi}$}.
Finally, let us turn to the four-point correlation functions;
using the physical vertex operators, they can be expanded as follows:\footnote{For notational convenience, we will make the simplifying assumption that all $\cN_{P,M_2}^{M_1}\cN_{M_3,M_4}^P\le 1$. More generally, one needs to sum over inequivalent choices of physical vertex operators ${\bf \Phi}$~\eqref{eqn:physical vertex op}.}
\begin{gather}
\Biggl\langle\prod_{i=1}^4\phi^i(z_i,\bar z_i)\Biggr\rangle
=\sum_{p=P\otimes {\tilde P}}d_{12p}\,d^{p}_{34} \mathcal{F}^{P}_{M_1,M_2,M_3,M_4}(z_1,z_2,z_3,z_4){\tilde{\cF}}^{\,{\tilde P}}_{{\tilde M}_1,{\tilde M}_2,{\tilde M}_3,{\tilde M}_4}(\bar z_1,\bar z_2,\bar z_3,\bar z_4) \nonumber
\\ \hphantom{\biggl\langle\prod_{i=1}^4\phi^i(z_i,\bar z_i)\biggr\rangle}{}
= z_{\rm 4-pt} \!\sum_{p=P\otimes {\tilde P}}(\cA_{1234})_{P\tilde P}\,\varphi^{P}_{M_1,M_2,M_3,M_4}(\tau){\tilde{\varphi}}^{\,{\tilde P}}_{{\tilde M}_1,{\tilde M}_2,{\tilde M}_3,{\tilde M}_4}(\bar \tau),
\label{eq:schan}
\end{gather}
where (cf.\ \eqref{eq:muij})
\begin{gather*} z_{\rm 4-pt} := \prod_{1\leq i< j\leq 4}z_{ij}^{\mu_{ij}}{\bar z}_{ij}^{\tilde\mu_{ij}}
\end{gather*}
and we have defined the matrix $\cA_{1234}$ by setting
\begin{equation}\label{def:cA}
(\cA_{1234})_{P\tilde P} := d_{12p}\,d^{p}_{34}.
\end{equation}
In the above, $p = P\otimes {\tilde P}$ runs through all internal channels, corresponding to physical states for which the spaces of chiral and anti-chiral blocks both have nonzero dimension.
Based on the $z_i$-dependence of the correlator~\eqref{eq:schan}, it is natural to define the {\em core correlator}
\begin{equation}\label{corr_def2}
{F}_{\phi^1,\phi^2,\phi^3,\phi^4}(\t,\bar\t):=\frac{\bigl\langle\prod_{i=1}^4{\phi^i}(z_i,\bar z_i)\bigr\rangle}{z_{\rm 4-pt}}.
\end{equation}
The fact that the physical correlators are single-valued functions of the location $z_i$ of the insertions, combined with the above and the identification~\eqref{lambda_function} between the $\lambda$-function and the cross ratio,
 then immediately leads to the fact that $\bigl\langle\prod_{i=1}^4\phi^i(z_i,\bar z_i)\bigr\rangle/z_{\rm 4-pt}$ is a non-holomorphic function on $\HH$ invariant under the action of {${\rm P}\Gamma(2)$}, and is moreover invariant under ${\rm P}\Gamma_0(2)$ when two of the four insertions are identical, and under the full modular group $\PSL_2(\ZZ)$ when at least three insertions are.

More generally, one has the following relation analogous to~\eqref{torus_mod_inv} for the torus partition functions:
\begin{Theorem}
The OPE coefficients~\eqref{def:cA} satisfy
\begin{gather*}
\cA_{1234} =(\rho_T)^{-1} \cA_{1243} \tilde \rho_{ T} = (\rho_S)^{-1} \cA_{1423} \tilde \rho_{S},
\end{gather*}
where $\rho_T$ and $\rho_S$ are as given in~\eqref{eq:tpq} and~\eqref{eq:spq} and similarly for $\tilde \rho$.
\end{Theorem}
\begin{proof}
Alternatively to the ``$s$-channel'' expansion above, we can choose to expand the correlator in the ``$t$-channel'' and ``$u$-channel'' and obtain
\begin{align*}
{F}_{\phi^1,\phi^2,\phi^3,\phi^4}(\t,\bar\t)
&{}=\sum_{p=P\otimes {\tilde P}}(\cA_{1234})_{P\tilde P} {\var}^{P}_{M_1,M_2,M_3,M_4}(\tau){\tilde{\var}}^{ {\tilde P}}_{{\tilde M}_1,{\tilde M}_2,{\tilde M}_3,{\tilde M}_4}(\bar \tau)\\
&{}=\sum_{p=P\otimes {\tilde P}}(\cA_{1243})_{P\tilde P} \var^{P}_{M_1,M_2,M_4,M_3}(\tau+1) {\tilde{\var}}^{ {\tilde P}}_{{\tilde M}_1,{\tilde M}_2,{\tilde M}_3,{\tilde M}_4}(\bar \tau+1)\\
&{}=
\sum_{p=P\otimes {\tilde P}}(\cA_{1423})_{P\tilde P}
\var^{P}_{M_1,M_4,M_3,M_2}\biggl(-{1\over\tau} \biggr)
{\tilde{\var}}^{ {\tilde P}}_{{\tilde M}_1,{\tilde M}_4,{\tilde M}_3,{\tilde M}_2}\biggl(-{1\over\bar\tau}\biggr).
\end{align*}
The desired statement then follows from the unitarity of $\rho$ and $\tilde \rho$, and upon applying~\eqref{STsigma}.
\end{proof}

\begin{Corollary}\label{cor_corecor}The core correlator satisfies
\begin{gather*}
{F}_{\phi^1,\phi^2,\phi^3,\phi^4}(\t,\bar\t)={F}_{\phi^1,\phi^2,\phi^3,\phi^4}\biggl(\frac{a\t+b}{c\t+d},{\frac{\overline{a\t+b}}{\overline{c\t+d}}}\biggr)
\end{gather*}
for all \smash{$\bigl(\begin{smallmatrix}a& b\\ c&d \end{smallmatrix}\bigr)\in \Gamma(2)$}. Moreover, the above holds for all \smash{$\bigl(\begin{smallmatrix}a& b\\ c&d \end{smallmatrix}\bigr)\in {\rm SL}_2(\ZZ)$} if $\phi_2=\phi_3=\phi_4$.
\end{Corollary}

\section{Examples}
\label{sec:examples}
In this section, we present a few examples to illustrate the theorems in the main text. While we organize the examples into various classes, we stress that our techniques apply uniformly to all examples, regardless of the class of theories they belong to. For all examples we choose, the conformal blocks will turn out to be elements of a one-dimensional space of vvmf's, and therefore we will be able to determine them with the braiding matrices of the CFT as the sole input. More general examples and more detailed analysis will be presented in the companion paper~\cite{followup}.
\subsection{Virasoro minimal models}
Our first class of examples arises in Virasoro minimal models. It should be emphasized that these are misleadingly simple: there is no difference between Virasoro-primary and VOA-primaries, the spaces of VOA-primaries are all 1-dimensional, and all fusion coefficients are either $0$ or $1$; it is not necessary for these properties to hold in order for our methods to be applicable.

Recall that Virasoro minimal models are parametrized by two positive integers $ p>p' $. The~left and right chiral algebras are just two copies of the Virasoro algebra at central charge
 \begin{gather*}
 c = 1-6\frac{(p-p')^2}{p p'},
 \end{gather*}
possibly extended. More precisely, the left and right chiral algebras are isomorphic to the canonical VOA structure on the irreducible Virasoro module with $h=0$ and the given $c$. We denote {this chiral algebra} as $ \mathcal{M}(p,p') $. A minimal model is unitary when $p = p'+1$.

The minimal model $\mathcal{M}(p,p')$ possesses the following collection of irreducible modules:
\begin{equation}\label{notation_mmmodule}
M_{r,s}\qquad\text{with}\quad 1\leq r < p',\ 1\leq s<p,
\end{equation}
subject to the identification $M_{r,s} = M_{p'-r,p-s}$. The conformal weight of $M_{r,s}$ is given by
\begin{gather*}
h_{r,s} = \frac{(pr -p's)^2-(p-p')^2}{4p'p}.
\end{gather*}
Furthermore one has $M_{r,s}=M_{r,s}^*$,
that is, the charge conjugation matrix $\mathbf{C}$ is the identity matrix. We denote the corresponding Virasoro primaries by $\phi_{r,s}= \phi_{p'-r,p-s}$.

The fusion rules {for the $\cM(p,p')$ minimal model} are given as follows:
\begin{equation}
\label{fusionruleMM}
M_{r,s}\otimes M_{m,n} = \bigoplus_{\substack{k = 1+ |r - m|\\ k+r + m = 1 \text{ mod } 2}}^{k_{\max}}\bigoplus_{\substack{l = 1+|s- n|\\l+s+n=1 \text{ mod } 2}}^{l_{\max}}M_{k,l},
\end{equation}
where
\begin{align*}
&k_{\max}= \min (r+m-1,2p'-1-r-m),\\
&l_{\max}= \min (s+n-1,2p-1-s-n).
\end{align*}
In particular, all fusion coefficients are either $0$ or $1$.

Explicit expressions are known for the braiding and {fusing} matrices of the minimal models~\mbox{\cite{Dotsenko:1984nm,Dotsenko:1984ad}}. In the following examples, we employed the \textsc{Mathematica} code provided in~\cite{Esterlis:2016psv} to compute the braiding matrices required to determine the $\PSL_2(\mathbb{Z})$ matrices $\cS_\cD$.

For the $\cM(p,p')$, clearly chiral blocks and conformal blocks coincide. It hence suffices to consider here the insertions to be $\phi_i=\phi_{r_i,s_i}$.
Virasoro conformal blocks have been determined by a variety of techniques, including by making explicit use of the Virasoro algebra, by exploiting the Zamolodchikov recursion relations~\cite{Zamolodchikov84}, or for minimal models by solving the BPZ equation (see Appendix~\ref{app:BPZ}). More recently, Perlmutter has found an explicit solution to Zamolodchikov's recursion relations~\cite{Perlmutter:2015iya} which provides expressions at arbitrary orders in the perturbation series of the Virasoro conformal blocks.
For the discussion that follows, it will be sufficient to know the leading order term in the $q$-expansion of the core blocks. From~\eqref{eq:lambdainfty} and~\eqref{eq:coreLO} we immediately see that
\begin{equation}
\varphi^{N}_{\phi_1,\phi_2,\phi_3,\phi_4}(w) = C q^{-\frac{h}{6}+\frac{h_{N}}{2}}(1+\ccO(q)),
\label{eq:cfblockexp}
\end{equation}
where \smash{$C = c_{M_1,M_2,N} c_{N^\ast,M_3,M_4} 2^{8(-\frac{h}{6}+\frac{h_{N}}{2})}$}, and
$M_i=M_{r_i, s_i}$ and $N=M_{r_N,s_N}$.

The examples below demonstrate how the technology developed in the previous sections efficiently determines the Virasoro blocks in terms of vector-valued modular forms, often leading to exact, analytic expressions.

\subsubsection[M(4,3)]{$\boldsymbol{\cM(4,3)}$}

The Ising model $\cM(4,3)$ has three independent primaries, whose properties are summarized in the following table:
\begin{center}
\begin{tabular}{ccc}
\toprule
Primary & $(r,s)$ & $h$\\
\toprule
1& $(1,1)$ & 0\\
 $\sigma$ &$(2,2)$ &$\frac{1}{16}$ \\[1mm]
 $\epsilon$ &$(2,1)$ & $\frac{1}{2}$\\\bottomrule
\end{tabular}
\end{center}

Given this chiral theory, the only modular-invariant way to combine left- and right-movers is to take the diagonal theory:
\begin{gather*}
Z_{T^2} = \vert\chi_1\vert^2+\vert\chi_\sigma\vert^2+\vert\chi_\epsilon\vert^2.
\end{gather*}

We can arrange all nontrivial core blocks into three vector-valued modular forms:
\begin{enumerate}\itemsep=0pt
\item $\cD=(\epsilon,\epsilon,\epsilon,\epsilon)$.
 The space of core blocks is one-dimensional and has the following leading order behavior:
 \begin{gather*}
 \varphi^1_{\epsilon,\epsilon,\epsilon,\epsilon}(\tau)= q^{-{1\over3}}(N_1+\ccO(q)).
 \end{gather*}

It transforms as a one-dimensional representation of $\PSL_2(\mathbb{Z})$ with
\begin{gather*}
\cT = \biggl(\ex\biggl(-\frac{1}{3}\biggr)\biggr),\qquad \cS = (1).
\end{gather*}
The space of vvmf's transforming under this representation, and with prescribed singular behavior, $\cM^!_0\bigl(\rho_\cD,-\tfrac{1}{3}\bigr)$, is itself one-dimensional. A straightforward way to see this is by applying the method of Section \ref{sec:forms}: from equation~\eqref{eq:trace}, we obtain
\begin{gather*}
\operatorname{Tr}\Lambda = -\frac13 \ \Longrightarrow\ \Lambda = \biggl(-\frac13\biggr).
\end{gather*}
Therefore, the principal part of $ \varphi^1_{\epsilon,\epsilon,\epsilon,\epsilon}$ is
\begin{gather*}
\cP_\Lambda \varphi^1_{\epsilon,\epsilon,\epsilon,\epsilon}(\tau) = q^{-{1\over3}}N_1
\end{gather*}
and so $ \varphi^1_{\epsilon,\epsilon,\epsilon,\epsilon}$ is uniquely determined by its normalisation. The generator of $\cM^!_0(\rho_\cD)$ is a~famous modular form:
\begin{gather*}
j(\tau)^{\frac13} = \frac{E_4(\tau)}{\eta(\tau)^8} = q^{-{1\over3}}\bigl(1+248 q +4124 q^2+34752 q^3+\cdots\bigr) = \chi_{\bf{1}}^{E_{8,1}}(\tau),
\end{gather*}
which coincides with the unique character \smash{$\chi_{\bf{1}}^{E_{8,1}}(\tau)$} at level 1 of the $E_8$ Kac--Moody algebra corresponding to the trivial representation $\bf{1}$.
As we will see in Section \ref{sec:spheretorus} and in more details in~\cite{followup}, this is not a mere coincidence.

A straightforward application of the Rademacher sum (where we restrict the sum to ${c \leq 500}$) reproduces this result to very good approximation:
\begin{gather*}
 \varphi^1_{\epsilon,\epsilon,\epsilon,\epsilon}(\tau)\simeq N_1q^{-{1\over3}}\bigl(1.001+247.991 q+4124.003q^2+34752.002 q^3+\cdots\bigr).
\end{gather*}

The normalization of the conformal block is uniquely fixed according to equation~\eqref{eq:cfblockexp}:
\begin{gather*}
 \varphi^1_{\epsilon,\epsilon,\epsilon,\epsilon}(\tau)= 2^{-8/3} c_{1\epsilon\epsilon}c^{1}_{\epsilon\epsilon} j(\tau)^{1/3}= \biggl(\frac{j(\tau)}{256}\biggr)^{1/3},
\end{gather*}
where the structure constants $c_{1\epsilon\epsilon}=c^1_{\epsilon\epsilon}=1$ are fixed by the normalization of the $\epsilon$ primary. Indeed, if we re-write the core block as a function of the cross-ratio $w = \lambda(\tau)$ using equation~\eqref{eq:jlambda} we recover the well-known expression for the conformal block~\cite{AlvarezGaume:1989vk}:
\begin{equation}\label{core_epsilon}
\cF^1_{\epsilon,\epsilon,\epsilon,\epsilon}(z_1,z_2,z_3,z_4)=\biggl(\prod_{i<j}z_{ij}^{-{1\over3}}\biggr)\frac{1-w+w^2}{w^{2/3}(1-w)^{2/3}}.
\end{equation}
For the physical correlator (for the diagonal theory), one obtains
\begin{gather*}
\langle \epsilon(z_1,\bar{z_1})\cdots\epsilon(z_4,\bar{z_4})\rangle = {2^{-{16\over 3}}} \biggl(\prod_{i<j}\vert z_{ij}\vert^{-{2\over3}}\biggr) Z^{E_{8,1}}_{T^2}(\tau,\bar{\tau}).
\end{gather*}
In other words, the corresponding core correlator coincides {up to a numerical prefactor} with the torus partition function of the level-1 $E_8$ WZW model {(cf.\ \eqref{corr_def2})}:
\begin{gather*}
{F_{\epsilon,\epsilon,\epsilon,\epsilon}}(\tau,\bar\tau) ={2^{-{16\over 3}}} Z^{E_{8,1}}_{T^2}(\tau,\overline{\tau}) = {2^{-{16\over 3}}} \bigl\vert\chi_{\bf 1}^{E_{8,1}}(\tau)\bigr\vert^2.
\end{gather*}

\item $\cD=(\sigma,\sigma,\sigma,\sigma)$.
In this case, the core blocks assemble into a two-dimensional vvmf:
\begin{gather*}
\underline{\varphi}(\tau)
=
\begin{pmatrix}
\varphi^1_{\sigma,\sigma,\sigma,\sigma}(\tau)\\
\varphi^\epsilon_{\sigma,\sigma,\sigma,\sigma}(\tau)
\end{pmatrix}
 =
 \begin{pmatrix}
 q^{-{1\over24}}(N_1+\ccO(q))\\
 q^{{5\over24}}(N_\epsilon+\ccO(q))
\end{pmatrix}
.
\end{gather*}
The vvmf transforms as a representation $\rho_\cD$ of $\PSL_2(\mathbb{Z})$ determined by the matrices
\begin{gather*}
\cT_\cD = \begin{pmatrix}\mathrm{e}\bigl(-\frac{1}{24}\bigr)&0 \\ 0&\mathrm{e}\bigl(\frac{5}{24}\bigr)\end{pmatrix},\qquad \cS_\cD = \frac{1}{\sqrt{2}}\begin{pmatrix}1&\hphantom{-}{} 1\\1&-1\end{pmatrix}.
\end{gather*}
Again, according to the discussion in Section \ref{sec:forms}, $\dim \bigl(\cM^!_0\bigl(\rho_\cD,{\bigl(-\tfrac{1}{24},\tfrac{5}{24}\bigr)}\bigr)\bigr) = 1$: indeed, equation~\eqref{eq:trace} gives
\begin{gather*}
\operatorname{Tr}\Lambda = -\frac{5}{6}.
\end{gather*}
We can therefore pick
\begin{gather*}
\Lambda =
\begin{pmatrix}
-\frac{1}{24} & \hphantom{-} 0 \\
\hphantom{-}0 & -\frac{19}{24}
\end{pmatrix}
\end{gather*}
so that
\begin{gather*}
\cP_\Lambda\underline{\varphi}(\tau)
=
\ex(\Lambda \tau)
\begin{pmatrix}
N_1\\
0
\end{pmatrix},
\end{gather*}
and again the vvmf is determined up to an overall factor.
The single basis vector of $\cM^!_0\bigl(\rho_\cD,{\bigl(-\tfrac{1}{24},\tfrac{5}{24}\bigr)}\bigr)$ is a well-known vector-valued modular form: its components are the two characters at level 1 of the $A_1$ Kac--Moody algebra corresponding to the trivial ($\bf{1}$) and fundamental ($\bf{2}$) representations:
\begin{gather*}
\begin{pmatrix}
\chi_{\bf{1}}^{A_{1,1}}(\tau)\vspace{1mm}\\
\chi_{\bf{2}}^{A_{1,1}}(\tau)
\end{pmatrix}
=
\begin{pmatrix}
\dfrac{\theta_3(2\tau)}{\eta(\tau)}\vspace{1mm}\\
\dfrac{\theta_2(2\tau)}{\eta(\tau)}
\end{pmatrix}=
\begin{pmatrix}
q^{-\frac{1}{24}}\bigl(1+3 q+4 q^2+7 q^3+\cdots\bigr)\\
q^{\frac{5}{24}}\bigl(2+2 q+6 q^2+8 q^3+\cdots\bigr)
\end{pmatrix}.
\end{gather*}

As a result, the normalization of the core blocks is again determined straightforwardly by looking at the vacuum channel
\begin{gather*}
\varphi^1_{\sigma,\sigma,\sigma,\sigma}(\tau) = 2^{-{1\over3}}q^{-\frac{1}{24}}(1+\ccO(q)),
\end{gather*}
which gives
\begin{gather*}
\underline{\varphi}(\tau)
=
2^{-{1\over3}}
\begin{pmatrix}
\chi_{\bf{1}}^{A_{1,1}}(\tau)\vspace{1mm}\\
\chi_{\bf{2}}^{A_{1,1}}(\tau)
\end{pmatrix}.
\end{gather*}
Again, we could have reproduced this result to good approximation by performing the Rademacher sum up to $c\leq 500$, which gives
\begin{equation}
\label{eq:expl}
\underline{\varphi}(\tau)
\simeq
{2^{-1/3}
\begin{pmatrix}
q^{-{1\over24}}\bigl(1.002+2.999 q+4.001 q^2+7.000 q^3+\cdots\bigr)\\
q^{5\over24}\bigl(2.001+2.000 q+5.998 q^2+7.999 q^3+\cdots\bigr)
\end{pmatrix}}.
\end{equation}
Using equation~\eqref{eq:expl}, from the second entry of $\underline{\varphi}(\tau)$ one obtains the well-known value {(up to a sign)} of the structure constant
\begin{gather*}
(c_{\sigma\sigma\epsilon})^2=\frac{1}{2}.
\end{gather*}
It can be checked that the core blocks we obtained are equivalent to the standard expressions for the conformal blocks in terms of the cross-ratio $w$:
\begin{gather}
\cF^1_{\sigma,\sigma,\sigma,\sigma}(z_1,z_2,z_3,z_4) = \biggl(\prod_{i<j} z_{ij}^{-\frac{1}{24}}\biggr)\frac{\bigl(1+\sqrt{1-w}\bigr)^{\frac12}}{\sqrt{2}w^{-\frac{1}{12}}(1-w)^{-\frac{1}{12}}},\nonumber\\
\cF^\epsilon_{\sigma,\sigma,\sigma,\sigma}(z_1,z_2,z_3,z_4) = \biggl(\prod_{i<j} z_{ij}^{-\frac{1}{24}}\biggr)\frac{\bigl(1-\sqrt{1-w}\bigr)^{\frac12}}{\sqrt{2}w^{-\frac{1}{12}}(1-w)^{-\frac{1}{12}}}.\label{core_sigma}
\end{gather}
Finally, we note that the physical correlator
\begin{gather*}
\langle \sigma(z_1,\bar{z_1})\cdots\sigma(z_4,\bar{z_4})\rangle = 2^{-{2\over3}}\biggl({\prod_{1\leq i<j\leq 4}} \vert z_{ij}\vert^{-\frac{1}{12}}\biggr)Z^{A_{1,1}}_{T^2}(\tau,\overline{\tau})
\end{gather*}
coincides, up to a scalar prefactor, to the torus partition function of the level-1 $A_1$ WZW model:
\begin{gather*}
Z^{A_{1,1}}_{T^2}(\tau,\overline{\tau}) = \bigl\vert\chi^{A_{1,1}}_{\bf{1}}\bigr\vert^2+ \bigl\vert\chi^{A_{1,1}}_{\bf{2}}\bigr\vert^2.
\end{gather*}

\item $\cD= (\sigma,\sigma,\epsilon,\epsilon)$.
The last set of core blocks involves unequal external operators, and therefore according to the discussion in case (B) of Section \ref{subsec:mod} and to Appendix \ref{app:generalSL2Z} they can be organized in terms of the following vvmf {for $\PSL_2(\ZZ)$}:
\begin{gather*}
\underline{\varphi}(\tau)
=
\begin{pmatrix}
\varphi^1_{\sigma,\sigma,\epsilon,\epsilon}(\tau)\\
\varphi^\sigma_{\sigma,\epsilon,\sigma,\epsilon}(\tau)\\
\varphi^\sigma_{\sigma,\epsilon,\epsilon,\sigma}(\tau)
\end{pmatrix}
=
 \begin{pmatrix}
 q^{-{3\over16}}(N_1+\ccO(q))\\
 q^{-{5\over32}}(N_\sigma+\ccO(q))\\
 q^{-{5\over32}}(N_\sigma'+\ccO(q))
 \end{pmatrix}
.
\end{gather*}
The corresponding $\PSL_2(\mathbb{Z})$ representation has
\begin{gather*}
\cT_\cD
=
\begin{pmatrix}
\mathrm{e}\bigl(-\frac{3}{16}\bigr)&0&0\\
0&0&\mathrm{e}\bigl(-\frac{5}{32}\bigr)\\
0&\mathrm{e}\bigl(-\frac{5}{32}\bigr)&0
\end{pmatrix}
\qquad
\text{and}
\qquad
\cS_\cD
=
\begin{pmatrix}
0&\hphantom{-}0&1\\
0&-1&0\\
1&\hphantom{-}0&0
\end{pmatrix}.
\end{gather*}
To apply the methods of Section \ref{sec:RH}, one first needs to diagonalize $\cT_\cD$, which leads to the equivalent representation $\rho'_\cD$ for which
\begin{gather*}
\cT_\cD'
=
\begin{pmatrix}
\mathrm{e}\bigl(-\frac{3}{16}\bigr)&0&0\\
0&\mathrm{e}\bigl(\frac{11}{32}\bigr)&0\\
0&0&\mathrm{e}\bigl(-\frac{5}{32}\bigr)
\end{pmatrix}
\qquad
\text{and}
\qquad
\cS_\cD'
=\frac{1}{2}
\begin{pmatrix}
0&\sqrt{2}&\sqrt{2}\\
\sqrt{2}&-1&\hphantom{-}1\\
\sqrt{2}&\hphantom{-}1&-1
\end{pmatrix}.
\end{gather*}
Equation~\eqref{eq:trace} then implies that \smash{$\dim \bigl(\cM^!_0\bigl(\rho'_\cD,{\bigl(-\tfrac{3}{16},\tfrac{11}{32},-\tfrac{5}{32}\bigr)}\bigr)\bigr)=1$}.

By using the explicit transformation properties of the theta functions, equations~\eqref{eq:tht} and~\eqref{eq:ths}, one can easily check that the core blocks are given in terms of the following expression:
\begin{equation}\label{corethetas}
\underline{\varphi}(\tau)
=
2^{-{3\over2}}
\begin{pmatrix}
\left(\dfrac{2\eta(\tau)}{\theta_2(\tau)}\right)^{1\over4}
\dfrac{\theta_3(\tau)^4+\theta_4(\tau)^4}{2\eta(\tau)^4}\vspace{1mm}\\
\left(\dfrac{2\eta(\tau)}{\theta_3(\tau)}\right)^{1\over4}
\dfrac{-\theta_2(\tau)^4+\theta_4(\tau)^4}{2\eta(\tau)^4}\vspace{1mm}\\
\left(\dfrac{2\eta(\tau)}{\theta_4(\tau)}\right)^{1\over4}
\dfrac{\theta_2(\tau)^4+\theta_3(\tau)^4}{2\eta(\tau)^4}
\end{pmatrix}
,
\end{equation}
where the normalization is again fixed by imposing the correct normalization for the vacuum channel.
{Note that} the fractional powers here and elsewhere are not a problem, as explained in Section \ref{sec:gamma2}.

The expressions appearing in~\eqref{corethetas} are equivalent to the well-known expressions for the conformal blocks~\cite{AlvarezGaume:1989vk}
\begin{gather}
\varphi^1_{\sigma,\sigma,\epsilon,\epsilon}(w)= \frac{1-w/2}{w^{3/8}(1-w)^{5/16}},\nonumber \\
\varphi^\sigma_{\sigma,\epsilon,\sigma,\epsilon}(w)= \frac{1-2w}{2(w(1-w))^{5/16}},\nonumber \\
\varphi^\sigma_{\sigma,\epsilon,\epsilon,\sigma}(w) = \frac{1+w}{2w^{5/16}(1-w)^{3/8}}.
\label{wellknown}
\end{gather}
The corresponding physical correlator is given simply by
\begin{gather*}
\langle\sigma(z_1,\bar z_1)\sigma(z_2,\bar z_2)\epsilon(z_3,\bar z_3)\epsilon(z_4,\bar z_4)\rangle = \vert z_{4-pt}\vert^2 \varphi^1_{\sigma,\sigma,\epsilon,\epsilon}(w)\, \varphi^1_{\sigma,\sigma,\epsilon,\epsilon}(\overline{w}),
\end{gather*}
and similarly for other permutations of external operators.

This example can be handled more directly as a one-dimensional $\Gamma_0(2)$ vvmf. Recall~the discussion in Section \ref{sec:gamma02}. The ${\rm P}\Gamma_0(2)$ representation here is $\rho_{ \ex(-3/16),-1}$, and we requi\-re leading power to be \smash{$q^{-3/16}$} near ${\rm i} \infty$ and leading power to be \smash{$q_0^{-5/16}$} near $\tau=0$, where~$q_0=\ex(-\pi {\rm i}/\tau)$ is the local coordinate for $\Gamma_0(2)$ at $\tau=0$ (which is the square-root of the local coordinate there for $\PSL_2(\mathbb{Z})$). The core block will be \smash{$\kappa^{-5/16}\sqrt{\kappa+\smash{\frac{1}{4}}\vphantom{A^1}}$} times~some Laurent polynomial in $\kappa$. The leading power of $q$ is correct in \smash{$\kappa^{-5/16}\sqrt{\kappa+\smash{\frac{1}{4}}\vphantom{A^1}}$}, so no positive powers of $\kappa$ can appear in the polynomial. Likewise, the leading power of $q_0$ is also correct, so no negative powers of $\kappa$ can appear either. Thus the polynomial is constant, and the core block is
a scalar multiple of \smash{$\kappa^{-5/16}\sqrt{\kappa+\smash{\frac{1}{4}}\vphantom{A^1}}$}. The reader can verify this agrees with the first component of $\underline{\varphi}$ appearing in~\eqref{corethetas}.
\end{enumerate}

{Finally we remark that it is accidental that
the expressions such as~\eqref{core_epsilon},~\eqref{core_sigma} and~\eqref{wellknown} for conformal blocks are so simple as functions of the cross-ratio $w$, and does not hold for larger values of $p$ and $p'$. In general, the expressions as vvmf in $\tau$ will be much more accessible than the multivalued functions of $w$.}

\subsubsection[M(12,11)]{$\boldsymbol{\cM(12,11)}$}
\label{sec:mod4dim}
Our next example highlights the construction of physical correlators out of core blocks. We~consider the $\cM(12,11)$ Virasoro minimal model and focus on the correlators associated to the chiral datum $\cD=(\phi_{1,4},\phi_{1,4},\phi_{1,4},\phi_{1,4})$, {in the notation of~\eqref{notation_mmmodule}}. An important feature of this example is that there exists more than one crossing-symmetry-invariant physical correlator that can be constructed by combining left- and right-movers in inequivalent ways (corresponding to inequivalent physical theories). As discussed in~\cite{Maloney:2016kee}, in such cases the method of modular averages cannot be applied straightforwardly. On the other hand, working at the chiral level we will find that the conformal blocks are once again completely determined by their $\PSL_2(\mathbb{Z})$ transformation, and we will find that in each of the two physical theories built by combining left- and right-movers the physical correlators are also uniquely determined by imposing modular invariance on them.
Write $\chi_{r,s}(\tau)$ for the character (graded dimension) of $M_{r,s}$. According to the classification of~\cite{Cappelli:1986hf,Cappelli:1987xt}, one can construct three inequivalent physical theories by combining two copies of $\cM(12,11)$: the diagonal one of type $(A_{10},A_{11})$, which has torus partition function
\begin{gather*}
Z_{T^2}(\tau,\bar \tau) ={\frac{1}{2}} \sum_{r=1}^{10}\sum_{s=1}^{11} \vert \chi_{r,s}(\tau)\vert^2;
\end{gather*}
the one of type $(A_{10},D_7)$, which has
\begin{gather*}
Z_{T^2}(\tau,\bar \tau) =\sum_{r=1}^{10}\sum_{s\in\{1,3,5\}}|\chi_{r,s}(\tau)|^2+\sum_{r=1}^{10}\sum_{s\in\{2,4,6\}}\chi_{r,s}(\tau)\overline{\chi_{r,12-s}(\tau)};
\end{gather*}
and the one of type $(A_{10},E_6)$, which has
\begin{gather*}
Z_{T^2}(\tau,\bar\tau) = \frac{1}{2}\sum_{r=1}^{10}\bigl(\vert\chi_{r,1}(\tau)+\chi_{r,7}(\tau)\vert^2+\vert\chi_{r,4}(\tau)+\chi_{r,8}(\tau)\vert^2+\vert\chi_{r,5}(\tau)+\chi_{r,11}(\tau)\vert^2\bigr).
\end{gather*}
As in~\cite{Maloney:2016kee}, we consider the study the (non-chiral) four-point function corresponding to the chiral datum $\cD = (\phi_{1,4},\phi_{1,4},\phi_{1,4},\phi_{1,4})$
\begin{gather*}
\langle \phi_{e}(z_1,\bar z_1)\phi_{e}(z_2,\bar z_2)\phi_{e}(z_3,\bar z_3)\phi_{e}(z_4,\bar z_4)\rangle,
\end{gather*}
where
\begin{gather*}
\phi_{e}(z,\bar z) \equiv \phi_{1,4}(z)\phi_{1,4}(\bar z).
\end{gather*}
This four point function only arises in the $(A_{10},{A_{11}})$ and $(A_{10},E_{6})$ models, since the field $\phi_e(z,\bar z)$ is not part of the spectrum of the $(A_{10},D_{7})$ model.

Applying~\eqref{fusionruleMM} gives
\begin{gather*}
M_{1,4}\otimes M_{1,4} = M_{1,1}\oplus M_{1,3}\oplus M_{1,5}\oplus M_{1,7}.
\end{gather*}
At the level of chiral theory, there is therefore a four-dimensional vector of core blocks:
\begin{gather*}\renewcommand{\arraystretch}{1.1}
\underline{\varphi}(\tau)
=
\begin{pmatrix}
\varphi^{M_{1,1}}_{\phi_{1,4},\phi_{1,4},\phi_{1,4},\phi_{1,4}}(\tau)\\[1.5mm]
\varphi^{M_{1,3}}_{\phi_{1,4},\phi_{1,4},\phi_{1,4},\phi_{1,4}}(\tau)\\[1.5mm]
\varphi^{M_{1,5}}_{\phi_{1,4},\phi_{1,4},\phi_{1,4},\phi_{1,4}}(\tau)\\[1.5mm]
\varphi^{M_{1,7}}_{\phi_{1,4},\phi_{1,4},\phi_{1,4},\phi_{1,4}}(\tau)
\end{pmatrix} =
\begin{pmatrix}
q^{-\frac{31}{24}}\bigl(N_{(1,1)}+{\ccO}(q)\bigr)\\
q^{-\frac{7}{8}}\bigl(N_{(1,3)}+{\ccO}(q)\bigr)\\
q^{\frac{11}{24}}\bigl(N_{(1,5)}+{\ccO}(q)\bigr)\\
q^{\frac{65}{24}}\bigl(N_{(1,7)}+{\ccO}(q)\bigr)
\end{pmatrix}
,
\end{gather*}
whose transformations under $\PSL_2(\mathbb{Z})$ are given in terms of
\begin{gather*}
\cT_\cD
=
\begin{pmatrix}
\mathrm{e}\bigl(-\frac{7}{24}\bigr)&0&0&0\\
0&\mathrm{e}\bigl(-\frac{7}{8}\bigr)&0&0\\
0&0&\mathrm{e}\bigl(\frac{11}{24}\bigr)&0\\
0&0&0&\mathrm{e}\bigl(-\frac{7}{24}\bigr)
\end{pmatrix}
\end{gather*}
and
\begin{gather*}
\cS_\cD
=
\begin{pmatrix}
\frac{1}{\sqrt{2}}-\frac{1}{\sqrt{6}}& \sqrt{\frac{1}{\sqrt{3}}-\frac{1}{3}} & \frac{1}{\sqrt{3}} & \frac{1}{\sqrt{3}}\\
\sqrt{\frac{1}{\sqrt{3}}-\frac{1}{3}} & \frac{\sqrt{6}}{2}-\frac{1}{\sqrt{2}} & 0 & -\sqrt{\frac{2}{3} \bigl(\sqrt{3}-1\bigr)}\\
\frac{1}{\sqrt{3}} & 0 & -\frac{1}{\sqrt{2}} & \frac{1}{\sqrt{6}}\\
\frac{1}{\sqrt{3}} & -\sqrt{\frac{2}{3} \bigl(\sqrt{3}-1\bigr)} & \frac{1}{\sqrt{6}} &\frac{1}{\sqrt{2}}-\frac{\sqrt{6}}{3}
\end{pmatrix}.
\end{gather*}

The method of Section \ref{sec:RH} shows that the space of vvmf's transforming under this $\PSL_2(\mathbb{Z})$ representation and with the appropriate singular behavior is once again one-dimensional;
as discussed in Section \ref{sec:hybrid}, one may then efficiently determine the core blocks either recursively by our hybrid technique, or exactly by noticing that it corresponds to a reducible $\PSL_2(\mathbb{Z})$ representation. The goal of this example, however, is to illustrate the construction of $\PSL_2(\mathbb{Z})$-invariant four-point functions.

In the diagonal theory, the four-point function is the usual
\begin{gather*}
\langle \phi_{e}(z_1,\bar z_1)\phi_{e}(z_2,\bar z_2)\phi_{e}(z_3,\bar z_3)\phi_{e}(z_4,\bar z_4)\rangle_{(A_{10},A_{10})}
=
\vert z_{4-pt}\vert^2\,
\sum_{i=0}^3\bigl\vert \varphi^{M_{1,1+2i}}_{\phi_{1,4},\phi_{1,4},\phi_{1,4},\phi_{1,4}}(\tau)\bigr\vert^2,
\end{gather*}
which is obviously modular invariant.

On the other hand, in the non-diagonal $(A_{10}, E_6)$ theory, only the following internal channels appear:
\begin{gather*}
p_1=M_{1,1}\times \widetilde{M_{1,1}},\qquad
p_2=M_{1,5}\times \widetilde{M_{1,5}},\qquad
p_3=M_{1,3}\times \widetilde{M_{1,7}},\\
p_4=M_{1,7}\times \widetilde{M_{1,3}},\qquad
p_5=M_{1,7}\times \widetilde{M_{1,7}},
\end{gather*}
so we must find a different modular-invariant four-point function that only involves these conformal families. In other words, we ask that the four-point correlator take the form
\begin{gather*}
\langle \phi_{e}(z_1,\bar z_1)\phi_{e}(z_2,\bar z_2)\phi_{e}(z_3,\bar z_3)\phi_{e}(z_4,\bar z_4)\rangle_{(A_{10},E_6)}
=
\vert z_{4-pt}\vert^2\,
{\underline\varphi}(\tau)
\cdot
D
\cdot
{\underline\varphi}(\overline\tau),
\end{gather*}
where
\begin{gather*}
D
=
\begin{pmatrix}
1&0&0&0\\
0&0&0& d_{ee p_4}^2\\
0&0&d_{e e p_2}^2&0\\
0&d_{eep_3}^2&0&d_{e e p_5}^2
\end{pmatrix}.
\end{gather*}
Invariance under $\cT_\cD$ is automatic. On the other hand, invariance under $\cS_\cD$ implies that
\begin{gather*}
\cS_\cD^{-1}\cdot D\cdot \cS_\cD = D,
\end{gather*}
which determines for us the OPE coefficients
\begin{gather*}
d_{ee p_2}^2 = \frac{3}{2},\qquad d_{ee p_3}^2=d_{ee p_4}^2=\sqrt{\frac{1}{2}},\qquad d_{ee p_5}^2 = \frac{1}{2}.
\end{gather*}

\subsubsection[M(10,7)]{$\boldsymbol{\cM(10,7)}$}
\label{sec:a12}
Our next example is the non-unitary minimal model $\cM(10,7)$, whose central charge is $c = 8/35$. Choosing chiral datum {$\cD = (\phi_{1,2},\phi_{1,2},\phi_{1,2},\phi_{1,2})$}, we obtain two core blocks:
\begin{gather*}
\underline{\varphi}(\tau)
=
\begin{pmatrix}
\varphi^{M_{1,1}}_{\phi_{1,2},\phi_{1,2},\phi_{1,2},\phi_{1,2}}(\tau)\\[1.5mm]
\varphi^{M_{1,3}}_{\phi_{1,2},\phi_{1,2},\phi_{1,2},\phi_{1,2}}(\tau)
\end{pmatrix}
=
\begin{pmatrix}
q^{-{1\over60}}\bigl(N_{(1,1)}+{\ccO}(q)\bigr)\\[0.5mm]
q^{11\over60}\bigl(N_{(1,3)}+{\ccO}(q)\bigr)
\end{pmatrix}.
\end{gather*}
The $\PSL_2(\mathbb{Z})$ data is
\begin{gather*}
\cT_\cD
=
\begin{pmatrix}
\mathrm{e}\bigl(-\frac{1}{60}\bigr)&0 \\
0 & \mathrm{e}\bigl(\frac{11}{60}\bigr)
\end{pmatrix}
\qquad
\text{and}
\qquad
\cS_\cD
=
\begin{pmatrix}
\sqrt{\frac{5+\sqrt{5}}{10}}&\hphantom{-}\sqrt{\frac{5-\sqrt{5}}{10}} \vspace{2mm}\\
\sqrt{\frac{5-\sqrt{5}}{10}} & -\sqrt{\frac{5+\sqrt{5}}{10}}
\end{pmatrix}
.
\end{gather*}
The space of vvmf's $\cM^!_0\bigl(\rho_\cD,{\bigl(-\tfrac{1}{60},\tfrac{11}{60}\bigr)}\bigr)$ is one-dimensional. The core blocks are solutions of the two-dimensional MLDE~\eqref{eq:MLDE2} with $\mu = 11/900$. That is, they coincide, up to a numerical prefactor, {with} the level-1 characters of the \smash{$A_{{1\over2}}$}, the \emph{intermediate vertex subalgebra}~\cite{Shtepin1994} of $A_1$:
\begin{gather*}\renewcommand{\arraystretch}{1.15}
\underline{\varphi}(\tau)
=
2^{\frac{22}{15}}
\begin{pmatrix}
\chi_{A_{{1\over2},1}}^{\mathbf{1}}(\tau)\\[2mm]
\chi_{A_{{1\over2},1}}^{\mathbf{1'}}(\tau)
\end{pmatrix}
=
2^{\frac{2}{15}}
\begin{pmatrix}
q^{-\frac{1}{60}}\bigl(1+q+q^2+q^3+2q^4+2q^5+3q^6+\ccO\bigl(q^7\bigr)\bigr)\\[0.5mm]
q^{\frac{11}{60}}\bigl(1+q^2+q^3+q^4+q^5+2q^6+\ccO\bigl(q^7\bigr)\bigr)
\end{pmatrix}\!
.
\end{gather*}

\subsubsection[M(13,10)]{$\boldsymbol{\cM(13,10)}$}
\label{sec:e712}
Finally we look at the non-unitary minimal model $\cM(13,10)$, of central charge $c = 38/65$. Again we choose chiral datum $\cD ={ (\phi_{2,1},\phi_{2,1},\phi_{2,1},\phi_{2,1})}$ and obtain two core blocks:
\begin{gather*}
\underline{\varphi}(\tau)
=
\begin{pmatrix}
\varphi^{M_{1,1}}_{\phi_{2,1},\phi_{2,1},\phi_{2,1},\phi_{2,1}}(\tau)\\[1.5mm]
\varphi^{M_{3,1}}_{\phi_{2,1},\phi_{2,1},\phi_{2,1},\phi_{2,1}}(\tau)
\end{pmatrix}
=
\begin{pmatrix}
q^{-\frac{19}{60}}\bigl(N_{(1,1)}+{\ccO}(q)\bigr)\\[0.5mm]
q^{\frac{29}{60}}\bigl(N_{(3,1)}+{\ccO}(q)\bigr)
\end{pmatrix}.
\end{gather*}
The $\PSL_2(\mathbb{Z})$ data is
\begin{gather*}
\cT_\cD
=
\begin{pmatrix}
\mathrm{e}\bigl(-\frac{19}{60}\bigr)&0 \\
0 & \mathrm{e}\bigl(\frac{29}{60}\bigr)
\end{pmatrix}
\qquad
\text{and}
\qquad
\cS_\cD
=
\begin{pmatrix}
\sqrt{\frac{5+\sqrt{5}}{10}}&\hphantom{-}\sqrt{\frac{5-\sqrt{5}}{10}} \vspace{2mm}\\
\sqrt{\frac{5-\sqrt{5}}{10}} & -\sqrt{\frac{5+\sqrt{5}}{10}}
\end{pmatrix}
.
\end{gather*}
The space of vvmf's $\cM^!_0\bigl(\rho_\cD,{\bigl(-\tfrac{19}{60},{29\over 60}\bigr)}\bigr)$ is one-dimensional. The core blocks are solutions of the two-dimensional MDE~\eqref{eq:MLDE2} with $\mu = 551/900$, and coincide with the level-1 characters of the~\smash{$E_{7+\frac{1}{2}}$} intermediate vertex subalgebra of $E_8$ as given in~\cite{Kawasetsu}! That is,
\begin{align*}
\underline{\varphi}(\tau)
&{}=
2^{-\frac{38}{15}}
\begin{pmatrix}
\chi_{E_{7+{1\over2},1}}^{\bf{1}}(\tau)\\[1mm]
\chi_{E_{7+{1\over2},1}}^{\bf{57}}(\tau)
\end{pmatrix}\nonumber\\
&{}=
2^{-\frac{38}{15}}
\begin{pmatrix}
q^{-\frac{19}{60}}\bigl(1+190 q+2831 q^2+22306 q^3+129276 q^4+611724 q^5+\ccO\bigl(q^6\bigr)\bigr)\\[0.5mm]
q^{\frac{29}{60}}\bigl(57+1102 q+9367 q^2+57362 q^3+280459 q^4+1181838 q^5+\ccO\bigl(q^6\bigr)\bigr)
\end{pmatrix}
.
\end{align*}

\subsection[RCFTs at c=1]{RCFTs at $\boldsymbol{c=1}$}
\label{sec:c1}
For our next example, we consider the $\mathbb{Z}_2$-orbifold of the compact boson at radius $R=\sqrt{2N}$, for~${N\in\mathbb{Z}}$. This family of models possesses the following spectrum of chiral algebra primaries~\cite{Dijkgraaf:1989hb}:

\begin{center}
\begin{tabular}{c l l}
\toprule
$h_\Phi$& \multicolumn{2}{c}{$\Phi$ }\\
\toprule
0 & $\mathbf{1}$\\
1 & $j$\\
$\frac{N}{4}$ & $\phi^i_{N}$&$i=1,2$\\
$\frac{k^2}{4N}$ & $\phi_{k}$&$k=1,\dots,N-1$\\
$\frac{1}{16}$ & $\sigma_i$&$i=1,2$\\
$\frac{9}{16}$ & $\tau_i$&$i=1,2$\\
\bottomrule
\end{tabular}
\end{center}

We will consider the case where $N$ is even, in which case the $S$-matrix for this CFT is given~by\footnote{We have corrected a factor of 2 with respect to~\cite{Dijkgraaf:1989hb}.}
\begin{gather*}\setlength{\arraycolsep}{5pt}
\begin{matrix}
\phantom{x}\\
\phantom{x}\\
\cS =\frac{1}{\sqrt{8N}}\\
\phantom{x}
\end{matrix}
\substack{
\phantom{x}\\
\phantom{x}\\
\phantom{x}\\
\phantom{x}\\
\phantom{x}\\
\left(
\begin{matrix}
\\\\\\\\\\\\\\\\\
\end{matrix}
\right.
}
\begin{matrix}
\mathbf{1}&j&\phi^1_N&\phi^2_N&\phi_{k'}&\sigma_1&\sigma_2&\tau_1&\tau_2&\phantom{\frac{1}{\frac{1}{\frac{1}{2}}}}\\
1&1&1&1&2&\hphantom{-}\sqrt{N}&\hphantom{-}\sqrt{N}&\hphantom{-}\sqrt{N}&\hphantom{-}\sqrt{N}&\mathbf{1} \\
1&1&1&1&2&\!\!-\sqrt{N}\!\!&\!\!-\sqrt{N}\!\!&\!\!-\sqrt{N}\!\!&\!\!-\sqrt{N}\!\!& j \\
1&1&1&1&2(-1)^{k'}&\hphantom{-}\sqrt{N}&\!\!-\sqrt{N}\!\!&\hphantom{-}\sqrt{N}&\!\!-\sqrt{N}\!\!&\phi^1_N \\
1&1&1&1&2(-1)^{k'}&\!\!-\sqrt{N}\!\!&\hphantom{-}\sqrt{N}&\!\!-\sqrt{N}\!\!&\hphantom{-}\sqrt{N}&\phi^2_N \\
2&2&\!2(-1)^k\!&\! 2(-1)^k\! &\!\!4\cos\bigl(\frac{\pi k k'}{N}\bigr)\!\!&0&0&0&0&\phi_k \\
\!\!\sqrt{N}\!&\!\!-\sqrt{N}\!\!\!&\hphantom{-}\sqrt{N}&\!\!-\sqrt{N}\!\!&0&\sqrt{2N}&0&-\sqrt{2N}&0&\sigma_1\\
\!\!\sqrt{N}\!&\!\!-\sqrt{N}\!\!\!&\!\!-\sqrt{N}\!\!&\hphantom{-}\sqrt{N}&0&0&\hphantom{-}\sqrt{2N}&0&-\sqrt{2N}&\sigma_2\\
\!\!\sqrt{N}\!&\!\!-\sqrt{N}\!\!\!&\hphantom{-}\sqrt{N}&\!\!-\sqrt{N}\!\!&0&\!\!-\sqrt{2N}\!\!&0&\hphantom{-}\sqrt{2N}&0&\tau_1\\
\!\!\sqrt{N}\!&\!\!-\sqrt{N}\!\!\!&\!\!-\sqrt{N}\!\!&\hphantom{-}\sqrt{N}&0&0&\!\!-\sqrt{2N}\!\!&0&\hphantom{-}\sqrt{2N}&\tau_2
\end{matrix}
\!\!\!\!\!\!\!\!\!\!\!\!\!\!\!\!
\substack{
\phantom{x}\\
\phantom{x}\\
\phantom{x}\\
\phantom{x}\\
\phantom{x}\\
\left.
\begin{matrix}
\\\\\\\\\\\\\\\\\
\end{matrix}
\right)
}
\end{gather*}

For definiteness, let us focus on the four-point functions between four $\phi_{N/2}$ operators:
\begin{gather*}
\cD = (\phi_{N/2},\phi_{N/2},\phi_{N/2},\phi_{N/2}).
\end{gather*}
From the $S$-matrix and~\eqref{Verfor:3point}, one obtains the fusion rule
\begin{gather*}
\phi_{N/2} \times \phi_{N/2} = \mathbf{1} + j + \phi_N^1+ \phi_N^2,
\end{gather*}
so the chiral blocks have four internal channels. (In the above equation, by slight abuse of notation we denote the modules by the same name as the primaries.)
However, as the two operators $\phi_N^1$ and $\phi_N^2$ are charge conjugate, the corresponding conformal blocks will be identical. We can therefore assemble the core blocks into a three dimensional representation of $\PSL_2(\mathbb{Z})$:
\begin{gather*}
\underline{\varphi}=
\begin{pmatrix}
\var^{\mathbf{1}}_{\phi_{N/2},\phi_{N/2},\phi_{N/2},\phi_{N/2}}(\tau)\\[1.5mm]
\var^{j}_{\phi_{N/2},\phi_{N/2},\phi_{N/2},\phi_{N/2}}(\tau)\\[1.5mm]
\var^{\phi_N^i}_{\phi_{N/2},\phi_{N/2},\phi_{N/2},\phi_{N/2}}(\tau)\\
\end{pmatrix}
.
\end{gather*}
The following T-transformation for the core blocks is readily computed:
\begin{gather*}
\cT_{\cD}
=
\begin{pmatrix}
\ex\bigl(-\frac{N}{24}\bigr)&0&0\\
0&\ex\bigl(-\frac{N}{24}+\frac{1}{2}\bigr)&0\\
0&0&\ex\bigl(-\frac{N}{24}+\frac{N}{8}\bigr)
\end{pmatrix}
.
\end{gather*}
As $\dim \underline{\varphi}\,\leq 5$, $\cS_\cD$ is also uniquely determined and is given by
\begin{gather*}
\cS_{\cD}
=
\frac{1}{2}
\begin{pmatrix}
1&\hphantom{-}1&\hphantom{-}\sqrt{2}\\
1&\hphantom{-}1&-\sqrt{2}\\
\sqrt{2}&-\sqrt{2}&\hphantom{-}0
\end{pmatrix}
\end{gather*}
which indeed satisfies
\begin{gather*}
(\cS_\cD\cdot\cT_\cD)^3=\cS_\cD\cdot\cS_\cD=1.
\end{gather*}
The space of vvmf's associated to this class of representations is one-dimensional. We find that it is given by
\begin{equation}\renewcommand{\arraystretch}{1.2}
\underline{\varphi}
=
2^{-\frac{N}{3}}
\begin{pmatrix}
\dfrac{1}{2}\left(\dfrac{\theta_3(\tau)^N}{\eta(\tau)^N}+\dfrac{\theta_4(\tau)^N}{\eta(\tau)^N}\right)\vspace{2mm}\\
\dfrac{1}{2}\left(\dfrac{\theta_3(\tau)^N}{\eta(\tau)^N}-\dfrac{\theta_4(\tau)^N}{\eta(\tau)^N}\right)\vspace{2mm}\\
\dfrac{1}{\sqrt{2}}\dfrac{\theta_2(\tau)^N}{\eta(\tau)^N}
\end{pmatrix}
=
2^{-\frac{N}{3}}
\begin{pmatrix}
\chi_{\bf{1}}^{D_{N,1}}(\tau)\\[1mm]
\chi_{\bf{N}}^{D_{N,1}}(\tau)\\[1mm]
\sqrt{2}\chi_{\bf{S}}^{D_{N,1}}(\tau)
\end{pmatrix},
\label{eq:equality D_N boson}
\end{equation}
in other words, we find that up to normalization they coincide with the characters $\chi_{\mathbf{R}}^{D_{N,1}}(\tau)$ of the $D_N = {\rm SO}(2N)$ Kac--Moody algebra at level 1. More precisely, the $D_{N,1}$ Kac--Moody algebra possesses four inequivalent primaries, corresponding to the trivial ($\bf{1}$), vector ($\bf{N}$), spinor ($\bf{S^+})$ and conjugate spinor ($\bf{S^-})$) representations, but the (unrefined) characters of the two spinor representations coincide:
\begin{gather*}
\chi_{\bf{S^+}}^{D_{N,1}}(\tau)=\chi_{\bf{S^-}}^{D_{N,1}}(\tau)=\chi_{\bf{S}}^{D_{N,1}}(\tau),
\end{gather*}
allowing us to organize them into a three-dimensional representation of $\PSL_2(\mathbf{Z})$, in complete analogy with the sphere core blocks.

\subsection{WZW models}
We illustrate our techniques by applying them to the $G={\rm SU}(2)$ WZW model at level 1. This model admits a free-field realization in terms of a compact chiral boson $\phi(x)$. The chiral algebra is generated by the operators $H={\rm i}\partial \phi$ and $E^\pm = {:}{\rm e}^{\pm {\rm i}\sqrt{2}\phi}{:}$, which have the following non-vanishing OPE's:
\begin{gather*}
H(z)H(w)\sim \frac{1}{(z-w)^2},\qquad
H(z)E^{\pm}(w)\sim \frac{\pm\sqrt{2}E^{\pm}(w)}{z-w},\\
E^+(z)E^{-}(w)\sim \frac{1}{(z-w)^2}+ \frac{\sqrt{2}H(w)}{z-w}.
\end{gather*}

For a generic group $G$, the chiral algebra primaries are labeled by highest weights $\omega$ of the Lie algebra of $G$. The conformal weight of a primary is given by
\begin{gather*}
h_\omega
=\frac{C_2(\omega)}{h^\vee_G+k},
\end{gather*}
where $C_2(\omega)$ is the Casimir invariant of the representation of highest weight $\omega$. For a chiral block to be nonzero, the tensor product of the corresponding representations must contain a singlet.

For $G={\rm SU}(2)$, at level 1 the only primaries are the vacuum and the WZW-primary \smash{${\xi_{\omega_1}\equiv {:}{\rm e}^{\frac{{\rm i}}{\sqrt{2}}\phi}{:}}$} corresponding to the highest weight $\omega_1$ of the fundamental representation of ${\rm SU}(2)$.
The core block corresponding to the chiral datum $ \cD=(\xi_{\omega_1},\xi_{\omega_1},\xi_{\omega_1},\xi_{\omega_1})$ would be a (scalar) $\PSL_2(\mathbb{Z})$ modular form with $\cT_\cD=\bigl(\mathrm{e}\bigl(-{1\over6}\bigr)\bigr)$, {since the leading order $q$-power in the corresponding core blocks is \smash{$q^{-{1\over 6}}$}}; however, no such modular form exists (see the discussion around equation~\eqref{eq:1dimbasis}), and therefore the conformal block vanishes. In order to obtain non-vanishing blocks, we must take some of the external operators to be the operator $\xi_{-\omega_1}=E^-\xi_{\omega_1}$, which is a~Virasoro-primary but not a WZW-primary.

The core block $ \cD=(\xi_{-\omega_1},\xi_{\omega_1},\xi_{\omega_1},\xi_{\omega_1})$ still vanishes for the same reason; on the other hand, for $ \cD=(\xi_{\omega_1},\xi_{\omega_1},\xi_{-\omega_1},\xi_{-\omega_1})$ one obtains the three-dimensional $\PSL_2(\mathbb{Z})$ representation
\begin{gather*}
\underline{\varphi}(\tau)
=
\begin{pmatrix}
\var^1_{\xi_{\omega_1},\xi_{\omega_1},\xi_{-\omega_1},\xi_{-\omega_1}}\\[1.5mm]
\var^1_{\xi_{\omega_1},\xi_{-\omega_1},\xi_{\omega_1},\xi_{-\omega_1}}\\[1.5mm]
\var^1_{\xi_{\omega_1},\xi_{-\omega_1},\xi_{-\omega_1},\xi_{\omega_1}}
\end{pmatrix}{(\tau)}
\end{gather*}
(alternatively, each component of $\underline{\varphi}(\tau)$ can be viewed as scalar-valued modular form for $\Gamma(2)$).
Determining the modular transformation properties requires some care. The OPE between~$\xi_{\omega_1}$ and~$\xi_{-\omega_1}$ contains the identity operator, and as a result the leading order $q$-power in the core~blocks \smash{${\varphi}^1_{\xi_{\omega_1},\xi_{-\omega_1},\xi_{\omega_1},\xi_{-\omega_1}}(\tau)$} is \smash{$q^{-\frac{1}{6}(4h_{\omega_1})}=q^{-\frac16}$}. On the other hand, the OPE of $\xi_{\pm\omega_1}$ with itself contains descendants of the identity but not the identity itself:
\begin{gather*}
\xi_{\pm\omega_1}(z)\xi_{\pm\omega_1}(w)= E^{\pm}(z)(z-w)^{\frac12}+{\ccO}\bigl((z-w)^{\frac32}\bigr).
\end{gather*}
As a consequence, for the core block \raisebox{1pt}{\smash{${\varphi}^1_{\xi_{\omega_1},\xi_{\omega_1},\xi_{-\omega_1},\xi_{-\omega_1}}(\tau)$}} the operator of lowest dimension appearing in the vacuum channel is the operator $E^{+}$ which has conformal weight 1. With this information, we obtain
\begin{gather*}
\cT_\cD
=
\begin{pmatrix}
\mathrm{e}\bigl(\frac{1}{3}\bigr) & 0 &0 \\
0 & 0 & \mathrm{e}\bigl(-\frac{1}{6}\bigr)\\
0 & \mathrm{e}\bigl(-\frac{1}{6}\bigr) & 0
\end{pmatrix}
.\end{gather*}
As the dimension of the representation is less than $6$, there is a unique $\cS_\cD$ compatible with $\cT_\cD$. This is easily found to be
\begin{gather*}
\cS_\cD
=
\begin{pmatrix}
0 & 0 &1 \\
0 & 1 & 0\\
1 & 0 & 0
\end{pmatrix}
.
\end{gather*}

The relevant space of vvmf's is again one-dimensional, as can be checked by applying equation~\eqref{eq:trace} (after applying the appropriate conjugation to $\cT_\cD$ and $\cS_\cD$ to make $\cT_\cD$ diagonal). It~is easy to recognize that the conformal blocks must be given by
\begin{gather*}
\underline{\varphi}(\tau)
=
\frac{2^{-\frac43}}{\eta(\tau)^4}
\begin{pmatrix}
\theta_2(\tau)^4\\
\theta_4(\tau)^4\\
\theta_3(\tau)^4
\end{pmatrix}
,
\end{gather*}
which, once expressed in terms of the cross ratio $w$, is in agreement with the expressions for the conformal blocks obtained by Knizhnik and Zamolodchikov~\cite{Knizhnik:1984nr}.

\subsection{Liouville theory}

In our last example, we show how our approach also extends to the case of non-rational conformal field theories which contain null states. In particular we apply our method to the Liouville theory with central charge
\begin{gather*}
c = 1+6 Q^2,\qquad\text{where}\quad Q = b + \frac{1}{b},\ b\in\mathbb{R}.
\end{gather*}
As is well known, the spectrum of Liouville theory primaries consists of a continuum of states~$\phi_{\mathfrak{p}}$ (see~\cite{Curtright:1982gt}) of momentum
\begin{gather*}
\alpha_\mathfrak{p} = \frac{b}{2}+\frac{1}{2b}+\mathfrak{p},\qquad \mathfrak{p}\in {\rm i}\mathbb{R}
\end{gather*}
as well as degenerate operators $\phi_{m,n}$ of momentum
\begin{gather*}
\alpha_{m,n} = -\frac{b}{2}m-\frac{1}{2b} n,\qquad m,n\in \mathbb{Z}_+.
\end{gather*}
The conformal weight of an operator is given in terms of its momentum $\alpha$ as
\begin{gather*}
h = \alpha(Q-\alpha).
\end{gather*}
For definiteness, let us consider a core block involving one degenerate operator of momentum $\alpha_{1,0}= -{b\over2}$ and three non-degenerate operators of momentum $\frac{Q}{2}+\mathfrak{p}$:
\begin{gather*}
\cD = (\phi_{1,0},\phi_{\mathfrak{p}},\phi_{\mathfrak{p}},\phi_{\mathfrak{p}}).
\end{gather*}
In this case, there is a two-dimensional space of core blocks corresponding to the modules $M_{\mathfrak{p}\pm b/2}$ with primary $\phi_{\mathfrak{p}\pm b/2} $. One finds
\begin{gather*}
\cT_\cD
=
\begin{pmatrix}
\mathrm{e}\bigl(\frac{1}{12}-\frac{b \mathfrak{p}}{2}\bigr)	&	0						\\
0						&	\mathrm{e}\bigl(\frac{1}{12}+\frac{b \mathfrak{p}}{2}\bigr)
\end{pmatrix}
,
\qquad
\cS_\cD
=
\begin{pmatrix}
\dfrac{\csc(\pi b \mathfrak{p})}{2}			&	\sqrt{1-\dfrac{\csc(\pi b \mathfrak{p})^2}{4}}\vspace{2mm}	\\
\sqrt{1-\dfrac{\csc(\pi b \mathfrak{p})^2}{4}}	&	-\dfrac{\csc(\pi b \mathfrak{p})}{2}
\end{pmatrix},
\end{gather*}
where $\cS_\cD$ is uniquely determined by requiring
\begin{gather*}
\cS_\cD^4=1,
\qquad
\cS_\cD^2=(\cS_\cD\cdot\cT_\cD)^3
\end{gather*}
and by working with a basis of conformal blocks such that $\cS_\cD$ is symmetric (cf.\ equation~\eqref{eq:SMLDE}). The conformal blocks are the solutions to the two-dimensional MLDE{, namely equation~\eqref{eq:MLDE2} with $\mu = -\frac{1}{36}+b^2\mathfrak{p}^2$. They are given in terms of hypergeometric functions as in equations~\eqref{eq:MLDEs}:}
\begin{align*}
\varphi^{M_{\mathfrak{p}\mp\frac{b}{2}}}_{\phi_{1,0},\phi_{\mathfrak{p}},\phi_{\mathfrak{p}},\phi_{\mathfrak{p}}} (\tau)
=
N_{\mp}\bigl(2^{-4} \lambda(\tau)\bigr)^{\frac16\pm b\mathfrak{p}}(1-\lambda(\tau))^{\frac16\pm b\mathfrak{p}}{}_2F_1\biggl(\frac{1}{2}\pm b\mathfrak{p},\frac{1}{2}\pm 3b\mathfrak{p},1\pm 2b\mathfrak{p};\lambda(\tau)\biggr),
\end{align*}
where
\begin{gather*}
N_-=1,\qquad N_+= 2^{6b\mathfrak{p}}\sqrt{\frac{\Gamma(-b\mathfrak{p})\Gamma(1-2b\mathfrak{p})\Gamma\bigl(\frac{1}{2}+3b\mathfrak{p}\bigr)}{\Gamma(b\mathfrak{p})\Gamma(1+2b\mathfrak{p}) \Gamma\bigl(\frac{1}{2}-3b\mathfrak{p}\bigr)}}.
\end{gather*}
The core blocks have the following series expansion:
\begin{gather*}
\varphi^{M_{\mathfrak{p}\mp\frac{b}{2}}}_{\phi_{1,0},\phi_{\mathfrak{p}},\phi_{\mathfrak{p}},\phi_{\mathfrak{p}}}(\tau)
\\
\quad{}= N_{\mp}q^{\frac{1}{12}\pm\frac{b \mathfrak{p}}{2}}\biggl( 1-\frac{2(1\pm 6 b \mathfrak{p})(1\mp 5 b \mathfrak{p})}{1\pm b \mathfrak{p}}q+\frac{2(1\pm 6 b \mathfrak{p})\bigl(-2\pm 9 b\mathfrak{p}-25 b^2\mathfrak{p}^2\pm 300 b^3\mathfrak{p}^3\bigr)}{(1\pm b \mathfrak{p})(2\pm b\mathfrak{p})}q^2 \nonumber\\
\qquad{}+2\frac{(1\pm 6b\mathfrak{p})\bigl(6\mp 25 b \mathfrak{p} - 274 b^2 \mathfrak{p}^2 \pm 1255 b^3 \mathfrak{p}^3 + 550 b^4 \mathfrak{p}^4 \pm 3000 b^5 \mathfrak{p}^5\bigr)}{(1\pm b \mathfrak{p})(2\pm b\mathfrak{p})(3\pm b\mathfrak{p})}q^3+{\ccO}\bigl(q^4\bigr)\biggr). 
\end{gather*}
Of course, as for the minimal models the core blocks obtained by solving the modular differential equation are consistent with the chiral blocks obtained from the BPZ equation.

\section{A sphere-torus correspondence}
\label{sec:spheretorus}

\begin{figure}[t]\centering
\includegraphics[width=0.44\textwidth]{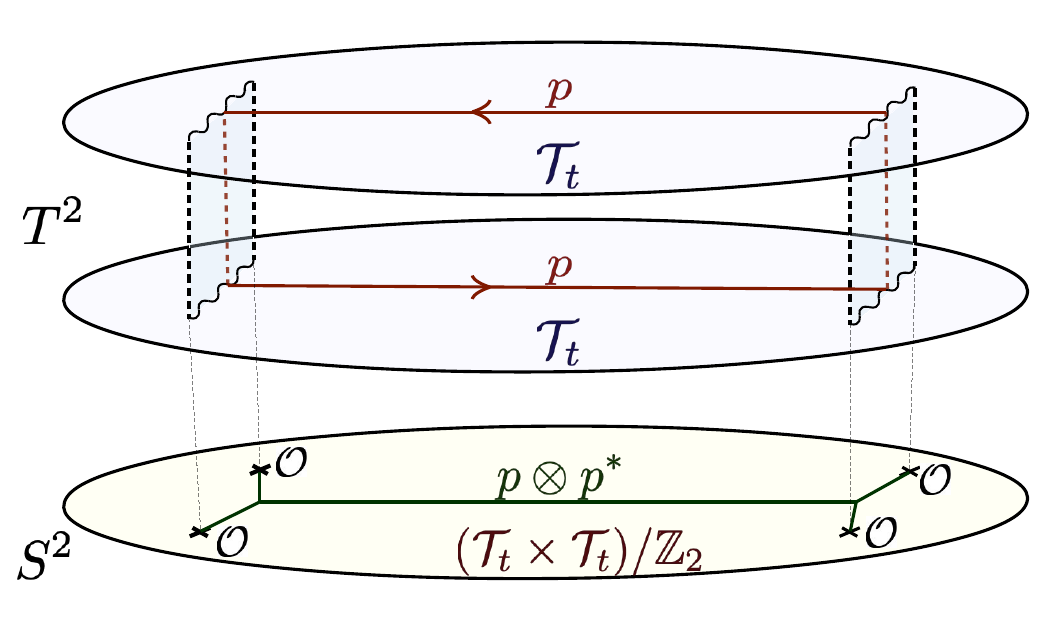}
\caption{Schematic depiction of the torus-sphere correspondence between $\cT_t$ and ${\rm Sym}^2\cT_t$.}\label{fig:st}
\end{figure}

In the previous section, we have encountered a number of examples in which the core blocks coincide with the characters of different VOAs. Note that this relation bears some similarity to another better-known relation between a theory $\cT_t$ on the torus and the corresponding symmetric product theory ${\rm Sym}^2\cT_t = (\cT_t\times\cT_t)/\mathbb{Z}_2$ on the sphere \cite{Hartman:2019pcd, Lunin:2000yv,Witten:2007kt}. We now briefly review the main ingredients of this correspondence. On one side, one considers the partition function of a torus theory $\cT_t$ of central charges $c=\tilde c$ \big(which for simplicity we take to be diagonal theory with chiral algebras $\cV_t= \tilde \cV_t$\big), and on the other side one considers a specific four-point correlator $\langle\ccO(z_1,\overline z_1),\dots,\ccO(z_4,\overline z_4) \rangle_{{\rm Sym}^2\cT_t}$ in the symmetric product theory ${\rm Sym}^2\cT_t $ of central charges $(2c, 2c)$. This correspondence can be understood geometrically by realizing the torus as the double cover of the four-punctured sphere, as mentioned in Section~\ref{section6.2}. The theory on the sphere can then be described starting from two copies of $\cT_t$, one for each of the two sheets. The~$\mathbb{Z}_2$ orbifold reflects the fact that in circling one of the branch points one moves between the two sheets, thus exchanging the two theories. This is implemented on the sphere side by inserting at every branch point a twist operator $\cO(z,\overline z)=O(z)\otimes \tilde O(\overline z)$, which is the twisted sector ground state of ${\rm Sym}^2\cT_t$ of conformal weights $\bigl(h_\cO,\tilde h_\cO\bigr)=\bigl(\frac{c}{16},\frac{c}{16}\bigr)$.\footnote{States in the twisted sector are in one-to-one correspondence with states of $\cT_t$, so the twisted sector ground state $\cO$ is unique as long as the CFT $\cT_t$ possesses a single vacuum.} It can be shown that the OPE of $\cO$ with itself includes every conformal family of $\cT_t$ \big(labeled by \smash{$p=M\otimes \tilde M\in \Upsilon_t\otimes\tilde\Upsilon_t$}\big) \cite{Hartman:2019pcd}. This can be used to argue that the torus partition function of $\cT_t$, which involves a sum over all $p$, can be equivalently viewed as a sphere four-point correlator of ${\rm Sym}^2\cT_t$, which involves a sum over the internal channels, with the conformal family labeled by $p\otimes p^*$ appearing as the internal channel (see Figure \ref{fig:st}). More precisely, one obtains the relation
\begin{gather*}
Z_{\cT_t}(\tau,\overline\tau) = 2^{\frac{2c}{3}}\,z_{\rm 4 pt}\, \langle \cO(z_1,\overline z_1),\dots,\cO(z_4,\overline z_4) \rangle_{{\rm Sym}^2\cT_t} = 2^{\frac{2c}{3}} F_{\cO,\cO,\cO,\cO}(\tau,\overline \tau)_{{\rm Sym}^2\cT_t},
\end{gather*}
where the factor in parentheses arises from the Weyl transformation between the variables~$\tau$ and~$z$ which parametrize the sphere and the torus respectively. If one takes $\cT_t$ to be the diagonal theory built out of two copies of a chiral algebra $\cA_t$, one can refine this correspondence and show that it holds at the chiral level, where it relates the characters of $\cV_t$ to the core blocks of ${\rm Sym}^2\cV_t$:%
\begin{equation}
\chi_M(\tau) = 2^{\frac{c}{3}} \varphi^{M\otimes M^*}_{O,O,O,O}(\tau)_{{\rm Sym}^2\cA_t}.
\label{eq:stcharold}
\end{equation}

We now turn to the chiral torus-sphere correspondence encountered in the present work. On the torus side, we take $\cV_t$ to be a Kac--Moody algebra at level 1 and $\cT_t$ to be the {corresponding (non-chiral) diagonal theory}. On the sphere side, we consider a theory $\cT_s$ which as we explain below is closely related to the coset theory $ G_1\times G_1/ G_2$, and has central charges
\begin{gather*}
c_s = \tilde c_s = 2\frac{\dim (G)}{h^\vee_G+1}-\frac{\dim (G)\cdot 2}{h^\vee_G+2} = \frac{2\dim (G)}{\bigl(h^\vee_G+1\bigr)\bigl(h^\vee_G+2\bigr)}.
\end{gather*}
Many of the ingredients of the correspondence appear to be strikingly similar to the $\cT_t-{\rm Sym}^2\cT_t$
correspondence.
However, note that the geometric intuition detailed in the previous paragraphs and in Figure \ref{fig:st} for the $\cT_t-{\rm Sym}^2\cT_t$ correspondence is not readily applicable in the present case.

Let us flesh out the ingredients of this chiral correspondence between $\cT_t$ and $\cT_s$ observables (see also Table \ref{tab:corr} in the introduction for a summary).
First, let us denote the respective chiral algebras as $\cV_t$ and $\cV_s$. We observe that there exists a module of $\cV_s$ whose primary $\phi_\ast\in M_\ast$ has conformal weight $h = \frac{c}{16}$; in analogy with the $\cT_t-{\rm Sym}^2\cT_t$ correspondence we will call this primary the twist operator. To each irreducible module $N$ of $\cV_t$, of conformal weight $h_N$, we associate a module $M(N)$ of $\cV_s$ of conformal weight $2\, h_N$. Then, in analogy with equation~\eqref{eq:stcharold}, we find an identity between the characters of the modules of $\cV_t$ and the core blocks of $\cV_s$ corresponding to chiral datum $\cD=(\phi_\ast,\phi_\ast,\phi_\ast,\phi_\ast)$:
\begin{gather*}
\chi_N(\tau) = 2^{c\over3} \varphi^{M(N)}_{\phi_\ast,\phi_\ast,\phi_\ast,\phi_\ast}(\tau)
,
\end{gather*}
where the quantity on the left-hand (resp.\ right-hand) side is defined for the chiral algebra $\cV_t$ (resp.\ $\cV_s$).
The above relation also implies that the ${\rm SL}_2(\mathbb{Z})$ matrices $\mathbf{S}$, $\mathbf{T}$ acting on the characters of $\cV_t$ coincide with the matrices $\cS_\cD$, $\cT_\cD$ acting on the core blocks of theory $\cV_s$. Finally, passing to the physical theories $\cT_t$ and $\cT_s$ obtained by combining the left- and right-movers, we find an identity between torus partition function and core correlators:
\begin{gather*}
\sum_{N,\tilde N} \cZ_{N,\tilde N} \chi_N(\tau)\chi_N(\overline\tau) = 2^{2 c\over3} \sum_{N,\tilde N} (\cA_\cD)_{M(N),\tilde M(\tilde N)} \varphi^{M(N)}_{\phi_\ast,\phi_\ast,\phi_\ast,\phi_\ast}(\tau)\varphi^{\tilde M(\tilde N)}_{\tilde\phi_\ast,\tilde\phi_\ast,\tilde\phi_\ast,\tilde\phi_\ast}(\overline \tau),
\end{gather*}
that is,
\begin{gather*}
Z_{T^2}(\tau,\overline{\tau}) = 2^{2 c\over3} F_{\phi_\ast\times\tilde\phi_\ast,\dots,\phi_\ast\times\tilde\phi_\ast}(\tau,\overline\tau),
\end{gather*}
where the quantity on the left-hand (resp.\ right-hand) side is defined for the chiral algebra $\cT_t$ (resp.\ $\cT_s$).
Note that the OPE coefficients $(\cA_\cD)_{M(N),\tilde M(\tilde N)}$ are identified with the multiplicities~$\mathcal{Z}_{N,\tilde N}$.

The sphere-torus correspondence equates the level-1 Kac--Moody algebra characters {to} specific core blocks of several well-known theories. For example, when $G$ belongs to the Deligne--Cvitanovi\'c family of exceptional Lie algebras~\cite{Cvitanovic:2008zz, Deligne} and $G\neq D_4$, the theory $\cT_s $ is a unitary Virasoro minimal model, as summarized in the following table, where we also provide the labels~$(r,s)$ of the operators $\phi_{r,s}$ playing the role of the twist operator in the $\cT_t-{\rm Sym}^2\cT_t$ correspondence:
\begin{center}
\begin{tabular}{lccccccccccc}
\toprule
$G$&&$A_1$&$A_2$&$G_2$&$F_4$&$E_6$&$E_7$&$E_8$\\\midrule
$\cT_s$&&$\cM(3,4)$&$\cM(5,6)$&$\cM(9,10)$&$\cM(10,11)$&$\cM(6,7)$&$\cM(4,5)$&$\cM(3,4)$\\\midrule
$(r,s)$&&$(1,2)$&$(1,2)$&$(1,2)$&$(2,1)$&$(2,1)$&$(2,1)$&$(2,1)$\\\midrule
$c$ && $\frac{1}{2}$ & $\frac{4}{5}$ & $\frac{14}{15}$ & $\frac{52}{55}$ & $\frac{6}{7}$ & $\frac{7}{10}$ & $\frac{1}{2}$ \\
\bottomrule
\end{tabular}
\end{center}
The missing entry for $G=D_4$ would correspond to a CFT at $c=1$ on the sphere and will be discussed shortly.

Note that the field content of (the chiral half of) the coset theory $G_1\times G_1/G_2$ in some cases consists of only a subset of the minimal model field content, and in particular the ``twist operator'' $\phi_{r,s}$ does not necessarily belong to the spectrum of the coset theory. This is the case for example when $G= A_2$, where the field content of the coset theory consists only of the Virasoro primaries $\phi_{1,1},$ $\phi_{1,3},$ $\phi_{1,5}$, $\phi_{2,1}$, $\phi_{2,3}$, $\phi_{2,5}$, and their descendants. The remaining modules, including the one associated to the primary $\phi_{1,2}$, can nevertheless still be realized as twisted modules in the coset theory.

As another example, we noted in Section \ref{sec:c1} that the torus characters for $G=D_N$ with~$N$ even coincide with core blocks of the $\mathbb{Z}_2$ orbifold of the boson at radius $R=\sqrt{2N}$, for the chiral datum $\cD= (\phi_{N/2},\phi_{N/2},\phi_{N/2},\phi_{N/2})$. See~\eqref{eq:equality D_N boson}. This in particular also includes the missing case~${G=D_4}$ in the Deligne--Cvitanovi\'c sequence.

There are a few ways in which the correspondence appears to generalize. First of all, it~is~possible to extend the Deligne--Cvitanovi\'c sequence by including the two \emph{intermediate Lie algebras}~\smash{$A_{\frac 12}$} and \smash{$E_{7+\frac 12}$}, which have dimensions $1$ and $190$ and dual Coxeter numbers $\tfrac{3}{2}$ and $24$, respectively. Mimicking the coset construction for ordinary Lie algebras suggests a relation between the characters of the two corresponding intermediate VOAs~\cite{Kawasetsu} and a pair of RCFTs on the sphere of central charge respectively
\begin{gather*}
\frac{2\cdot 1}{\bigl(\frac{3}{2}+1\bigr)\bigl(\frac{3}{2}+2\bigr)} = \frac{8}{35} \qquad \text{and} \qquad \frac{2\cdot 190}{(24+1)(24+2)} = \frac{38}{65},
\end{gather*}
which are respectively the central charges of the Virasoro minimal models
$\cM(10,7)$, $\cM(13,10)$.
Indeed, for the former minimal model we have found in Section~\ref{sec:a12} that for the choice of chiral datum {$\cD = (\phi_{1,2},\phi_{1,2},\phi_{1,2},\phi_{1,2})$}
 one has
\begin{align*}
2^{\frac{2}{15}}\varphi^{M_{1,1}}_{\phi_{1,2},\phi_{1,2},\phi_{1,2},\phi_{1,2}}(\tau)= \chi^{\mathbf{1}}_{A_{{1\over2},1}}(\tau),\qquad
2^{\frac{2}{15}}\varphi^{M_{1,3}}_{\phi_{1,2},\phi_{1,2},\phi_{1,2},\phi_{1,2}}(\tau)= \chi^{\mathbf{2}}_{A_{{1\over2},1}}(\tau),
\end{align*}
while for the latter minimal model we found in Section~\ref{sec:e712} that for $\cD = (\phi_{2,1},\phi_{2,1},\phi_{2,1},\phi_{2,1})$ one has
\begin{align*}
2^{\frac{38}{15}}\varphi^{M_{1,1}}_{\phi_{2,1},\phi_{2,1},\phi_{2,1},\phi_{2,1}}(\tau)= \chi^{\mathbf{1}}_{E_{7+{1\over2},1}}(\tau),\qquad
2^{\frac{38}{15}}\varphi^{M_{3,1}}_{\phi_{2,1},\phi_{2,1},\phi_{2,1},\phi_{2,1}}(\tau)= \chi^{\mathbf{57}}_{E_{7+{1\over2},1}}(\tau).
\end{align*}
To the best of our knowledge, this is the first, albeit quite indirect, appearance of the intermediate algebra $E_{7+\frac 12}$ in a physical theory!

Another intriguing relation is obtained if one takes the torus theory to be the affine algebra~$G$ belonging to the Deligne--Cvitanovi\'c sequence, at the negative level
\begin{gather*}
k = -\frac{h^\vee_G}{6}-1.
\end{gather*}
Notably, this class of theories, which has been the object of considerable interest in the context of representation theory~\cite{Arakawa:2016hkg,2015arXiv150600710A}, also appears prominently in the context of 4d $\cN=2$ SCFT, where it captures the chiral algebra of the rank-one SCFTs describing a D3 brane probing a~singular $F$-theory background~\cite{ Beem:2013sza,Beem:2019tfp,Beem:2017ooy}.

The characters for this class of CFTs are solutions to a two-dimensional MLDE. In the ca\-ses~\smash{$G=A_{\frac 12},A_1,A_2,G_2$} and $F_4$, both characters transform as modular forms, and admit a~well-defined $q$-series expansion. For \smash{$G=D_4,E_6,E_7,E_{7+\frac 12}$}, and $E_8$, the solution corresponding to the vacuum character admits a~$q$-series expansion, while the other solution includes logarithmic terms, and the two characters transform as a quasi-modular form \cite{KNS}. The appearance of logarithms can be interpreted as resulting from the regularization of the refined character (which also depends on a parameter $\textbf{m}\in\mathfrak{h}^G_{\mathbb{C}}$ valued in the complexification of the Cartan subalgebra of $G$), which in the unrefined limit is divergent~\cite{Beem:2017ooy,Creutzig}.

We find that the characters of $G_{k}$ coincide with certain Virasoro core blocks at central charge
\begin{gather*}
c = 2\frac{\dim (G)\cdot k}{h^\vee_G+k}-\frac{\dim (G)\cdot 2 k}{h^\vee_G+2 k} = \frac{2 \dim (G)k^2}{\bigl(h^\vee_G+k\bigr)\bigl(h^\vee_G+2 k\bigr)},
\end{gather*}
which we compute by using the solution of~\cite{Perlmutter:2015iya} to Zamolodchikov's recurrence formula. Specifically, in the core blocks we take the external operators to have conformal weight $\smash{h_{e} = -\frac{h^\vee_G+1}{8}}$ and the internal operator to have conformal weight $0$ or $\smash{h_p=-\frac{h^\vee_G}{3}}$. For
\begin{gather*}
G \in\big\{ G_2,D_4,F_4,E_6,E_7,E_{7+{1\over2}},E_8\big\},
\end{gather*}
these values are realized in Liouville theory at
\begin{equation}
\label{eq:b}
b=\sqrt{\dfrac{2}{\frac{1}{3}h^\vee_G-1}},
\end{equation}
if we take all external operators to be the degenerate operator $\phi_{1,0}$ and the intermediate primaries to be the vacuum and $\phi_{2,0}$. For $G=A_{{1\over2}},A_1$, $b$ as given in equation~\eqref{eq:b} is imaginary and the core blocks are realized in timelike Liouville theory. Finally, $G=A_2$ corresponds to taking~${c\to\infty}$,~${h_e= -\frac{1}{2}}$ and~$h_p=0 $ or $1$ in the Virasoro core blocks.

For all $G$, the vacuum character $\chi_{G_{k}}^{\mathbf{vac}}(\tau)$ is non-singular, and we find the following relation:
\begin{gather*}
\varphi_{\phi_\ast,\phi_\ast,\phi_\ast,\phi_\ast}^{M(\mathbf{vac})}(\tau) = 2^{\frac{16}{3}h_e} \chi_{G_{k}}^{\mathbf{vac}}(\tau).
\end{gather*}

For \smash{$G = A_{{1\over2}},A_1,A_2,G_2, F_4$}, we also find agreement between the non-vacuum character
\begin{gather*}
\chi_{G_{k}}^{\mathbf{non~vac}}(\tau)
\end{gather*}
 and the other conformal block:
\begin{gather*}
\varphi_{\phi_\ast,\phi_\ast,\phi_\ast,\phi_\ast}^{M(\mathbf{non~vac})}(\tau) = 2^{\frac{16}{3}h_e} \chi_{G_{k}}^{\mathbf{non~vac}}(\tau).
\end{gather*}
For \smash{$G=D_4,E_6,E_7,E_{7+\frac 12}$}, and $E_8$, as discussed above, the $q$-series expansion of the non-va\-cuum~character involves logarithmic terms. Consistent with this, we find that certain coefficients of the solution of~\cite{Perlmutter:2015iya} to Zamolodchikov's recurrence formula have poles at \smash{$h_p=-\frac{h^\vee_G}{3}$} and therefore the $q$-series expansion is ill-defined.
 We provide some details for this sphere-torus relation in~Table~\ref{tab:2nddeligne}.

\begin{table}[t!]
\resizebox{\textwidth}{!}{\begin{tabular}{lrrrrl}
\toprule
\small{Torus VOA}&$b^2$&$c$ & $h_e$ & $h_p$ &\hspace{.8in} $ 2^{-\frac{16}{3}h_e}q^{\frac{2}{3}h_e-\frac{1}{2}h_p}f_{h_e,h_e,h_e,h_e}^{h_p}(\tau)$\\
\midrule
\multirow{2}{*}{$A_{{1\over2},-5/4}$} &\multirow{2}{*}{$-4$}&\multirow{2}{*}{$-\frac{25}{2}$} & \multirow{2}{*}{$-\frac{5}{16}$} & 0 & $1 + q + 3 q^2 + 4 q^3 + 7 q^4 + 10 q^5+\cdots$\\
&&&&$-\frac{1}{2}$& $1 +3 q+4q^2 + 7q^3+13q^4+19q^5+\cdots$\\
 \midrule
\multirow{2}{*}{$A_{1,-4/3} $}&\multirow{2}{*}{$-6$}&\multirow{2}{*}{$-24$} & \multirow{2}{*}{$-\frac{3}{8}$} & 0 & $1 + 3 q + 9 q^2 + 19 q^3 + 42 q^4 + 81 q^5+\cdots$\\
& && & $-\frac{2}{3}$ & $1 + 8q+17q^2 + 46q^3+98q^4+198q^5+\cdots$\\
\midrule
\multirow{2}{*}{$A_{2,-3/2} $}&\multirow{2}{*}{$\infty$}&\multirow{2}{*}{$\infty$} & \multirow{2}{*}{$-\frac{1}{2}$} & $0$ & $1 + 8 q + 36 q^2 + 128 q^3 + 394 q^4 + 1088 q^5 +\cdots$\\
& && & $-1$ & $1 +28q+ 134q^2 + 568q^3+1809q^4+5316 q^5+\cdots$\\
\midrule
\multirow{2}{*}{$G_{2,-5/3} $}&\multirow{2}{*}{$6$}&\multirow{2}{*}{$50$} & \multirow{2}{*}{$-\frac{5}{8}$} & $0$ & $1 + 14 q + 92 q^2 + 456 q^3 + 1848 q^4 + 6580 q^5+\cdots$\\
& && & $-\frac{4}{3}$ & $1 + 78 q + 729 q^2 + 4382 q^3+19917q^4+77274q^5+\cdots$\\
\midrule
\multirow{2}{*}{$D_{4,-2} $}&\multirow{2}{*}{$2$}&\multirow{2}{*}{$28$} & \multirow{2}{*}{$-\frac{7}{8}$} & $0$ & $1 + 28 q + 329 q^2 + 2632 q^3 + 16380 q^4 + 85764 q^5+\cdots$\\
& && & $-2$ & $-\,$\\
\midrule
\multirow{2}{*}{$F_{4,-5/2} $}&\multirow{2}{*}{$1$}&\multirow{2}{*}{$25$} & \multirow{2}{*}{$-\frac{5}{4}$} & $0$ & $1 + 52 q + 1106 q^2 + 14808 q^3 + 147239 q^4 + 1183780 q^5+\cdots$\\
& && & $-3$ & $1\! -\!272 q\!-\!34696q^2\!-\!1058368q^3\!-\!17332196q^4\!-\!197239456q^5\!+\!\cdots$\\
\midrule
\multirow{2}{*}{$E_{6,-3} $}&\multirow{2}{*}{$\frac{2}{3}$}&\multirow{2}{*}{$26$} & \multirow{2}{*}{$-\frac{13}{8}$} & $0$ & $1 + 78 q + 2509 q^2 + 49270 q^3 + 698425 q^4 + 7815106 q^5 +\cdots$\\
& && & $-4$ & $-\,$\\
\midrule
\multirow{2}{*}{$E_{7,-4} $}&\multirow{2}{*}{$\frac{2}{5}$}&\multirow{2}{*}{$\frac{152}{5}$} & \multirow{2}{*}{$-\frac{19}{8}$} & $0$ & $1 \!+\! 133 q\! +\! 7505 q^2\! +\! 254885 q^3 + 6093490 q^4 + 112077998 q^5+\cdots$\\
& && & $-6$ &$-\,$\\
\midrule
\multirow{2}{*}{$E_{7+{1\over2},-5} $}&\multirow{2}{*}{$\frac{2}{7}$}&\multirow{2}{*}{$\frac{250}{7} $}& \multirow{2}{*}{$-\frac{25}{8}$} & $0$ & $1 \!+\! 190 q \!+\! 15695 q^2 \!+\! 783010 q^3 \!+\! 27319455 q^4 \!+\! 725679750 q^5\!+\!\cdots$\\
& && & $-8$ & $-\,$\\
 \midrule
\multirow{2}{*}{$E_{8,-6} $}&\multirow{2}{*}{$\frac{2}{9}$}&\multirow{2}{*}{$\frac{124}{3}$} & \multirow{2}{*}{$-\frac{31}{8}$} & $0$ & ${1\! +\! 248 q\! +\! 27249 q^2\! +\! 1821002 q^3\! +\! 85118126 q^4\!+\! 3017931282 q^5\!+\!\cdots}$\\
& && & $-10$ & $-\,$\\
\bottomrule
\end{tabular}}
\caption{Correspondence between the characters of Kac--Moody algebras at level $k=-\frac{h^\vee_G}{6}-1$ and Virasoro core blocks, normalized here so that the leading order term is $1$. Note that for $G= \smash{D_4,E_6,E_7,E_{7+{1\over2}}},E_8$ the $q$-expansion of the Virasoro blocks corresponding to the non-vacuum character is ill-defined.}
\label{tab:2nddeligne}
\end{table}

Finally, additional relations appear to exist between the level-1 characters of affine Kac--Moody algebras and core blocks of minimal models with three identical external operators. For instance, corresponding to the chiral datum $\cD = (\phi_{2,1},\phi_{2,3},\phi_{2,3},\phi_{2,3})$ in the $\cM(4,5)$ minimal model are two core blocks associated to the internal channels $M_{2,1}$ and $M_{2,3}$. It turns out that these core blocks are again proportional to the characters of $A_{1,1}$:
\begin{align*}
\bem
\varphi^{M_{2,1}}_{\phi_{2,1},\phi_{2,3},\phi_{2,3},\phi_{2,3}}\\[1.5mm]
\varphi^{M_{2,3}}_{\phi_{2,1},\phi_{2,3},\phi_{2,3},\phi_{2,3}} \eem (\tau)= N \bem \chi^{\mathbf{3}}_{A_{1,1}} \\[1mm] \chi^{\mathbf{1}}_{A_{1,1}}\eem (\tau)
\end{align*}
for some $N\in \CC$.
These additional relations will also be addressed in more detail in the companion paper~\cite{followup}.

We conclude this section with a few additional remarks on the physical interpretation of the sphere-torus correspondence. Given an affine algebra $G_1$ at level 1, whose central charge~is given~by \smash{$c = \frac{\dim G}{h^\vee_G+1}$}, one can construct the following two distinct VOAs: the symmetric product,~${\rm Sym}^2 G_1$, and the diagonal coset \smash{$\frac{G_1\times G_1}{G_2}$}. Both theories contain twist operators $\eta$ of the same conformal dimension $h_\eta = \tfrac{c}{16}$, and these turn out to have identical fusion rules
\begin{gather*}
\eta\times \eta = \sum_i \phi_i
\end{gather*}
in the two theories, involving primaries $\phi_i$ of conformal dimension $h_i = 2 h^G_i$, where $h^G_i$ are the dimensions of certain $G_1$ primaries. This observation is sufficient, given the low dimensionality of the ${\rm SL}(2,\mathbb{Z})$ representations involved, to guarantee that the four-point core blocks with four insertions of $\eta$ are identical in the two theories ${\rm Sym}^2 G_1$ and $\frac{G_1\times G_1}{G_2}$, although their interpretation in the two theories is different since they are defined with respect to the respective extended chiral algebras, which are distinct. In turn, the ${\rm Sym}^2 G_1$ core blocks coincide with $G_1$ characters by virtue of the relation~\eqref{eq:stcharold}, which has a clear physical interpretation. At present, however, it remains unclear to us whether a direct physical interpretation of the sphere-torus correspondence can be given, and whether the correspondence generalizes to further classes of examples.

\appendix

\section{MLDE and BPZ}
\label{app:BPZ}
The components $f^1(\tau),\dots,f^n(\tau)$ of a weight-zero $\PSL_2(\mathbb{Z})$ vector-valued modular form are necessarily solutions to a degree-$n$ modular differential equation (MLDE), that is, an equation which takes the form
\begin{gather*}
\bigl(D_{(0)}^n+g_{n-1}D_{(0)}^{n-1}+\cdots+g_{1}D_{(0)}+g_0\bigr) f=0
\end{gather*}
using notation explained in section 6 (cf.\ \eqref{def:differential operator}). The coefficient $g_k(\tau)$ (namely $b_k/b_n$ in the notation of Section~\ref{section6.3.3}) is a meromorphic modular form of weight $2(n-k)$ for $\PSL_2(\mathbb{Z})$. When the Wronskian $b_d(\tau)$ is a power of the $\eta$-function \cite{Ma}, then the $g_k$ will be holomorphic and hence could be expressed as polynomials in $E_4(\tau)$, $E_6(\tau)$. We will restrict to this case in the following.

Modular differential equations have been applied to the context of RCFT since the 1980s~\cite{Anderson:1987ge}; in particular, characters of a RCFT are solutions to MLDEs whose $q$-expansion coefficients are positive integers, and this strategy was used early on to classify unitary RCFTs with one or two characters~\cite{Mathur:1988na,Naculich:1988xv} -- see also~\cite{Chandra:2018pjq} and references therein for more recent developments.\looseness=1

Order 2 MLDEs in particular depend on a single parameter $\mu$:
\begin{equation}
\biggl\{D^2-\frac{E_2(\tau)}{6}D-\frac{\mu}{4} E_4(\tau)\biggr\}f(\tau) = 0.\label{eq:MLDE2}
\end{equation}
where $D = \frac{1}{2\pi {\rm i}}{{\rm d}\over {\rm d}\tau}$. Two solutions to this MLDE can in general be expressed in terms of hypergeometric functions~\cite{Mathur:1988gt}:
\begin{gather}
f_1(\tau) = \bigl(2^{-4}\lambda(\tau)(1-\lambda(\tau))\bigr)^{\frac{1-x}{6}}{}_2F_1\biggl(\frac{1}{2}-\frac{x}{6},\frac{1}{2}-\frac{x}{2},1-\frac{x}{3},\lambda(\tau)\biggr) = q^{\frac{1-x}{12}}(1+\ccO(q)),\!\label{eq:MLDEs}
\\
f_2(\tau) = N \bigl(2^{-4}\lambda(\tau)(1-\lambda(\tau))\bigr)^{\frac{1+x}{6}}{}_2F_1\biggl(\frac{1}{2}+\frac{x}{6},\frac{1}{2}+\frac{x}{2},1+\frac{x}{3},\lambda(\tau)\biggr) = N q^{\frac{1+x}{12}}(1+\ccO(q)), \nonumber
\end{gather}
where $x = \sqrt{1+36\mu}$ and
\begin{gather*}
N = 2^{x}\sqrt{\frac{\Gamma\bigl(1-\frac{x}{3}\bigr)\Gamma\bigl(-\frac{x}{6}\bigr)\Gamma\bigl(\frac{1}{2}+\frac{x}{2}\bigr)}{\Gamma\bigl(1+\frac{x}{3}\bigr)\Gamma\bigl(\frac{x}{6}\bigr)\Gamma\bigl(\frac{1}{2}-\frac{x}{2}\bigr)}}.
\end{gather*}
The normalization is such that the modular $S$ matrix is symmetric:
\begin{gather}
\mathbf{S}
=
\begin{pmatrix}
\!\dfrac{\Gamma\bigl(1\!-\!\frac{x}{3}\bigr)\Gamma\bigl(\frac{x}{3}\bigr)}{\Gamma\bigl(\frac{1}{2}\!-\!\frac{x}{6}\bigr)\Gamma\bigl(\frac{1}{2}\!+\!\frac{x}{6}\bigr)}\!\! & \!\!\!\!\sqrt{\dfrac{\Gamma\bigl(1\!+\!\frac{x}{3}\bigr)\Gamma\bigl(\frac{x}{3}\bigr)\Gamma\bigl(1\!-\!\frac{x}{3}\bigr)\Gamma\bigl(-\frac{x}{3}\bigr)}{\Gamma\bigl(\frac{1}{2}\!+\!\frac{x}{6}\bigr)\Gamma\bigl(\frac{1}{2}\!+\!\frac{x}{2}\bigr)\Gamma\bigl(\frac{1}{2}\!-\!\frac{x}{6}\bigr)\Gamma\bigl(\frac{1}{2}\!-\!\frac{x}{2}\bigr)}} \vspace{2mm}\\
\!\!\sqrt{\dfrac{\Gamma\bigl(1\!+\!\frac{x}{3}\bigr)\Gamma\bigl(\frac{x}{3}\bigr)\Gamma\bigl(1\!-\!\frac{x}{3}\bigr)\Gamma\bigl(-\frac{x}{3}\bigr)}{\Gamma\bigl(\frac{1}{2}\!+\!\frac{x}{6}\bigr)\Gamma\bigl(\frac{1}{2}\!+\!\frac{x}{2}\bigr)\Gamma\bigl(\frac{1}{2}\!-\!\frac{x}{6}\bigr)\Gamma\bigl(\frac{1}{2}\!-\!\frac{x}{2}\bigr)}}	&	\dfrac{\Gamma\bigl(1\!+\!\frac{x}{3}\bigr)\Gamma\bigl(-\frac{x}{3}\bigr)}{\Gamma\bigl(\frac{1}{2}\!-\!\frac{x}{6}\bigr)\Gamma\bigl(\frac{1}{2}\!+\!\frac{x}{6}\bigr)}
\end{pmatrix}\!
.
\label{eq:SMLDE}
\end{gather}

The values of $\mu$ for which solutions to equation~\eqref{eq:MLDE2} have a $q$-expansion with positive integer coefficients were determined in~\cite{Mathur:1988na}; the solutions to the MLDE equation in these cases are the characters of a sequence of RCFTs labeled by a Lie algebra $G$ belonging to the Deligne--Cvitanovi\'c sequence:
\begin{center}
\begin{tabular}{lccccccccccc}
\toprule
$\mu$&\hspace{.5in}&$\frac{11}{900}$&$\frac{5}{144}$&$\frac{1}{12}$&$\frac{119}{900}$&$\frac{2}{9}$&$\frac{299}{900}$&$\frac{5}{12}$&$\frac{77}{144}$&$\frac{551}{900}$&$\frac{2}{3}$\\\midrule
$G$&&$A_{\frac{1}{2}}$&$A_1$&$A_2$&$G_2$&$D_4$&$F_4$&$E_6$&$E_7$&$E_{7\frac{1}{2}}$&$E_8$\\
\bottomrule
\end{tabular}
\end{center}

For $G$ a simple Lie algebra, the RCFT is simply the level-1 $G$ Kac--Moody algebra, while in the cases $G=A_{\frac{1}{2}}$ and $E_{7\frac{1}{2}}$, one finds the characters of the \emph{intermediate vertex subalgebra} associated to the intermediate Lie algebra $G$~\cite{LM2004, Shtepin1994}.

In Section \ref{sec:forms}, we have argued that the core blocks are components of a weight-zero {weakly} holomorphic vector-valued modular form; in particular, they must be solutions of a MLDE. On the other hand, four-point chiral blocks of Virasoro minimal models are known to be solutions to the BPZ differential equation~\cite{Belavin:1984vu}, due to the presence of null states. In what follows, we verify explicitly that the BPZ equation satisfied by the chiral block (in the simplest case where the null descendant occurs at level 2) can be turned into the order-$2$ MLDE for the core block.

Let us recall briefly how the BPZ equation arises. Consider a chiral block between Virasoro primaries $\phi_1(z_1),\phi_2(z_2),\dots$ of conformal weight $h_1,h_2,\dots$, and assume $\phi_1(z_1)$ has a null Virasoro descendant at level $ n $. A generic level-$ n $ descendant of a Virasoro primary operator $ \phi_1(z_1) $ can written in the form
\begin{gather*}
\phi' (z) = \sum_{\substack{Y \\ |Y| = n}}\alpha_{Y} L_{-Y} \phi(z)
\end{gather*}
for $\alpha_Y\in \mathbb{C}$; here,
\begin{gather*} Y = \{r_1,\dots,r_k\},\qquad 1\leq r_1 \leq r_2\leq \dots \leq r_k \end{gather*}
is a Young diagram of size $ n $, and $ L_{-Y} = L_{-r_1}L_{-r_2}\dots L_{-r_k}$. When such a descendant is a null state, any physical correlator involving it vanishes:
\begin{gather*}
\bigl\langle \bigl(\phi^1\bigr)'(z_1,\bar z_1)\phi^2(z_2,\bar z_2)\dots\bigr\rangle = 0,
\end{gather*}
where $\phi^i(z,\bar z)=\phi_i(z)\times\tilde\phi_i(\bar z)$. This implies that physical correlators involving the primary field~$\phi^1(z,\bar z)$ satisfy a holomorphic order $n$ differential equation:
\begin{equation}
\sum_{\substack{Y\\\vert Y \vert = n}} \alpha_{Y}\mathcal{L}_{-Y}(z_1)\langle \phi(z_1,\bar z_1)\phi_1(z_2,\bar z_2)\dots \rangle = 0,\label{eq:BPZ}
\end{equation}
where
\begin{gather*}
\mathcal{L}_{-Y}(z_1) = \mathcal{L}_{-r_1}(z_1)\mathcal{L}_{-r_2}(z_1)\dots\mathcal{L}_{-r_k}(z_1)
\end{gather*}
and
\begin{gather*}
\mathcal{L}_{-r}(z_1) = \sum_{i\geq 2} \biggl(\frac{(r-1)h_i}{z_{i1}^r}-\frac{1}{z_{i1}^{r-1}}\partial_{z_i}\biggr).
\end{gather*}
Let us from now on focus on four-point blocks. The BPZ equation~\eqref{eq:BPZ} then translates to the following differential equation for the chiral blocks:
\begin{equation}
\Biggl[\partial_{z_1}^2-t \sum_{i=2}^4 \biggl(\frac{h_i}{z_{i1}^2}-\frac{1}{z_{i1}}\partial_{z_i}\biggr)\Biggr]\mathcal{F}^{P}_{\phi_1,\phi_2,\phi_3,\phi_4}(z_1,z_2,z_3,z_4) = 0,
\label{eq:BPZ2}
\end{equation}
\newline
where $P$ is an internal channel. We claim that the BPZ equation lifts to a modular differential equation for the core of the conformal block. Let us explicitly verify this in the case where the null state is at level $2$. Such a null state is given by $ \phi_1'(z_1) = \bigl(L_{-1}^2-t L_{-2}\bigr)\phi(z_1) $, where
\begin{gather*}
t=\frac{4}{3}h_1+\frac{2}{3}.
\end{gather*}
We now express the holomorphic cross ratio $w$ as $w = \lambda(\tau)$ in terms of the modular parameter on the torus. The relation between conformal and core block is given by
\begin{equation} \mathcal{F}^{P}_{\phi_1,\phi_2,\phi_3,\phi_4}(z_1,z_2,z_3,z_4) = \varphi^{P}_{\phi_1,\phi_2,\phi_3,\phi_4}(w)\prod_{1\le j< k\le 4}z_{jk}^{\mu_{jk}}, \label{eq:blocktocore} \end{equation}
where $ \mu_{jk} $ is defined in equation~\eqref{eq:muij}. To obtain the differential equation for the core block, we express the differential operator appearing in equation~\eqref{eq:BPZ2} in terms of derivatives with respect to the cross ratio $ w = {z_{12}z_{34} \over z_{13}z_{24}}$. Thus, for example
\begin{gather*}
 \partial_{z_1} = \partial_{z_1}( w) \partial_{ w} +\frac{\mu_{1 2}}{{z_{12}}}+\frac{\mu_{1 3}}{{z_{13}}}+\frac{\mu_{1 4}}{{z_{14}}},
\end{gather*}
where
\begin{gather*}
\partial_{z_1}( w) = \frac{z_{23}z_{34}}{z_{13}^2z_{24}},
\end{gather*}
and similarly for $ \partial_{z_2}( w)$, $\partial_{z_3}( w)$, $\partial_{z_4}( w) $. Likewise, one finds
\begin{align*} \partial_{z_1}^2 &{}= -\biggl(\frac{\mu_{12}}{z_{12}^2}+\frac{\mu_{13}}{z_{13}^2}+\frac{\mu_{14}}{z_{14}^2}\biggr)+\biggl(\frac{\mu_{12}}{z_{12}}+\frac{\mu_{13}}{z_{13}}+\frac{\mu_{14}}{z_{14}}\biggr)^2\nonumber\\
&\quad{}+ 2\biggl(\frac{\mu_{12}}{z_{12}}+\frac{\mu_{13}}{z_{13}}+\frac{\mu_{14}}{z_{14}}\biggr)\partial_{z_1}( w)\partial_{ w}+\partial_{z_1}^2( w)\partial_{ w}+\partial_{z_1}( w)^2\partial_{ w}^2.
\end{align*}
Taking the limit $ (z_{1},z_2,z_3,z_4) \to (\infty,1, w,0)$ and using relation~\eqref{eq:blocktocore}, we obtain the following differential equation for the core block:
\begin{gather}
\biggl\{{{\rm d}^2 \over {\rm d}w^2}+\biggl[2\biggl(\frac{\mu_{12}}{ w}+\frac{\mu_{14}}{ w-1}\biggr)+t\frac{2 w-1}{ w( w-1)}\biggr]{{\rm d} \over {\rm d}w}+\biggl[\biggl(\frac{\mu_{12}}{ w}+\frac{\mu_{14}}{ w-1}\biggr)^2-\frac{\mu_{12}}{{ w}^2}-\frac{\mu_{14}}{( w-1)^2}\biggr]\nonumber\\
\qquad{}+t\biggl(\frac{\mu_{12}-h_2}{{ w}^2}+\frac{\mu_{14}-h_4}{( w-1)^2}-\frac{\mu_{24}}{ w({ w}-1)}\biggr)\biggr\} \varphi^{P}_{\phi_1,\phi_2,\phi_3,\phi_4}( w)=0.\label{eq:coreBPZ}
\end{gather}
Next, we use $ w = \lambda(\tau) = \bigl(\frac{\theta_2(\tau)}{\theta_3(\tau)}\bigr)^4$ to rewrite equation~\eqref{eq:coreBPZ} in terms of the variable $ q=\mathrm{e}(\tau)$. In particular, we find
\begin{gather*}
{{\rm d} \over {\rm d}w}= \frac{2}{\lambda(\tau)(1-\lambda(\tau))\theta_3(\tau)^4}D,
 \end{gather*}
and
\begin{gather*}
{{\rm d}^2 \over {\rm d}w^2} = \frac{4}{\lambda(\tau)^2(1-\lambda(\tau))^2\theta_3(\tau)^8}\biggl\{D^2+\biggl(\frac{2\lambda(\tau)-1}{2}\theta_3(\tau)^4-\frac{D\bigl(\theta_3(\tau)^4\bigr)}{\theta_3(\tau)^4}\biggr)D\biggr\}.
\end{gather*}
Making also use of the identities
\begin{gather*}
\bigl(\lambda(\tau)^2-\lambda(\tau)+1\bigr)\theta_3(\tau)^8= E_4(\tau),\\
\frac{2\lambda(\tau)-1}{6}\theta_3(\tau)^4-\frac{D\bigl(\theta_3(\tau)^4\bigr)}{\theta_3(\tau)^4} = -\frac{1}{6} E_2(\tau),
\end{gather*}
we can rewrite equation~\eqref{eq:coreBPZ} as
\begin{gather*}
\Biggl\{D^2+\biggl[-\frac{E_2(\tau)}{6}+\biggl(\theta_2(\tau)^4\frac{h_4-h_3}{3}+\theta_3(\tau)^4\frac{h_4-h_2}{3}+\theta_4(\tau)^4\frac{h_3-h_2}{3}\biggr)\biggr]D\\
\qquad{}-\frac{1}{18}E_4(\tau)(2h_1 (h-1)+h)-\frac{1}{36}\sum_{i=2}^4\Lambda_i\theta_i(\tau)^8\Biggr\} \varphi^{P}_{\phi_1,\phi_2,\phi_3,\phi_4}(\lambda(\tau))=0,
\end{gather*}
where
\begin{gather*}
\Lambda_i = (4h_1+1)(3\mu_{1i}+2h_1)+\prod_{\substack{j=2\\j\neq i}}^4(3\mu_{1j}+2h_1).
\end{gather*}
Let us now further specialize to the case $ \phi_2=\phi_3=\phi_4$ with {$h_2=h_3=h_4=h_\ast$}, corresponding to $ \PSL_2(\mathbb{Z}) $ symmetry. Note that in this case $\Lambda_i = 0$, so we obtain
\begin{gather*}
\biggl\{ D^2-\frac{E_2(\tau)}{6}D-\frac{\mu(h_1,h_*)}{4}E_4(\tau)\biggr\}\varphi^{P}_{\phi_1,\phi_2,\phi_2,\phi_2}(\lambda(\tau)) = 0,
\end{gather*}
which takes precisely the form of the modular differential equation~\eqref{eq:MLDE2}, with
\begin{gather*}
\mu(h_1,h_*)= \frac{2}{9}\bigl(2h_1^2+6h_1h_*+3h_*-h_1\bigr).
\end{gather*}
This establishes that the second-order BPZ equations for four-point chiral blocks are equivalent to modular linear differential equations satisfied by the core blocks.
It should be straightforward to establish analogous relations for the higher-order cases.

\section{Braiding and fusing}\label{sec:brafu}

In this appendix, we collect a few useful properties of the braiding and fusing operators (the latter are also called 6j-symbols)
\begin{gather*}
\begin{split}
&B\left[{\begin{matrix}N_2&N_3\\ N_1&N_4\end{matrix}}\right]\!(\varepsilon)\colon \ \bigoplus_P \mathcal{Y}^{N_1}_{N_2,P}\otimes \mathcal{Y}^{P}_{N_3,N_4}\longrightarrow \bigoplus_Q\mathcal{Y}^{N_1}_{N_3,Q}\otimes \mathcal{Y}^{Q}_{N_2,N_4},\\
&F\left[{\begin{matrix}N_2&N_3\\ N_1&N_4\end{matrix}}\right]\!\colon \ \bigoplus_P\mathcal{Y}^{N_1}_{N_2,P}\otimes \mathcal{Y}^{P}_{N_3,N_4}\longrightarrow \bigoplus_Q\mathcal{Y}^{N_1}_{Q,N_4}\otimes \mathcal{Y}^{Q}_{N_2,N_3},
\end{split}
\end{gather*}
from the classic papers~\cite{Moore:1988qv,Moore:1989vd}, to which we refer for details. The choice of sign $\varepsilon=\pm$ is explained in Section \ref{sec:bkg}, and admits a natural generalization to any $\varepsilon\in\mathbb{Z}$. Picking bases \smash{$\Phi^{N_1,a}_{N_2,N_3}$}, for the spaces \smash{$\mathcal{Y}^{N_1}_{N_2,N_3}$}, we write the action of braiding and fusing as
\begin{gather*}
\Phi^{N_1,a}_{N_2,P}\Phi^{N_1,b}_{N_2,P}=\sum_Q\sum_{c,d}B_{P,Q}\begin{bmatrix}N_2&N_3\\ N_1&N_4\end{bmatrix}^{a,b}_{c,d}\!(\varepsilon) \Phi^{N_1,c}_{N_3,Q}\Phi^{Q,d}_{N_2,N_4},\\
\Phi^{N_1,a}_{N_2,P}\Phi^{N_1,b}_{N_2,P}=\sum_Q\sum_{c,d}F_{PQ}\begin{bmatrix}N_2&N_3\\ N_1&N_4\end{bmatrix}^{a,b}_{c,d} {\Phi^{N_1,c}_{Q,N_4}\ \Phi^{Q,d}_{N_2,N_3}}.
\end{gather*}
Exchange of $N_2$, $N_3$ in the intertwiner \smash{$\Phi^{N_1}_{N_2,N_3}$} defines an isomorphism, which is the skew symmetry relation:
\begin{equation} \label{def:xi}
\xi^{N_1}_{N_2,N_3}\colon\ \mathcal{Y}^{N_1}_{N_2,N_3}\to \mathcal{Y}^{N_1}_{N_3,N_2},
\end{equation}
which squares to the identity operator. We write its action on the basis \smash{$\Phi^{N_1,a}_{N_2,N_3}$} as
\begin{equation*}
\Phi^{N_1,a}_{N_2,N_3}= \sum_b \bigl(\xi^{N_1}_{N_2,N_3}\bigr)^a_b \Phi^{N_1,b}_{N_3,N_2}.
\end{equation*}
Similarly, exchange of $N_1$, $N_2$ defines an isomorphism which sends an intertwiner to its adjoint:%
\begin{equation} \label{def:zeta}
\zeta^{N_1}_{N_2,N_3}\colon \ \mathcal{Y}^{N_1}_{N_2,N_3}\to \mathcal{Y}^{N_2^*}_{N_1^*,N_3},
\end{equation}
and we write its action on the basis \smash{$\Phi^{N_1,a}_{N_2,N_3}$} as
\begin{gather*}
\Phi^{N_1,a}_{N_2,N_3}= \sum_b \bigl(\zeta^{N_1}_{N_2,N_3}\bigr)^a_b \Phi^{N_2^*,b}_{N_1^*,N_3}.
\end{gather*}
These maps are carefully studied in~\cite[Section 7]{HL:1995}.

It is clear that the matrices $\zeta$ and $\xi$ are involutions in the sense that
\begin{equation}\label{ord2zetaxi}
\xi_{N_i,N_j}^{N_k}\xi_{N_j,N_i}^{N_k} = \zeta_{N_i,N_j}^{N_k}\zeta_{N_k^*,N_j}^{N_i^*} = I.
\end{equation}
When \smash{$\cN^{N_1}_{N_2,N_3}=1$}, as will be the case in the main text, $\zeta$ and $\xi$ are simply phases.

When one of the $N_i$ is the vacuum module one obtains simple expressions for the braiding matrices:
\begin{gather}
B_{Q R}\begin{bmatrix} N_1&N_2\\ \cV&P\end{bmatrix}^{\cdot,b}_{\cdot,d}\!(+) = \bigl(\zeta_{N_2,P}^{N_1^*}\bigr)^{b}_{d} \delta_{N_1^*, Q}{\delta_{N_2^*,R}}\,\mathrm{e}\biggl(\frac{h_{N_1}+h_{N_2}-h_P}{2}\biggr),\nonumber\\
B_{Q R}\begin{bmatrix} N_3&N_4\\ P&\cV\end{bmatrix}^{a,\cdot}_{c,\cdot}\!(+) = \bigl(\xi_{N_3,N_4}^{P}\bigr)^a_c\delta_{N_4, Q}{\delta_{N_3, R}}\,\mathrm{e}\biggl(\frac{h_{N_3}+h_{N_4}-h_P}{2}\biggr).\label{eq:B}
\end{gather}
One can further show that braiding and {fusing} are related as follows:
\begin{gather}
B_{PQ}\begin{bmatrix}N_2&N_3\\N_1&N_4\end{bmatrix}^{a,b}_{c,d}(\varepsilon)\nonumber\\
\qquad{}=\sum_{e,f}\bigl(\xi_{N_3,N_4}^P\bigr)^b_e\bigl(\xi_{Q,N_3}^{N_1}\bigr)^f_c\,\mathrm{e} \biggl(\frac{\varepsilon}{2}(h_{P}+h_{Q}-h_{N_1}-h_{N_4})\biggr)F_{PQ}\begin{bmatrix}N_2&N_4\\N_1&N_3\end{bmatrix}^{a,e}_{f,d}.\label{eq:BF}
\end{gather}
As a consequence,
\begin{gather}
\sum_Q\sum_{c,d} B_{PQ} \begin{bmatrix}N_2&N_3\\N_1&N_4\end{bmatrix}^{a,b}_{c,d} (\varepsilon)\,\mathrm{e} \biggl(\varepsilon\biggl(h_{N_1}+h_{N_4} -\frac{h_{P}}{2}-h_{Q}-\frac{h_{R}}{2}\biggr)\biggr) B_{QR} \begin{bmatrix}N_3&N_2\\N_1&N_4\end{bmatrix}^{c,d}_{e,f} (\varepsilon)\nonumber \\
\qquad{}={\delta_{P, R}}\delta^{a}_{e}\delta^{b}_f\label{eq:BB}
\end{gather}
and
\begin{gather*}
\sum_Q{\sum_{b,c,d,e,h,i}} F_{PQ}\begin{bmatrix}N_2&N_3\\N_1&N_4\end{bmatrix}^{a,b}_{c,d}F_{QR}\begin{bmatrix}N_3&N_4\\N_1&N_2\end{bmatrix}^{h,i}_{e,f} \bigl(\xi_{N_4,N_3}^P\bigr)^g_b\bigl(\xi_{Q,N_4}^{N_1}\bigr)^c_h\bigl(\xi_{N_4,N_2}^{Q}\bigr)^d_i\bigl(\xi_{R,N_4}^{N_1}\bigr)^e_j\nonumber\\
\qquad{}={\delta_{P,R}}\delta^{a}_{j}\delta^g_f.
\end{gather*}

\section{Modular group representations for general cases}
\label{app:generalSL2Z}

According to Theorem \ref{thm:sl2z}, in cases (B) respectively (C) the vector of core blocks \raisebox{1pt}{\smash{$\underline{\var}_{\phi_1,\phi_2,\phi_3,\phi_3}(\widetilde{w})$}} in~\eqref{eq:phiA} only transforms with respect to {${\rm P}\Gamma_0(2)$} respectively {${\rm P}\Gamma(2)$}. At the cost of increasing the dimensions, we can repackage the core blocks in those cases so that they too transform with respect to ${\rm PSL}_2(\mathbb{Z})$. In representation theory terminology, this corresponds to inducing the representation from the subgroups ${\rm P}\Gamma_0(2)$ and ${\rm P}\Gamma(2)$ up to ${\rm PSL}_2(\mathbb{Z})$. In practice though, for these cases it is usually easier to work with the core blocks as vvmf's for ${\rm P}\Gamma_0(2)$ or ${\rm P}\Gamma(2)$, {though both options are given} in Theorem \ref{thm:sl2z}.
\begin{itemize}\itemsep=0pt
\item[(B)] $\cD=(M_1,M_2,M_3,M_3)$, with $\phi_1\in M_1$, $\phi_2 \in M_2$, and $ \phi_3=\phi_4\in M_3$.

We can also arrange the core blocks in a vector as follows:
\begin{equation}
\label{eq:phiB}
\underline{\varphi}_{B;\phi_1,\phi_2,\phi_3}=
\begin{pmatrix}
\var_{\phi_1,\phi_2,\phi_3,\phi_3}^{P_{1}}\\
\vdots\\
\var_{\phi_1,\phi_2,\phi_3,\phi_3}^{P_{i_P}}\\
\var_{\phi_1,\phi_3,\phi_2,\phi_3}^{Q_{1}}\\
\vdots\\
\var_{\phi_1,\phi_3,\phi_2,\phi_3}^{Q_{i_Q}}\\
\var_{\phi_1,\phi_3,\phi_3,\phi_2}^{R_{1}}\\
\vdots\\
\var_{\phi_1,\phi_3,\phi_3,\phi_2}^{R_{i_R}}
\end{pmatrix},
\end{equation}
where $P_{i}, Q_{i},R_{i}\in \Upsilon(\cV)$. We also define the following matrices, written in block form
\begin{gather*}
\cT_{B} =
\begin{pmatrix}
T\bigl[\begin{smallmatrix}2&3\\1 &3\end{smallmatrix}\bigr]&0&0\\
0&0&T\bigl[\begin{smallmatrix}3&2\\1 &3\end{smallmatrix}\bigr]\\
0&T\bigl[\begin{smallmatrix}3&3\\1 &2\end{smallmatrix}\bigr]&0
\end{pmatrix},
\qquad
\cS_{B}=
\begin{pmatrix}
0 & 0 & S\bigl[\begin{smallmatrix}2&3\\1 &3\end{smallmatrix}\bigr]\\
0&S\bigl[\begin{smallmatrix}3&2\\1 &3\end{smallmatrix}\bigr]&0\\
S\bigl[\begin{smallmatrix}3&3\\1 &2\end{smallmatrix}\bigr]&0&0
\end{pmatrix}.
\end{gather*}
Then \raisebox{1.5pt}{\smash{$\underline{\var}_{B;\phi_1,\phi_2,\phi_3}$}} has three times the number of rows of \smash{$\underline{\var}_{\phi_1,\phi_2,\phi_3,\phi_3}$}, and transforms with respect to ${\rm PSL}_2(\mathbb{Z})$.

\item[(C)] $\cD=(M_1,M_2,M_3,M_4)$ with the $M_i$ arbitrary (e.g., some may be equal), but with all states $\phi_i\in M_i$ distinct.
We can arrange the core blocks corresponding to various permutations of the external operators as follows:
\begin{equation}
\label{eq:phiC}
\underline{\varphi}_{C;\phi_1,\phi_2,\phi_3,\phi_4}=
\begin{pmatrix}
\var_{\phi_1,\phi_2,\phi_3,\phi_4}^{P_{1}}\\
\vdots\\
\var_{\phi_1,\phi_2,\phi_3,\phi_4}^{P_{i_P}}\\
\var_{\phi_1,\phi_2,\phi_4,\phi_3}^{Q_{1}}\\
\vdots\\
\var_{\phi_1,\phi_2,\phi_4,\phi_3}^{Q_{i_Q}}\\
\var_{\phi_1,\phi_3,\phi_2,\phi_4}^{R_{1}}\\
\vdots\\
\var_{\phi_1,\phi_3,\phi_2,\phi_4}^{R_{i_R}}\\
\var_{\phi_1,\phi_3,\phi_4,\phi_2}^{U_{1}}\\
\vdots\\
\var_{\phi_1,\phi_3,\phi_4,\phi_2}^{U_{i_U}}\\
\var_{\phi_1,\phi_4,\phi_2,\phi_3}^{V_{1}}\\
\vdots\\
\var_{\phi_1,\phi_4,\phi_2,\phi_3}^{V_{i_V}}\\
\var_{\phi_1,\phi_4,\phi_3,\phi_2}^{W_{1}}\\
\vdots\\
\var_{\phi_1,\phi_4,\phi_3,\phi_2}^{W_{i_W}}
\end{pmatrix}
\end{equation}
and define
\begin{gather*}
\cT_{C}=
\begin{pmatrix}
0& T\bigl[\begin{smallmatrix}2&3\\1 &4\end{smallmatrix}\bigr]&0&0&0&0\\
T\bigl[\begin{smallmatrix}2&4\\1 &3\end{smallmatrix}\bigr]&0&0 & 0 &0&0\\
0 & 0 &0& T\bigl[\begin{smallmatrix}3&2\\1 &4\end{smallmatrix}\bigr]&0&0 \\
0 & 0 & T\bigl[\begin{smallmatrix}3&4\\1 &2\end{smallmatrix}\bigr]&0&0&0\\
0 & 0 &0&0&0& T\bigl[\begin{smallmatrix}4&2\\1 &3\end{smallmatrix}\bigr] \\
0 & 0 &0&0& T\bigl[\begin{smallmatrix}4&3\\1 &2\end{smallmatrix}\bigr]&0\\
\end{pmatrix},
\\ 
\cS_{C}=
\begin{pmatrix}
0 & 0 &0&0&0& S\bigl[\begin{smallmatrix}2&3\\1 &4\end{smallmatrix}\bigr] \\
 0 & 0 &0& S\bigl[\begin{smallmatrix}2&4\\1 &3\end{smallmatrix}\bigr]&0&0 \\
 0 & 0 &0&0& S\bigl[\begin{smallmatrix}3&2\\1 &4\end{smallmatrix}\bigr]&0\\
0& S\bigl[\begin{smallmatrix}3&4\\1 &2\end{smallmatrix}\bigr]&0&0&0&0\\
0 & 0 & S\bigl[\begin{smallmatrix}4&2\\1 &3\end{smallmatrix}\bigr]&0&0&0\\
S\bigl[\begin{smallmatrix}4&3\\1 &2\end{smallmatrix}\bigr]&0&0 & 0 &0&0
\end{pmatrix}.
\end{gather*}
Then \raisebox{1.5pt}{\smash{$\underline{\var}_{C;\phi_1,\phi_2,\phi_3,\phi_4}$}} has six times {the number of rows of \smash{$\underline{\var}_{\phi_1,\phi_2,\phi_3,\phi_4}$}, and transforms with respect to ${\rm PSL}_2(\mathbb{Z})$.}
\end{itemize}

\subsection*{Acknowledgements}

We thank Vassilis Anagiannis, Christopher Beem, Francesca Ferrari, Alex Maloney, Greg Moore, and Gim Seng Ng for useful conversations. We are also grateful to the anonymous referees for their helpful suggestions. This project has received funding from the European Union’s Horizon 2020 research and innovation programme under the Marie Sklodowska-Curie grant agreement No 708045. The work of M.C.\ and G.L.\ is supported by ERC starting grant H2020 \#640159. The work of M.C.\ has also received support from NWO vidi grant (number 016.Vidi.189.182).
The work of T.G.\ is supported by an NSERC Discovery grant.

\pdfbookmark[1]{References}{ref}
\LastPageEnding

\end{document}